\documentclass[11pt]{article}%
\usepackage{amsfonts}
\usepackage{amsmath}
\usepackage{amssymb}
\usepackage{amsfonts,setspace}
\usepackage{caption}
\usepackage[caption=false]{subfig}
\usepackage{epsfig,latexsym,graphicx}
\usepackage{appendix}
\usepackage{graphicx}%
\usepackage[numbers, sort, compress]{natbib}
\providecommand{\U}[1]{\protect\rule{.1in}{.1in}}
\newtheorem{theorem}{Theorem}[section]

\newtheorem{corollary}[theorem]{Corollary}

\newtheorem{definition}[theorem]{Definition}

\newtheorem{remark}{Remark}

\begin{document}

\title{Robust Hypothesis Testing and Model Selection for \\Parametric Proportional Hazard Regression Models}
\author{Amarnath Nandy$^{1}$,  Abhik Ghosh$^{1}$, Ayanendranath Basu$^{1}$, Leandro Pardo$^{2}$\\$^{1}$
	{\small Interdisciplinary Statistical Research Unit, Indian Statistical Institute, Kolkata, India}\\
	$^{2}${\small Department of Statistics and O.R. I, Complutense University of Madrid, 28040 Madrid, Spain}
}
\date{}
\maketitle

\begin{abstract}
The semi-parametric Cox proportional hazards regression model has been widely used for many years in several applied sciences. 
However, a fully parametric proportional hazards model, if appropriately assumed, can often lead to more efficient inference. 
To tackle the extreme non-robustness of the traditional maximum likelihood estimator in the presence of outliers in the data
under such fully parametric proportional hazard models, a robust estimation procedure has recently been proposed 
extending the concept of the minimum density power divergence estimator (MDPDE) under this set-up. 
In this paper, we consider the problem of statistical inference under the parametric  proportional hazards model 
and develop robust Wald-type hypothesis testing and model selection procedures using the MDPDEs. 
We have also derived the necessary asymptotic results which are used to construct the testing procedure 
for general composite hypothesis and study its asymptotic powers. 
The claimed robustness properties are studied theoretically via appropriate influence function analyses. 
We have studied the finite sample level and power of the proposed MDPDE based Wald-type test through extensive simulations 
where comparisons are also made with the existing semi-parametric methods. 
The important issue of the selection of appropriate robustness tuning parameter is also discussed.  
The practical usefulness of the proposed robust testing and model selection procedures is finally 
illustrated through three interesting real data examples.
\end{abstract}

\noindent\underline{\textbf{Keywords}}: Cox Regression; Minimum Density Power Divergence Estimator (MDPDE); 
Parametric Survival Model, Robust Wald-type Test, Divergence Information Criterion (DIC).

\section{Introduction}
\label{Sec 1}

Survival analysis is a major branch of statistics dealing with the time-to-event data where observations are mostly right censored. 
There are different censoring mechanisms that can be assumed depending on the study design, its operation and data collection strategies. 
Among them, the most common and simplest one is random censoring  where the censoring time ($C$) is assumed 
to be distributed independently of the actual lifetime (time to the event under consideration) variable ($T$). 
Such a censoring mechanism is commonly assumed for medical studies where the reason for patient dropouts is often unknown. 
Given $n$ independent randomly censored observations $x_1, \ldots, x_n$ from such a study, 
we have $x_{i}= \min \left(t_{i},c_{i}\right)$, for each $i=1,\ldots,n$, 
where $t_1, \ldots, t_n$ are the independent and identically distributed (IID) realizations of $T$, 
and  $c_{1},\ldots,c_{n}$ are the respective IID realizations of  $C$. 
If we assume that the distribution functions of $T$ and $C$ are $G_T$ and $G_C$, respectively, 
then the observed lifetime variable $X= \min (T,C)$ has distribution $G_{X}=1-\left(1-G_{T}\right)\left(1-G_{C}\right)$. 
Additionally, it is generally known whether an observation is censored or not; 
we denote this censoring information on $n$ observations, respectively, by $\delta_{i}=I_{\left(t_{i} \leq c_{i}\right)}$ 
for each $i=1,\ldots,n$, and the associated random variable  by $\delta=I_{\left(T \leq C\right)}$, 
where $I_{S}$ denotes the indicator of an event $S$. 
In most practical applications, we additionally have several related covariates 
and  our aim is to explain (and often predict) the lifetime $T$ from given values of the auxiliary variables 
$\boldsymbol{Z} \in \mathbb{R}^{p}$. 
Let $\boldsymbol{z}_{i}$ denotes the observed value of $\boldsymbol{Z}$ on the $i$-th observation for $i=1, \ldots, n$. 
Then the widely used semi-parametric Cox proportional hazard  model relates the lifetimes with the observed covariate values 
via its hazard rate. For the $i$-th observation, this is given by  
\begin{eqnarray}
\lambda_{i}\left(t\right)=\lambda \left(t|\boldsymbol{Z}=\boldsymbol{z}_{i}\right)
=\lambda_{0}\left(t\right)e^{\boldsymbol{\beta}'\boldsymbol{z}_{i}},~~~~~~i=1, \ldots, n,
\label{EQ:1}
\end{eqnarray}
where the (unknown) regression coefficient $\boldsymbol{\beta} \in \mathbb{R}^{p}$ takes into account the effect of covariates 
and $\lambda_0$ is the (unknown) hazard rate of $T$ in the absence of covariates, known as the baseline hazard. 
Cox \cite{Cox:1972,Cox:1975} proposed the partial likelihood method for estimating the regression coefficient in model (\ref{EQ:1})
based on the observed data $\left\{(x_{i},\delta_{i},\boldsymbol{z_{i}}): i=1,2,3,...,n\right\}$; 
but the resulting estimator and the subsequent inference are seen to be highly unstable under data contamination. 
Later several others \cite{Bednarski:1993,Minder/Bednarski:1996} have proposed 
appropriate robust estimates of $\boldsymbol{\beta}$ under the semi-parametric Cox model (\ref{EQ:1}).

In this paper, we consider a more efficient fully parametric version of (\ref{EQ:1}) as given by 
\begin{eqnarray}
\label{EQ:2}
\lambda_{i,\boldsymbol{\theta}}\left(t\right)=\lambda_{\boldsymbol{\theta}}\left(t|\boldsymbol{Z=z_{i}}\right)=\lambda \left(t,\boldsymbol{\gamma}\right)e^{\boldsymbol{\beta'z_{i}}}~~~~~~i=1, \ldots,n
\end{eqnarray}
where $\lambda(t; \boldsymbol{\gamma})$ is a known parametric hazard function, 
defined in terms of the unknown $q$-dimensional parameter $\boldsymbol{\gamma}$, 
and hence the full parameter vector is now given by $\boldsymbol{\theta}=\left(\boldsymbol{\gamma',\beta'}\right)' \in \mathbb{R}^{p+q}$. 
Such a fully parametric version of the Cox model is often referred to as the parametric proportional hazards model
and has several advantages as noted in \cite{Hjort:1992,Borgan:1984,Hosmer:2008,ghosh}. 
In practice, the parametric form for the baseline hazard in (\ref{EQ:2}) may often be assumed 
based on empirical investigations of the given data or from previous  experiences with similar studies. 
The classical inference procedures for the parametric model in (\ref{EQ:2}), based on the randomly censored observations, 
are formed using the maximum likelihood estimators (MLEs) which are extremely non-robust against data contamination. 
However, modern complex datasets are frequently prone to outliers, and 
a robust procedure for inference would be of extreme practical value under the parametric model (\ref{EQ:2}). 
However, such a robustness consideration for fitting a regression model with randomly censored responses was lacking in the literature 
for a long time except for a few scattered attempts for the accelerated failure time (AFT) models; 
see, e.g., \cite{Agostinelli:2017,Shen et al.:2018,Asadzadeh:2018}. 
For the parametric proportional hazard model (\ref{EQ:2}), 
Ghosh and Basu \cite{ghosh} have recently developed a robust parameter estimation procedure 
extending the minimum density power divergence (DPD) approach of \cite{Basu:1998,Basu:2011}. 
In this paper, we follow up their work to further develop robust inference for hypothesis testing and model selection 
for the parametric model in (\ref{EQ:2}) based on randomly censored observations.

The MLE based classical Wald test is also extremely non-robust in the presence of outliers in the data
which has previously been noted by several authors \cite[e.g.,][]{Lin/Wei:1989,Jones:1991} particularly for the Cox regression model. 
However, in the context of censored data with covariates, the robust testing procedures are available only for 
the traditional semi-parametric Cox model in (\ref{EQ:1}) and not for its parametric version given in (\ref{EQ:2}). 
When no additional covariate is available, a robust Wald-type testing procedure based on the randomly censored observations 
has been developed more recently in \cite{Ghosh et.al.:2019}. 
Our aim, in this paper, is to present similar robust Wald-type testing procedures for testing  general parametric  hypothesis 
under the fully parametric proportional hazard regression model in (\ref{EQ:2}), 
along with detailed investigation of their asymptotic and robustness properties. 
To define such tests, we will utilize the robust  minimum DPD estimator (MDPDE) of the parameter vector $\boldsymbol{\theta}$
developed in \cite{ghosh} for model (\ref{EQ:2}).  
Once the general theory is developed, we will also illustrate our proposals for some special cases, such as 
the test for significance of any covariate and the test for selecting a proper parametric baseline hazard.

Another important aspect of statistical inference is \textit{model selection}, 
where we choose an appropriate parametric model from a pool of available candidates that best fits the observed data.
In case of our parametric Cox regression via models of the form (\ref{EQ:2}), 
we often need to select the best fitted model both in terms of the covariates to be included  
(from a larger set of possible covariates) and the form of the baseline hazard function to use. 
The most common likelihood-based model selection criteria, such as AIC and BIC, are also severely non-robust under data contamination. 
Recently, a few robust model selection criteria have been discussed in the literature \cite[see, e.g.,][]{Bednarski:2006,Wang:2016} 
but they all considered the semi-parametric Cox model (\ref{EQ:1}).
So, in this paper we also discuss a robust model selection criterion for the parametric Cox regressions (\ref{EQ:2}) 
extending the idea of divergence information criterion (DIC) based on the robust MDPDE of the model parameters.
The DIC was originally introduced in \cite{Mattheou/Lee:2009} for the simple IID observations with complete data 
and later extended for non-homogeneous (but still complete) data in \cite{Mantalos et al:2010}. 
Here, we extend the concept of DIC further for the randomly censored response to select 
the best parametric proportional hazards regression model from the available set of all possible models.

The rest of the paper is organized as follows. In Section \ref{Sec2}, for the sake of completeness, 
we present the MDPDE of $\boldsymbol{\theta}$ under the parametric proportional hazards regression model (\ref{EQ:2}) 
following \cite{ghosh}. The Wald-type test statistics, based on these MDPDEs, are presented in Section \ref{Sec3}, 
together with their asymptotic properties.  
In Section \ref{Sec 4}, we have derived the theoretical robustness properties of the proposed Wald-type test statistics
along with the stability of its level and power via the influence function analysis. 
The model selection criterion, namely the MDPDE based DIC, is described  in Section \ref{Sec 5}. 
In Section \ref{Sec 6} and \ref{Sec 7},  we have provided the empirical results to illustrate our proposed inference methodologies 
based on extensive simulation studies and interesting real data examples, respectively. 
In Section \ref{Sec 8}, we have discussed the important issue of selecting the robustness tuning parameter in our proposal. 
Finally, Section \ref{Sec 9} concludes the paper.                                
For brevity, we present the proofs of all results in the Appendix.

\section{The MDPDE for the parametric Cox regression model}
\label{Sec2}

In the parametric proportional hazards model (\ref{EQ:2}), given the covariate value $\boldsymbol{Z}=\boldsymbol{z}_{i}$
and the censoring indicator $\delta=\delta_i$, the density function
for the $i$-th observed lifetime $x_{i}$ is given by
\begin{equation}
f_{i,\boldsymbol{\theta}}(x)=f_{\boldsymbol{\theta}}(x|\delta_i, \boldsymbol{z}_{i})
=\left(  \lambda\left(  x,\boldsymbol{\gamma}\right)
e^{\boldsymbol{\beta}'\boldsymbol{z}_{i}}\right)  ^{\delta_i}%
\exp\left(  -\Lambda_{\boldsymbol{\gamma}}(x)e^{\boldsymbol{\beta}'\boldsymbol{z}_{i}}\right),
\label{2.1}%
\end{equation}
where $\Lambda_{\boldsymbol{\gamma}}(x)={\displaystyle\int_{0}^x}\lambda\left(s,\boldsymbol{\gamma}\right)  ds$.
Clearly, given $\boldsymbol{z}_{i}$ and $\delta_i$, the variables $X_{i}$s are independent but non-homogeneous;
our aim is to model their true densities, say $h_{i}(x)=h(x|\delta=\delta_i, \boldsymbol{Z}=\boldsymbol{z}_{i})$, 
by the parametric density function of the form $f_{i,\boldsymbol{\theta}}(x)$ given in (\ref{2.1}), 
respectively, for every $i=1, \ldots, n$. 
So, we follow the approach of \cite{Ghosh/Basu:2013} to define the MDPDE of $\boldsymbol{\theta}$ in the present scenario as
\begin{equation}
\widehat{\boldsymbol{\theta}}_{n,\alpha}=\left(  \widehat{\boldsymbol{\gamma}}_{n,\alpha}'%
,\widehat{\boldsymbol{\beta}}_{n,\alpha}'\right)  '=\arg\min_{\boldsymbol{\theta}}\frac{1}{n}%
{\textstyle\sum\limits_{i=1}^{n}}
d_{\alpha}(\widehat{h}_{i},f_{i,\boldsymbol{\theta}})\label{2.2}%
\end{equation}
where $d_{\alpha}$ is the DPD measure \cite{Basu:2011,Basu:1998} and 
$\widehat{h}_i$ is an empirical estimate of $h_i$ for each $i=1,..,n$. 
Upon simplification, the MDPDE can be obtained minimizing the simpler objective function
\begin{eqnarray}
H_{n,\alpha}\left(  \boldsymbol{\theta}\right)  =\frac{1}{n}{\textstyle\sum\limits_{i=1}^{n}}
\left[ {\displaystyle\int} f_{i,\boldsymbol{\theta}}(x)^{1+\alpha}dx -\frac{1+\alpha}{\alpha}
\lambda_{\boldsymbol{\theta}}(x_{i}|\boldsymbol{z}_i)^{\alpha\delta_{i}}S_{\boldsymbol{\theta}}(x_{i}|\boldsymbol{z}_i)^{\alpha}
+\frac{1}{\alpha}\right]
\label{EQ:H_na}
\end{eqnarray}
with $\lambda_{\boldsymbol{\theta}}(t|\boldsymbol{z}_i) = \lambda_{\boldsymbol{\theta}}(t|\boldsymbol{Z}=\boldsymbol{z}_i)$
being as given in (\ref{EQ:2}) and
$
S_{\boldsymbol{\theta}}(t|\boldsymbol{z}) =
S_{\boldsymbol{\theta}}(t|\boldsymbol{Z}=\boldsymbol{z})
=\exp\left(  -\Lambda_{\mathbf{\gamma}}(t)e^{\boldsymbol{\beta}'\boldsymbol{z}}\right).
$

\begin{remark}
It is known that the DPD measure at $\alpha=0$ coincides with the Kullback-Leibler divergence ($d_{\mbox{KL}}$).
On the other hand, if $L_{n}(\boldsymbol{\theta})$ denotes the likelihood function associated to the random
sample $\left(  x_{i},\delta_{i},\boldsymbol{z}_{i}\right)$, $i=1, \ldots, n$, it is not difficult to establish that
\[
\frac{1}{n}\log L_{n}(\boldsymbol{\theta}) = k-\frac{1}{n}{\textstyle\sum\limits_{i=1}^{n}}	d_{KL}(\widehat{h}_{i},f_{i,\boldsymbol{\theta}}),
\]
where $k$ is a constant independent of $\boldsymbol{\theta}$. 
Therefore, the MDPDE at $\alpha=0$ coincides with the MLE. 
In this sense the MDPDE can be considered as a natural extension of the MLE, 
obtained replacing the Kullback-Leibler divergence by the DPD measure with some $\alpha>0$.
\label{Remark:2.1}
\end{remark}

Ghosh and Basu \cite{ghosh} studied the detailed properties of the MDPDE, 
obtained by minimizing $H_{n,\alpha}\left(  \boldsymbol{\theta}\right)$, 
under the parametric proportional hazards model (\ref{EQ:2}) both theoretically and empirically. 
For completeness, we here present their main theoretical results that would be need in the subsequent parts of the paper. 
Firstly, note that the score functions corresponding to the density 
$f_{i,\boldsymbol{\theta}}(x)=f_{\boldsymbol{\theta}}(x|\delta_i, \boldsymbol{z}_{i})$ in (\ref{2.1})
are given by 
\begin{eqnarray}
\boldsymbol{u}_{\boldsymbol{\theta}}^{(1)}(x|\delta_{i}, \boldsymbol{z}_i) &=&
\psi_{\boldsymbol{\gamma}}(x)\delta_i  - \Psi_{\boldsymbol{\gamma}}(x)  e^{\boldsymbol{\beta}'\boldsymbol{z}_{i}},
~~~\mbox{ and }~~
\boldsymbol{u}_{\boldsymbol{\theta}}^{(2)}(x|\delta_{i}, \boldsymbol{z}_i) =
 \boldsymbol{z}_{i}\left[\delta_i  - \Lambda_{\boldsymbol{\gamma}}(x)  e^{\boldsymbol{\beta}'\boldsymbol{z}_{i}}\right],
\nonumber
\end{eqnarray}
where
\[
\psi_{\boldsymbol{\gamma}}(x)=\frac{\partial\log\lambda\left(
	x,\boldsymbol{\gamma}\right)  }{\partial\boldsymbol{\gamma}},
~~~\mbox{ and }~~
\Psi_{\boldsymbol{\gamma}}(x)
={\displaystyle\int_{0}^{x}}\frac{\partial\lambda\left(  s,\boldsymbol{\gamma}\right)  }{\partial	\boldsymbol{\gamma}}ds
={\displaystyle\int_{0}^{x}}\psi_{\boldsymbol{\gamma}}(s)\lambda\left(  s,\boldsymbol{\gamma}\right)  ds.
\]
Then, under standard differentiability assumptions, 
the MDPDE of $\boldsymbol{\theta}$, under the model (\ref{EQ:2}), can be obtained 
by solving the following system of equations \cite{ghosh}:
\begin{equation}
\frac{1}{n}{\textstyle\sum\limits_{i=1}^{n}}
\boldsymbol{u}_{n}^{(1,\alpha)}(\boldsymbol{\theta}|x_i, \delta_i, \boldsymbol{z}_i)=\boldsymbol{0}_{q},
~~~~~~
\frac{1}{n}{\textstyle\sum\limits_{i=1}^{n}}
\boldsymbol{u}_{n}^{(2,\alpha)}(\boldsymbol{\theta}|x_i, \delta_i, \boldsymbol{z}_i)=\boldsymbol{0}_{p},
\label{2.3}%
\end{equation}
where 
\begin{align}
\boldsymbol{u}_{n}^{(1,\alpha)}(\boldsymbol{\theta}|x, \delta, \boldsymbol{z}) =&
 \left\{  
\psi_{\boldsymbol{\gamma}}(x)\lambda_{\boldsymbol{\theta}}(x|\boldsymbol{z})^{\alpha}S_{\boldsymbol{\theta}}(x|\boldsymbol{z})^{\alpha}
-\left(\lambda_{\boldsymbol{\theta}}(x|\boldsymbol{z})^{\alpha}-1\right) \Psi_{\boldsymbol{\gamma}}(x)
S_{\boldsymbol{\theta}}(x|\boldsymbol{z})^{\alpha}e^{\boldsymbol{\beta}'\boldsymbol{z}}\right\}\delta
 \nonumber\\
&  ~~~~~~~~~~~~ 
-\Psi_{\boldsymbol{\gamma}}(x)S_{\boldsymbol{\theta}}(x|\boldsymbol{z})^{\alpha}e^{\boldsymbol{\beta}'\boldsymbol{z}}
-\boldsymbol{\xi}^{(1,\alpha)}(\boldsymbol{\theta}|\delta, \boldsymbol{z}),  
\label{2.4}%
\end{align}
and
\begin{align}
\boldsymbol{u}_{n}^{(2,\alpha)}(\boldsymbol{\theta}|x, \delta, \boldsymbol{z}) =&
 \left\{  \lambda_{\boldsymbol{\theta}}(x|\boldsymbol{z})^{\alpha}S_{\boldsymbol{\theta}}(x|\boldsymbol{z})^{\alpha}
 -\left(  \lambda_{\boldsymbol{\theta}}(x|\boldsymbol{z})^{\alpha}-1\right)  \Lambda_{\boldsymbol{\gamma}}(x)
 S_{\boldsymbol{\theta}}(x|\boldsymbol{z})^{\alpha}e^{\boldsymbol{\beta}'\boldsymbol{z}}\right\}  \boldsymbol{z}\delta
 \nonumber\\
&  ~~~~~~~~~~~~ 
-\boldsymbol{z}\Lambda_{\boldsymbol{\gamma}}(x)S_{\boldsymbol{\theta}}(x|\boldsymbol{z})^{\alpha}e^{\boldsymbol{\beta}'\boldsymbol{z}}
-\boldsymbol{\xi}^{(2,\alpha)}(\boldsymbol{\theta}|\delta, \boldsymbol{z}),
\label{2.5}%
\end{align}
with
\[
\boldsymbol{\xi}^{(j,\alpha)}(\boldsymbol{\theta}|\delta, \boldsymbol{z})
= {\displaystyle\int} \boldsymbol{u}_{\boldsymbol{\theta}}^{(j)}(x|\delta, \boldsymbol{z})
f_{\boldsymbol{\theta}}(x|\delta, \boldsymbol{z})^{1+\alpha}dx,
~~~~\text{ for }~ j=1,2.
\]
Let us denote the MDPDE, obtained by solving (\ref{2.3}), as $\widehat{\boldsymbol{\theta}}_{n,\alpha}$.
Its asymptotic distribution has been studied in \cite{ghosh} under some appropriate assumptions;
we will refer to their Assumptions (A)--(E) as the \textit{Ghosh-Basu Conditions} throughout the rest of the paper.
Besides the  technical conditions, it includes a crucial assumption about the structure of the underlying true distribution; 
Assumption (A), in particular, states that the true hazard rate of the $i$-th observation, 
given the covariate value $\boldsymbol{Z}=\boldsymbol{z}_{i}$, is of form 
$\lambda_{i}(s)=\lambda_{0}(s)h_{0}(\boldsymbol{z}_{i})$ for some positive functions $\lambda_{0}$ and $h_{0}$.	
Under these conditions, the limiting form of the MDPDE estimating equations in (\ref{2.3}), as $n\rightarrow\infty$, 
is given by \cite{ghosh}
\begin{equation}
\boldsymbol{u}_{0}^{(1,\alpha)}\left(  \boldsymbol{\theta}\right) =\boldsymbol{0}_{q},
~~~~~~
\boldsymbol{u}_{0}^{(2,\alpha)}\left(  \boldsymbol{\theta}\right) =\boldsymbol{0}_{p},
\label{2.3F}%
\end{equation}
where we assume that the process is observed till the maximum time-point $T_{\max}$ and define
\begin{align*}
\boldsymbol{u}_{0}^{(1,\alpha)}\left(  \boldsymbol{\theta}\right)   &  =
{\displaystyle\int_{0}^{T_{\max}}} \left[  \psi_{\boldsymbol{\gamma}}(s)\lambda\left(s,\boldsymbol{\gamma}\right)^{\alpha}
\left\{r_{\alpha,\alpha}^{(0)}(s)\lambda_{0}(s)-\widetilde{q}_{\alpha,\alpha}^{(0)}(s)\lambda\left(s,\boldsymbol{\gamma}\right)\right\}  
-\Psi_{\boldsymbol{\gamma}}(s)\left\{ q_{0,\alpha}^{(0)}(s)-\widetilde{q}_{\alpha,\alpha}^{(0)}(s)\right\}  \right.
\\
&  \left.  ~~~~~~~~~~~~~ - \lambda\left(  s,\boldsymbol{\gamma}\right)^{\alpha}\Psi_{\boldsymbol{\gamma}}(s) \left\{
r_{\alpha+1,\alpha}^{(0)}(s)\lambda_{0}(s)-\widetilde{q}_{\alpha+1,\alpha}^{(0)}(s)\lambda\left(s,\boldsymbol{\gamma}\right)  \right\}  
\right]  ds,
\end{align*}
and
\begin{align*}
\boldsymbol{u}_{0}^{(2,\alpha)}\left(  \boldsymbol{\theta}\right)   &  =
{\displaystyle\int_{0}^{T_{\max}}}\left[  \lambda\left(s,\boldsymbol{\gamma}\right)^{\alpha}
\left\{r_{\alpha,\alpha}^{(1)}(s)\lambda_{0}(s)-\widetilde{q}_{\alpha,\alpha}^{(1)}(s)\lambda\left(s,\boldsymbol{\gamma}\right)\right\}  
- \Lambda_{\boldsymbol{\gamma}}(s)\left\{  q_{0,\alpha}^{(1)}(s)-\widetilde{q}_{\alpha,\alpha}^{(1)}(s)\right\}  \right.  
\\
&  \left.  ~~~~~~~~~~~~~ - \lambda\left(  s,\boldsymbol{\gamma}\right)  ^{\alpha}\Lambda_{\boldsymbol{\gamma}}(s)\left\{  
r_{\alpha+1,\alpha}^{(1)}(s)\lambda_{0}(s)-\widetilde{q}_{\alpha+1,\alpha}^{(1)}(s)\lambda\left(s,\boldsymbol{\gamma}\right)  \right\}  
\right]ds,
\end{align*}
with 
\begin{eqnarray}
\widetilde{q}_{\alpha_1,\alpha_2}^{(j)}(s) &=& E\left[\left(\boldsymbol{Z}\right)^{j}
e^{\left(\alpha_1+1\right)\boldsymbol{\beta}'\boldsymbol{z}}S_{\boldsymbol{\theta}}(s|\boldsymbol{Z})^{\alpha_{2}+1}\right]
\nonumber\\
r_{\alpha_{1},\alpha_{2}}^{(j)}(s) &=& E\left[\left(\boldsymbol{Z}\right)^{j}I\left(  X\geq s\right)  
e^{\alpha_{1}\boldsymbol{\beta}'\boldsymbol{z}} S_{\boldsymbol{\theta}}(s|\boldsymbol{Z})^{\alpha_{2}}h_{0}(\boldsymbol{Z})\right]  
\nonumber \\
q_{\alpha_{1},\alpha_{2}}^{(j)}(s) &=& E\left[\left(\boldsymbol{Z}\right)^{j}I\left(  X\geq s\right)  
e^{\left(  \alpha_{1}+1\right)  \boldsymbol{\beta}'\boldsymbol{z}}S_{\boldsymbol{\theta}}(s|\boldsymbol{Z})^{\alpha_{2}}\right]
\nonumber
\end{eqnarray}
for each $j=0,1 $ and any $\alpha_{1},\alpha_{2}\geq 0$.
Here $(\boldsymbol{Z})^0$ denotes 1 and hence $\widetilde{q}_{\alpha_1,\alpha_2}^{(0)}$, $r_{\alpha_{1},\alpha_{2}}^{(0)}$
and $q_{\alpha_{1},\alpha_{2}}^{(0)}$ are all scalars; however, $\widetilde{q}_{\alpha_1,\alpha_2}^{(1)}$, 
$r_{\alpha_{1},\alpha_{2}}^{(1)}$ and $q_{\alpha_{1},\alpha_{2}}^{(1)}$ are $p$-dimensional vectors. 
Furthermore, when the true underlying distribution has the hazard rate as the assumed model in (\ref{EQ:2}), 
we always have $\widetilde{q}_{\alpha_1,\alpha_2}^{(j)}=r_{\alpha_{1},\alpha_{2}}^{(j)}=q_{\alpha_{1},\alpha_{2}}^{(j)}$  for $j=0, 1$. 
Accordingly, we define the corresponding Fisher consistent statistical functional as the solution of (\ref{2.3F}). 
More precisely, if the observations $(x_i, \delta_{i}, z_i)$, $ i = 1, \ldots,  n$,  
are assumed to be IID realizations of the random variable $(X, \delta, \boldsymbol{Z})$ having joint distribution $H$,
the minimum DPD functional $\boldsymbol{U}_\alpha(\boldsymbol{H})$, corresponding to the MDPDE $\widehat{\boldsymbol{\theta}}_{n,\alpha}$,
as a solution to the limiting MDPDE estimating equations given in (\ref{2.3F}).

The following theorem, taken from \cite{ghosh}, present the asymptotic distribution of the MDPDE $\widehat{\boldsymbol{\theta}}_{n,\alpha}$
under the parametric proportional hazards regression model (\ref{EQ:2}).

\begin{theorem}[\cite{ghosh}]
\label{Theo:1}
Suppose that the observed data $\left(  x_{i},\delta_{i},\boldsymbol{z}_{i}\right)
,$ $i=1,...,n,$ are IID realizations of the random triplet $(X,\delta,\boldsymbol{Z})$ having the true joint distribution $H$ 
such that the minimum DPD functional $\boldsymbol{U}_\alpha(\boldsymbol{H})=\boldsymbol{\theta}_0$ exists and is unique.
Then, under Ghosh-Basu Conditions, there exists a consistent sequence of MDPDEs 
$\widehat{\boldsymbol{\theta}}_{n,\alpha}=\left(\widehat{\boldsymbol{\gamma}}_{n,\alpha}', \widehat{\boldsymbol{\beta}}_{n,\alpha}'\right)'$, as a solution to the estimating equations in (\ref{2.3}),
which satisfies
\[
\sqrt{n}\left(  \widehat{\boldsymbol{\theta}}_{n,\alpha}-\boldsymbol{\theta}_0\right)  
\underset{n\longrightarrow\infty}{\overset{\mathcal{D}}{\longrightarrow}}
\mathcal{N}\left(  \boldsymbol{0}_{q+p}, \boldsymbol{\Sigma}_{\alpha}(\boldsymbol{\theta}_0) \right)  ,
\]
where $\boldsymbol{0}_{d}$ denotes the $d$-vector of zeros and 
$\boldsymbol{\Sigma}_{\alpha} \left(\boldsymbol{\theta} \right)
=\boldsymbol{J}_{\alpha}^{-1}\left(  \boldsymbol{\theta}\right)  \boldsymbol{K}_{\alpha}\left(  \boldsymbol{\theta}\right)  
\boldsymbol{J}_{\alpha}^{-1}\left(\boldsymbol{\theta}\right)
$
with
\[
\boldsymbol{J}_{\alpha}\left(  \boldsymbol{\theta}\right)  =-\left(
\begin{array}
[c]{cc}%
\frac{\partial\boldsymbol{u}_{0}^{(1,\alpha)}\left(  \boldsymbol{\theta
}\right)  }{\partial\boldsymbol{\gamma}'} & \frac{\partial\boldsymbol{u}%
_{0}^{(1,\alpha)}\left(  \boldsymbol{\theta}\right)  }{\partial
\boldsymbol{\beta}'}\\
\frac{\partial\boldsymbol{u}_{0}^{(2,\alpha)}\left(  \boldsymbol{\theta
}\right)  }{\partial\boldsymbol{\gamma}'} & \frac{\partial\boldsymbol{u}%
_{0}^{(2,\alpha)}\left(  \boldsymbol{\theta}\right)  }{\partial
\boldsymbol{\beta}'}%
\end{array}
\right)
~~~\mbox{and }~~
\boldsymbol{K}_{\alpha}\left(  \boldsymbol{\theta}\right)  =\mbox{Var}_{_H}\left[
\begin{array}
[c]{c}%
\boldsymbol{u}_{n}^{(1,\alpha)}(\boldsymbol{\theta}|X, \delta, \boldsymbol{Z})  \\
\boldsymbol{u}_{n}^{(2,\alpha)}(\boldsymbol{\theta}|X, \delta, \boldsymbol{Z})
\end{array}
\right].
\]
\end{theorem}

We will utilize the asymptotic distribution of the MDPDE, as stated in the above theorem, 
to define and study the robust testing and model selection procedures for the parametric proportional hazards model (\ref{EQ:2}).
For this purpose, we would need a consistent estimator of the asymptotic variance matrix 
$\boldsymbol{\Sigma}_{\alpha}(\boldsymbol{\theta}_0)$ of the MDPDE $\widehat{\boldsymbol{\theta}}_{n,\alpha}$.
Such an variance estimate can be constructed as 
\begin{equation}
\boldsymbol{\Sigma}_{n,\alpha} = 
\boldsymbol{J}_{n,\alpha}^{-1}\left(\widehat{\boldsymbol{\theta}}_{n,\alpha}\right)  
\boldsymbol{K}_{n,\alpha}\left(\widehat{\boldsymbol{\theta}}_{n,\alpha}\right)  
\boldsymbol{J}_{n,\alpha}^{-1}\left(\widehat{\boldsymbol{\theta}}_{n,\alpha}\right),
\label{EQ:Asym_var_est}
\end{equation}
where $\boldsymbol{J}_{n,\alpha}(\boldsymbol{\theta})$ and $\boldsymbol{K}_{n,\alpha}(\boldsymbol{\theta})$ 
are some consistent (and continuous) estimator of the matrices $\boldsymbol{J}_{\alpha}(\boldsymbol{\theta})$ 
and $\boldsymbol{K}_{\alpha}(\boldsymbol{\theta})$, respectively.
Following the discussion in \cite{ghosh}, we will particularly use their empirical  estimators as given by
\[
\boldsymbol{J}_{n,\alpha}\left( \boldsymbol{\theta}\right)  = - \frac{1}{n}{\textstyle\sum\limits_{i=1}^{n}} \left[
\begin{array}
[c]{cc}%
   \frac{\partial\boldsymbol{u}_{n}^{(1,\alpha)}(\boldsymbol{\theta}|x_i, \delta_i, \boldsymbol{z}_i)}{\partial\boldsymbol{\gamma}'} 
& \frac{\partial\boldsymbol{u}_{n}^{(1,\alpha)}(\boldsymbol{\theta}|x_i, \delta_i, \boldsymbol{z}_i)}{\partial\boldsymbol{\beta}'}
\\
   \frac{\partial\boldsymbol{u}_{n}^{(2,\alpha)}(\boldsymbol{\theta}|x_i, \delta_i, \boldsymbol{z}_i)}{\partial\boldsymbol{\gamma}'} 
   & \frac{\partial\boldsymbol{u}_{n}^{(2,\alpha)}(\boldsymbol{\theta}|x_i, \delta_i, \boldsymbol{z}_i) }{\partial\boldsymbol{\beta
	}'}%
\end{array}
\right],
\]
and
\[
\boldsymbol{K}_{n,\alpha}\left(\boldsymbol{\theta}\right)   = \frac{1}{n}{\textstyle\sum\limits_{i=1}^{n}}
\left[
\begin{array}
[c]{cc}%
\boldsymbol{u}_{n}^{(1,\alpha)}(\boldsymbol{\theta}|x_i, \delta_i, \boldsymbol{z}_i)
\boldsymbol{u}_{n}^{(1,\alpha)}(\boldsymbol{\theta}|x_i, \delta_i, \boldsymbol{z}_i)'
& \boldsymbol{u}_{n}^{(1,\alpha)}(\boldsymbol{\theta}|x_i, \delta_i, \boldsymbol{z}_i)
\boldsymbol{u}_{n}^{(2,\alpha)}(\boldsymbol{\theta}|x_i, \delta_i, \boldsymbol{z}_i)'
\\
\boldsymbol{u}_{n}^{(2,\alpha)}(\boldsymbol{\theta}|x_i, \delta_i, \boldsymbol{z}_i)
\boldsymbol{u}_{n}^{(1,\alpha)}(\boldsymbol{\theta}|x_i, \delta_i, \boldsymbol{z}_i)'
& \boldsymbol{u}_{n}^{(2,\alpha)}(\boldsymbol{\theta}|x_i, \delta_i, \boldsymbol{z}_i)
\boldsymbol{u}_{n}^{(2,\alpha)}(\boldsymbol{\theta}|x_i, \delta_i, \boldsymbol{z}_i)'%
\end{array}
\right].
\]
\bigskip

\section{Robust hypothesis testing using the MDPDEs}
\label{Sec3}

\subsection{The Wald-type tests}
\label{Sec 3.1}

We now consider the problem of testing the general composite hypothesis
given by%
\begin{equation}
H_{0}:\boldsymbol{\theta}\in\Theta_{0} ~~~ \text{against} ~~~ H_{1}:\boldsymbol{\theta}\notin\Theta_{0},
\label{2.7}%
\end{equation}
where $\Theta_{0}$ is a given subset of the parameter space $\Theta\subseteq \mathbb{R}^{p+q}$. 
This null parameter space $\Theta_{0}$ is often
defined by a set of $r (\leq p+q)$ restrictions of the form
\begin{equation}
\boldsymbol{m}(\boldsymbol{\theta})=\boldsymbol{0}_{r},\label{2.8}%
\end{equation}
where $\boldsymbol{m}:\mathbb{R}^{p+q}\rightarrow\mathbb{R}^{r}$ is a known function;
that is,  $\Theta_{0}=\{ \boldsymbol{\theta} \in\Theta:  \boldsymbol{m}(\boldsymbol{\theta})=\boldsymbol{0}_{r}\}$. 
We assume that the $\left(p+q\right)\times r$ matrix
$\boldsymbol{M}(\boldsymbol{\theta})=\frac{\partial\boldsymbol{m}'(\boldsymbol{\theta})}{\partial\boldsymbol{\theta}}$
exists, is continuous in $\boldsymbol{\theta}$, and $\mathrm{rank}\left(
\boldsymbol{M}(\boldsymbol{\theta})\right)  =r$. 

An important example is the test for significance of the regression model with the null hypothesis 
$H_{0}:\boldsymbol{\beta}=\boldsymbol{0}_{p}$, indicating no covariate effect. 
This is a special case of (\ref{2.7}) with $r=p$ and $\boldsymbol{m}(\boldsymbol{\theta})=\boldsymbol{\beta}$, 
so that $\Theta_{0}=\{\left(  \boldsymbol{\gamma,\beta}\right)  \in\Theta: \boldsymbol{m}(\boldsymbol{\gamma,\beta})=\boldsymbol{\beta}=\boldsymbol{0}_{p}\}$ 
and $\boldsymbol{M}=[\mathbb{I}_p ; \mathbb{O}]'$; 
here $\mathbb{I}_p$ denotes the $p\times p$ identity matrix and $\mathbb{O}$ is a null matrix of appropriate order.

In the following  definition we present the Wald-type test statistics for testing (\ref{2.7}),
utilizing the MDPDEs  and the consistent estimator of their asymptotic variance matrix.

\begin{definition}
The Wald-type test statistics for testing (\ref{2.7}) under the parametric proportional hazard model (\ref{EQ:2}) is given by
\begin{equation}
W_{n}\left(\widehat{\boldsymbol{\theta}}_{n,\alpha}\right) = n\boldsymbol{m}'(\widehat{\boldsymbol{\theta}}_{n,\alpha})\left(  
\boldsymbol{M}'(\widehat{\boldsymbol{\theta}}_{n,\alpha})\boldsymbol{\Sigma}_{n,\alpha}
\boldsymbol{M}(\widehat{\boldsymbol{\theta}}_{n,\alpha})\right)  ^{-1}
\boldsymbol{m}(\widehat{\boldsymbol{\theta}}_{n,\alpha}),
\label{2.10}%
\end{equation}
where $\widehat{\boldsymbol{\theta}}_{n,\alpha}$ is the MDPDE of $\boldsymbol{\theta}$ 
obtained based on a randomly censored sample of size $n$ and tuning parameter $\alpha \geq 0$, 
and the matrix
$\boldsymbol{\Sigma}_{n,\alpha}$ is  as defined in (\ref{EQ:Asym_var_est}).
\end{definition}

The following theorem presents the asymptotic null distribution of our Wald-type test statistics
which can be used to determine the required critical values for the testing procedures;
the proof is given in the Appendix.

\begin{theorem}\label{THM:asymp_null}
Suppose that the assumptions of Theorem \ref{Theo:1} hold true under the parametric proportional hazards model (\ref{EQ:2})
and the matrix $\boldsymbol{\Sigma}_{\alpha} \left(\boldsymbol{\theta} \right)$ is continuous in $\boldsymbol{\theta}\in\Theta_0$. 
Then, the asymptotic distribution of the Wald-type test statistic $W_{n}(\widehat{\boldsymbol{\theta}}_{n,\alpha})$, 
under the composite null hypothesis in (\ref{2.7}), is a chi-square with $r$ degrees of freedom.
\end{theorem}


Using the above theorem, the null hypothesis in (\ref{2.7}) will be rejected, 
based on the Wald-type test statistics $W_{n}(\widehat{\boldsymbol{\theta}}_{n,\alpha})$, if
\begin{equation}
W_{n}(\widehat{\boldsymbol{\theta}}_{n,\alpha})>\chi_{r,\tau}^{2},
\label{2.101}%
\end{equation}
where $\chi_{r,\tau}^{2}$ is the $100(1-\tau)$ percentile of a chi-square distribution with $r$ degrees of freedom.

\subsection{Asymptotic power Analysis}
\label{Sec 3.2}

Let us now study the power of the MDPDE based Wald-type tests for testing the hypothesis in (\ref{2.7})
under the parametric proportional hazards regression model (\ref{EQ:2}).
Our first theorem helps to obtain an approximation to the power function of the
Wald-type tests of the form (\ref{2.101}); see Appendix for its proof.

\begin{theorem}\label{THM:app_power}
Let {$\boldsymbol{\theta}$}$^{\ast}\in\Theta \setminus \Theta_{0}$ be the true value of
the parameter so that the MDPDE $\widehat{\boldsymbol{\theta}}_{n,\alpha}%
\underset{n\longrightarrow\infty}{\overset{\mathcal{P}}{\longrightarrow}}%
${$\boldsymbol{\theta}$}$^{\ast}.$ Denote,%
\[
\ell^{\ast}({\boldsymbol{\theta}}_{1},{\boldsymbol{\theta}}_{2}%
)=n\boldsymbol{m}'\left(  {\boldsymbol{\theta}}_{1}\right)  \left(
\boldsymbol{M}'({\boldsymbol{\theta}}_{2})\boldsymbol{\Sigma}_{\alpha}
(\boldsymbol{\theta}_{2})\boldsymbol{M}({\boldsymbol{\theta}}_{2})\right)
^{-1}\boldsymbol{m}\left(  {\boldsymbol{\theta}}_{1}\right)  .
\]
Then%
\[
\sqrt{n}\left(  \ell^{\ast}(\widehat{\boldsymbol{\theta}}_{n,\alpha
},\widehat{\boldsymbol{\theta}}_{n,\alpha})-\ell^{\ast}({\boldsymbol{\theta}%
}^{\ast},{\boldsymbol{\theta}}^{\ast})\right)  \underset{n\longrightarrow
\infty}{\overset{\mathcal{L}}{\longrightarrow}}\mathcal{N}\left(
\boldsymbol{0},\sigma_{W_{n}}^{2}\left(  {\boldsymbol{\theta}}^{\ast}\right)
\right)
\]
where%
\[
\sigma_{W_{n}}^{2}\left(  {\boldsymbol{\theta}}^{\ast}\right)  =\left(
\frac{\partial\ell^{\ast}({\boldsymbol{\theta}},{\boldsymbol{\theta}}^{\ast}%
)}{\partial{\boldsymbol{\theta}}}\right)  _{{\boldsymbol{\theta}%
}={\boldsymbol{\theta}}^{\ast}}'\boldsymbol{\Sigma}_{\alpha}({\boldsymbol{\theta}%
}^{\ast})\left(  \frac{\partial\ell^{\ast}({\boldsymbol{\theta}}%
,{\boldsymbol{\theta}}^{\ast})}{\partial{\boldsymbol{\theta}}}\right)
_{{\boldsymbol{\theta}}={\boldsymbol{\theta}}^{\ast}}'.
\]
\end{theorem}
%

\begin{remark}
Based on Theorem \ref{THM:app_power}, 
we approximate the power function, say $\beta_{W_{n}}({\boldsymbol{\theta}}^{\ast})$, of the Wald-type test statistics in (\ref{2.10}), 
at the parameter value $\boldsymbol{\theta}^{\ast}\in\Theta \setminus \Theta_{0}$ as 
\begin{equation}
\beta_{W_{n}}\left(  {\boldsymbol{\theta}}^{\ast}\right)  \cong1-\Phi\left(
\frac{\sqrt{n}}{\sigma_{W_{n}}\left(  {\boldsymbol{\theta}}^{\ast}\right)
}\left(  \frac{\chi_{r,\alpha}^{2}}{n}-\ell^{\ast}\left(  {\boldsymbol{\theta
}}^{\ast},{\boldsymbol{\theta}}^{\ast}\right)  \right)  \right)  .
\label{2.11}%
\end{equation}
Here $\Phi\left(  x\right)  $ is the distribution function of a normal random
variable with mean zero and variance 1. It implies that
\[
\lim_{n\rightarrow\infty}\beta_{W_{n}}({\boldsymbol{\theta}}^{\ast})=1,
\]
i.e. the Wald-type test statistics considered in (\ref{2.10}) are consistent
in the sense of Fraser \cite{Fraser:1957}.
\end{remark}

\begin{remark}
Based on (\ref{2.11}), we can further compute  the required sample size in order to
achieve a targeted power value. If we want a power of $\beta_{W_{n}}\left(
{\boldsymbol{\theta}}^{\ast}\right)  =\pi,$ we need a sample of size $n=\left[  n^{\ast}\right]  +1$, where
$\left[  x\right]  $ is the largest integer less than or equal to $x$, and
\[
n^{\ast}=\frac{A+B+\sqrt{A(A+2B)}}{2\ell^{\ast}\left(  {\boldsymbol{\theta}}^{\ast},{\boldsymbol{\theta}}^{\ast}\right)  },
\]
with $A=\sigma_{W_{n}}\left(  {\boldsymbol{\theta}}^{\ast}\right)  ^{2}\left(\Phi^{-1}(1-\pi)\right)^{2}$ 
and $B=\frac{1}{2}\chi_{r,\alpha}^{2}\ell^{\ast}\left(  {\boldsymbol{\theta}}^{\ast},{\boldsymbol{\theta}}^{\ast}\right).$
\end{remark}

Next, we derive the asymptotic power of the Wald-type test statistics considered in (\ref{2.10})  at
a contiguous sequence of alternative hypotheses close to the null hypothesis. 
Let $\boldsymbol{\theta}_{n}\in\Theta \setminus \Theta_{0}$ be a given contiguous alternative, 
and let $\boldsymbol{\theta}_{0}$ be the element in $\Theta_{0}$ closest to $\boldsymbol{\theta}_{n}$ in terms of the Euclidean distance. 
One possibility to introduce such contiguous alternative hypotheses, in this context, 
is to consider a fixed non-zero $\boldsymbol{d}\in\mathbb{R}^{p+q}$ and permit $\boldsymbol{\theta}_{n}$ to
move towards $\boldsymbol{\theta}_{0}\in \Theta_0$ as $n$ increases, through the relation
\begin{equation}
H_{1,n}:\boldsymbol{\theta} = \boldsymbol{\theta}_{n}, ~~~\mbox{ where }~~\boldsymbol{\theta}_{n}=\boldsymbol{\theta}_{0}+n^{-1/2}\boldsymbol{d}. 
\label{22.12}%
\end{equation}
A second approach is the relaxation of the condition $\boldsymbol{m}\left(\boldsymbol{\theta}\right)  =\boldsymbol{0}_{r}$ that defines $\Theta_{0}$.
Let $\boldsymbol{\delta}\in\mathbb{R}^{r}$ and consider the sequence of parameters $\{\boldsymbol{\theta}_{n}\}$ moving towards
$\boldsymbol{\theta}_{0}\in \Theta_0$ according to the set up
\begin{equation}
H_{1,n}^{\ast}: \boldsymbol{\theta} = \boldsymbol{\theta}_{n}, ~~~\mbox{ where }~~
\boldsymbol{m}\left(  \boldsymbol{\theta}_{n}\right)=n^{-1/2}\boldsymbol{\delta}.
\label{22.19}%
\end{equation}
Note that a Taylor series expansion of $\boldsymbol{m}\left(
\boldsymbol{\theta}_{n}\right)  $ around $\boldsymbol{\theta}$$_{0}$ yields
\begin{equation}
\boldsymbol{m}\left(  \boldsymbol{\theta}_{n}\right)  =\boldsymbol{m}\left(
\boldsymbol{\theta}_{0}\right)  +\boldsymbol{M}'(\boldsymbol{\theta}%
_{0})\left(  \boldsymbol{\theta}_{n}-\boldsymbol{\theta}_{0}\right)  +o\left(
\left\Vert \boldsymbol{\theta}_{n}-\boldsymbol{\theta}_{0}\right\Vert \right)
.\label{22.20}%
\end{equation}
By substituting $\boldsymbol{\theta}$$_{n}=$$\boldsymbol{\theta}$%
$_{0}+n^{-1/2}\boldsymbol{d}$ in (\ref{22.20}) and taking into account
that\textbf{\ }$\boldsymbol{m}($$\boldsymbol{\theta}$$_{0})=\boldsymbol{0}%
_{r}$, we get
\begin{equation}
\boldsymbol{m}\left(  \boldsymbol{\theta}_{n}\right)  =n^{-1/2}\boldsymbol{M}%
'(\boldsymbol{\theta}_{0})\boldsymbol{d}+o\left(  \left\Vert
\boldsymbol{\theta}_{n}-\boldsymbol{\theta}_{0}\right\Vert \right)
.\label{22.21}%
\end{equation}
So, the equivalence relationship between the hypotheses $H_{1,n}$ and
$H_{1,n}^{\ast}$ is obtained as
\begin{equation}
\boldsymbol{\delta}=\boldsymbol{M}'(\boldsymbol{\theta}_{0})\boldsymbol{d},~~
\text{ as }n\rightarrow\infty.\label{22.22}%
\end{equation}
In the following theorem we present the asymptotic distribution of the Wald-type test
statistics under these contiguous alternative hypotheses $H_{1,n}$ and $H_{1,n}^{\ast}$, 
which can be used to compute the corresponding  asymptotic power functions of the proposed Wald-type tests.
Here, $\chi_{r}^{2}(b)$ will denote a non-central chi-square random variable with degrees of freedom $r$ 
and non-centrality parameter $b$.

\begin{theorem}\label{THM:Cont_power}
Assume that the conditions of Theorem \ref{Theo:1} hold.
Then, the asymptotic distribution of $W_{n}({\boldsymbol{\widehat{\boldsymbol{\theta}}_{n,\alpha}}})$ is given by
\begin{equation}
W_{n}({\boldsymbol{\widehat{\boldsymbol{\theta}}_{n,\alpha}}}%
)\underset{n\rightarrow\infty}{\overset{\mathcal{L}}{\longrightarrow}}\chi
_{r}^{2}\left(  \boldsymbol{d}'\boldsymbol{M}(\boldsymbol{\theta}%
_{0})\left(  \boldsymbol{M}'(\boldsymbol{\theta}_{0})\boldsymbol{\Sigma
}_{\alpha}(\boldsymbol{\theta}_{0})\boldsymbol{M}(\boldsymbol{\theta}_{0})\right)
^{-1}\boldsymbol{M}'(\boldsymbol{\theta}_{0})\boldsymbol{d}\right),
\label{22.23}%
\end{equation}
under $H_{1,n}$ given in (\ref{22.12}), and by
\begin{equation}
W_{n}({\boldsymbol{\widehat{\boldsymbol{\theta}}_{n,\alpha}}}%
)\underset{n\rightarrow\infty}{\overset{\mathcal{L}}{\longrightarrow}}\chi
_{r}^{2}\left(  \boldsymbol{\delta}'\left(  \boldsymbol{M}'%
(\boldsymbol{\theta}_{0})\boldsymbol{\Sigma}_{\alpha}(\boldsymbol{\theta}%
_{0})\boldsymbol{M}(\boldsymbol{\theta}_{0})\right)  ^{-1}\boldsymbol{\delta
}\right),  \label{22.24}%
\end{equation}
under $H_{1,n}^{\ast}$ given in (\ref{22.19}).
\end{theorem}
%

\begin{remark}
The result presented under the contiguous alternative hypotheses $H_{1,n}$ in (\ref{22.12}) 
can also be used to give an approximation of the power function at any ${\boldsymbol{\theta}}^{\ast}.$ 
For this purpose, we express $\boldsymbol{\theta}^{\ast}$ as 
\[
{\boldsymbol{\theta}}^{\ast}=\boldsymbol{\theta}_{0}+\left(
{\boldsymbol{\theta}}^{\ast}-\boldsymbol{\theta}_{0}\right)
=\boldsymbol{\theta}_{0}+\frac{1}{\sqrt{n}}\sqrt{n}\left(  {\boldsymbol{\theta
}}^{\ast}-\boldsymbol{\theta}_{0}\right)  .
\]
and then apply the results from Theorem \ref{THM:Cont_power} with 
$d=\sqrt{n}\left(  {\boldsymbol{\theta}}^{\ast}-\boldsymbol{\theta}_{0}\right)  $ 
to obtain the desired power approximation.
\end{remark}


\section{Robustness analysis of Wald-type tests}
\label{Sec 4}

\subsection{Influence function of the Wald-type test statistics}
\label{Sec 4.1}

The influence function (IF) is a classical tool  for studying local robustness of any statistical functional 
under infinitesimal contamination at a distant outlying point \cite{Ronchetti:1979,Ronchetti:1981,Hampel:1986}.
Here, we will use it to theoretically justify the claimed robustness of our proposed Wald-type test statistics 
under the parametric proportional hazards model (\ref{EQ:2}).
For this purpose, we first need to define the statistical functional associated with the Wald-type test statistics defined in (\ref{2.10}). 

Let us continue with the notation and assumptions of Section \ref{Sec2}.
In particular, assume that the observations $(x_i;\delta_{i}; \boldsymbol{z}_{i})$, $i = 1, \ldots, n$, 
are IID realizations of $(X, \delta, \boldsymbol{Z})$ having true joint distribution $H$,
and $\boldsymbol{U}_{\alpha}(H)$ denotes the MDPDE functional at $H$ having a tuning parameter $\alpha\geq 0$. 
Using it, we define the statistical functional associated with the Wald-type test statistics for testing the composite hypothesis (\ref{2.7}), 
evaluated at $H$, as given by (ignoring the multiplier $n$) 
\begin{eqnarray}
W_{\alpha}\left(H\right)=\boldsymbol{m}'\left(\boldsymbol{U}_{\alpha}(H)\right)\left(\boldsymbol{M}'\left(\boldsymbol{\theta}_{0}\right)
\boldsymbol{\Sigma} _{\alpha} \left(\boldsymbol{\theta}_{0}\right)\boldsymbol{M}\left(\boldsymbol{\theta}_{0}\right)\right)^{-1} 
\boldsymbol{m}\left(\boldsymbol{U}_{\alpha}(H)\right).
\label{4.5}
\end{eqnarray}
Next, we consider the contaminated distribution $H_{\epsilon}=\left(1-\epsilon\right)H+\epsilon\wedge_{\boldsymbol{y}_{t}}$,
where there is $100\epsilon\%$ contamination by the degenerate distribution $\wedge_{\boldsymbol{y}_{t}}$
at a contamination point $\boldsymbol{y}_{t}= (x_{t},\delta_{t},\boldsymbol{z_{t}})$. 
Then, the $r$-th order IF of our proposed Wald-type test statistic is defined as ($r=1, 2, \ldots$)
$$
IF_r(\boldsymbol{y}_{t},W_{\alpha},\boldsymbol{H})=\frac{\partial^r}{\partial \epsilon^r} W_{\alpha} \left( H_{\epsilon} \right) |_{\epsilon=0}. 
$$

We can follow the general arguments presented in \cite{Ghosh et.al.:2016} to compute the above IF of the MDPDE based Wald-type tests 
for the present case of parametric Cox regression model. 
In particular, for any null parameter value $\boldsymbol{\theta}_0\in\Theta_0$, 
let us denote the joint distribution of $(X, \delta, \boldsymbol{Z})$ to be $H_{\boldsymbol{\theta}_0}$, 
under which $x_i$, given $\delta_i$ and $z_i$, follows the model density $f_{i, \boldsymbol{\theta}_0}$ as in (\ref{2.1}), for each $i=1, \ldots, n$. 
Then, one can easily show that, at this null distribution $H=H_{\boldsymbol{\theta}_{0}}$, 
the first order IF of the Wald-type test functional $W_{\alpha}$ is identically zero and the second order $IF$ has the form
\begin{eqnarray}
\label{4.6}
IF_2\left(\boldsymbol{y}_{t}, W_{\alpha},H_{\boldsymbol{\theta}_{0}}\right) = 
2IF'\left(\boldsymbol{y}_{t},\boldsymbol{U}_{\alpha},H_{\boldsymbol{\theta_{0}}}\right) 
\boldsymbol{N}\left(\boldsymbol{\theta}_{0}\right)IF\left(\boldsymbol{y}_{t},\boldsymbol{U}_{\alpha},H_{\boldsymbol{\theta_{0}}}\right), 
\end{eqnarray}
where $\boldsymbol{N}\left(\boldsymbol{\theta}_{0}\right)=\boldsymbol{M}\big(\boldsymbol{\theta}_{0}\big)
\big[\boldsymbol{M}'\big(\boldsymbol{\theta}_{0}\big) \boldsymbol{\Sigma}_{\alpha}\big(\boldsymbol{\theta}_{0}\big)
\boldsymbol{M}\big(\boldsymbol{\theta}_{0}\big)\big]^{-1} \boldsymbol{M}'\big(\boldsymbol{\theta}_{0}\big)$ and 
$IF( \boldsymbol{y}_{t},\boldsymbol{U}_{\alpha},H_{\boldsymbol{\theta_0}})$ denotes the $IF$ of the MDPDE functional
$\boldsymbol{U}_{\alpha}$ having the form \cite{ghosh}
\begin{eqnarray}
\label{4.1}
IF\left(\boldsymbol{y}_{t},\boldsymbol{U}_{\alpha},H_{\boldsymbol{\theta}_{0}}\right)
=\boldsymbol{J}_{\alpha}^{-1}\left(\boldsymbol{\theta_{0}}\right)
\begin{pmatrix}
\boldsymbol{u}_{n}^{(1,\alpha)}(\boldsymbol{\theta}_0|\boldsymbol{y}_t)\\
\boldsymbol{u}_{n}^{(2,\alpha)}(\boldsymbol{\theta}_0|\boldsymbol{y}_t)\\
\end{pmatrix},
\end{eqnarray}
with $\boldsymbol{u}_{n}^{(1,\alpha)}(\boldsymbol{\theta}|x, \delta, \boldsymbol{z})$,
$\boldsymbol{u}_{n}^{(2,\alpha)}(\boldsymbol{\theta}|x, \delta, \boldsymbol{z})$ and $\boldsymbol{J}_{\alpha}(\boldsymbol{\theta})$ 
being as defined in Section \ref{Sec2}.

We can clearly observe that, due to the presence of the terms 
$\lambda\left(x_{t},\boldsymbol{\gamma}\right)^{\alpha}$, $e^{\alpha \boldsymbol{\beta'z_{t}}}$ and 
$S_{\boldsymbol{\theta}}\left(x_{t}|\boldsymbol{z}_{t}\right)^{\alpha}$ in 
$\boldsymbol{u}_{n}^{(j,\alpha)}(\boldsymbol{\theta}|x, \delta, \boldsymbol{z})$, $j=1,2$,
the above IF of the MDPDE remains bounded at contamination points for any $\alpha >0$.
Therefore the second order IF of the Wald-type test statistic also remains bounded in $\boldsymbol{y}_t$ for all $\alpha>0$,
which indicates the desired robustness  properties for our proposed class of tests. 

\subsection{Level and power influence functions}
\label{Sec 4.2}

We now study the robustness of the level and power of the proposed Wald-type tests 
through the corresponding influence functions \cite{Heritier:1994}.
These were studied for general MDPDE based Wald-type tests with IID complete data in \cite{Ghosh et.al.:2016}
which we will extend here for the randomly censored data.

To start with, we compute the asymptotic level and power against the contiguous alternatives of the form (\ref{22.12}), 
along with the additional contamination, respectively, through the distributions
$
H_{\epsilon,\boldsymbol{y}_{t}}^{L}=\left(1-\frac{\epsilon}{\sqrt{n}}\right)H_{\boldsymbol{{\theta}_{0}}}
+\frac{\epsilon}{\sqrt{n}}\wedge_{\boldsymbol{y}_{t}}$ 
and  
$H_{\epsilon,\boldsymbol{y}_{t}}^{P}=\left(1-\frac{\epsilon}{\sqrt{n}}\right)H_{\boldsymbol{{\theta}_{n}}}
+\frac{\epsilon}{\sqrt{n}}\wedge_{\boldsymbol{y}_{t}}$,
where $ \boldsymbol{\theta}_{0} \in \boldsymbol{\Theta}_{0}$ is the assumed true (null) parameter value and 
$\boldsymbol{y}_{t}=(x_{t},\delta_{t},\boldsymbol{z_{t}})$ is the contamination point as in the previous subsection. 
Also, let us define the following quantities 
\[
\mathcal{T}_{W_{n}}(\epsilon, \boldsymbol{y}_{t})= \lim_{n\rightarrow\infty} P_{{H}^{L}_{\epsilon,\boldsymbol{y}_{t}}}
\left(W_{n}\left(\boldsymbol{\theta}_{0}\right) > \chi^{2}_{r,\tau}\right),
\]
\[
\beta_{W_{n}}(\boldsymbol{{\theta}_{n}},\epsilon,\boldsymbol{y}_{t})= \lim_{n\rightarrow\infty}P_{H^{P}_{\epsilon,\boldsymbol{y}_{t}}}
\left(W_{n}\left( \boldsymbol{\theta}_{0}\right) > \chi^{2}_{r,\tau}\right).
\]
Then, the level influence function (LIF) and the power influence function (PIF),
for our proposed robust Wald-type test statistic $W_{n}(\boldsymbol{\theta_{0}})$ for testing the hypothesis in (\ref{2.7})  
at the level of significance $\tau$, are defined as
\[
LIF\left(\boldsymbol{y}_{t},W_{n},H_{\boldsymbol{\theta}_{0}}\right)
=\frac{\partial}{\partial \epsilon}\mathcal{T}_{W_{n}}(\epsilon, \boldsymbol{y}_{t})|_{\epsilon=0},
\]
and
\[
PIF\left(\boldsymbol{y}_{t},W_{n},H_{\boldsymbol{\theta}_{0}}\right)
=\frac{\partial}{\partial \epsilon}\beta_{W_{n}}(\boldsymbol{{\theta}_{n}},\epsilon,\boldsymbol{y}_{t})|_{\epsilon=0}.
\]

So, we first need to find the asymptotic distribution of the Wald-type test statistic $W_{n}$ 
under the contiguous contaminated distribution $H^{P}_{\epsilon,\boldsymbol{y}_{t}}$
which is presented in the following theorem; its proof is presented in Appendix.

\begin{theorem}
\label{Theo:10}
Under the assumptions of Theorem \ref{Theo:1}. the asymptotic distribution of $W_{n}(\widehat{\boldsymbol{\theta }}_{n,\alpha})$ 
under the distribution $H_{\epsilon,\boldsymbol{y_{t}}}^{P}$ is a non-central $\chi^{2}$ with degrees of freedom $r$ 
and non-centrality parameter 
\begin{equation}
d^{*} = \boldsymbol{d}_{\epsilon,\boldsymbol{y}_{t},\alpha}'\left(\boldsymbol{\theta_{0}}\right)
\boldsymbol{N}\left(\boldsymbol{\theta_{0}}\right)\boldsymbol{d}_{\epsilon,\boldsymbol{y}_{t},\alpha}\left(\boldsymbol{\theta_{0}}\right),
\label{4.13}
\end{equation}
where $\boldsymbol{d}_{\epsilon,\boldsymbol{y}_{t},\alpha}\left(\boldsymbol{\theta_{0}}\right)
= \boldsymbol{d}+\epsilon IF(\boldsymbol{y}_{t},\boldsymbol{U}_{\alpha},H_{\boldsymbol{\theta}_{0}})$.
\end{theorem}

\bigskip
Now, using the infinite series expansion of a non-central chi-square distribution function in terms of that of the central chi-square variables, 
we can deduce from Theorem \ref{Theo:10} that
\begin{eqnarray}
\beta_{W_{n}}\left(\boldsymbol{\theta_{n}},\epsilon, \boldsymbol{y_{t}}\right) 
&=& \lim_{n\rightarrow\infty} P_{\boldsymbol{H^{P}_{\epsilon,\boldsymbol{y}_{t}}}}
\left(W_{n}\left( \boldsymbol{\theta}_{0}\right) > \chi^{2}_{r,\tau}\right) \cong P\left(\chi^{2}_{r}\left(d^{*}\right) > \chi^{2}_{r,\tau} \right)
 \nonumber\\
&=&\sum_{v=0}^{\infty}\boldsymbol{C_{v}}\left(\boldsymbol{M}'(\boldsymbol{\theta}_{0})\boldsymbol{d}_{\epsilon,\boldsymbol{y_{t}},\alpha}\left(\boldsymbol{\theta_{0}}\right),\left(  \boldsymbol{M}'(\boldsymbol{\theta}_{0})\boldsymbol{\Sigma
	 }_{\alpha}(\boldsymbol{\theta}_{0})\boldsymbol{M}(\boldsymbol{\theta}_{0})\right)
	 ^{-1}\right) P\left(\chi^{2}_{r+2v} > \chi^{2}_{r,\tau}\right),~~
\label{4.14}
\end{eqnarray} 
where 
\[
\boldsymbol{C}_{v}(\boldsymbol{t},\boldsymbol{A}) = \frac{(\boldsymbol{t}'\boldsymbol{A}\boldsymbol{t})^{v}}{v!2^{v}}
e^{-\frac{1}{2}\boldsymbol{t}'\boldsymbol{A}\boldsymbol{t}}.
\]

\begin{remark}
If we put $\epsilon = 0$ in (\ref{4.14}), 
we get the asymptotic power of the proposed Wald-type tests under the contiguous alternatives in (\ref{22.12}) 
as given by
\[
\beta_{\alpha}(\boldsymbol{\theta}_{n}) = \beta_{W_{n}}(\boldsymbol{\theta}_{n},0,\boldsymbol{y}_{t}) \cong 
\sum_{v=0}^{\infty}\boldsymbol{C}_{v} \left(\boldsymbol{M}'(\boldsymbol{\theta}_{0})\boldsymbol{d},\left(  \boldsymbol{M}'(\boldsymbol{\theta}_{0})
\boldsymbol{\Sigma}_{\alpha}(\boldsymbol{\theta}_{0})\boldsymbol{M}(\boldsymbol{\theta}_{0})\right)^{-1}\right) 
P\left(\chi^{2}_{r+2v} > \chi^{2}_{r,\tau}\right).
\]
This result can also be obtained using the result (\ref{22.23}) stated in Theorem \ref{THM:Cont_power}.
\end{remark}

Next, substituting $\boldsymbol{d} = \boldsymbol{0}$ in Theorem \ref{Theo:10}, we get the following corollary in a straightforward manner. 

\begin{corollary}
\label{Remark:13}
Under the assumptions of Theorem \ref{Theo:1}, the asymptotic distribution of $W_{n}(\widehat{\boldsymbol{\theta }}_{n,\alpha})$ 
under the distribution $\boldsymbol{H^{L}_{\epsilon,\boldsymbol{y}_{t}}}$ is  non-central chi-square with degrees of freedom $r$ 
and non-centrality parameter 
\[
\epsilon^{2}IF\left(\boldsymbol{y}_{t},\boldsymbol{U}_{\alpha},H_{\boldsymbol{\theta}_{0}}\right)'
\boldsymbol{N}\left(\boldsymbol{\theta}_{0}\right) IF\left(\boldsymbol{y}_{t},\boldsymbol{U}_{\alpha},H_{\boldsymbol{\theta}_{0}}\right).
\]
Thus, the asymptotic level of our MDPDE-based Wald-type tests, under contiguous contamination, is given by
\begin{eqnarray}
\mathcal{T}_{W_{n}}\left(\epsilon, \boldsymbol{y}_{t}\right)&=&\beta_{W_{n}}\left(\boldsymbol{\theta}_{0}, \epsilon, \boldsymbol{y}_{t}\right) 
\nonumber\\
&\cong& \sum_{v=0}^{\infty}\boldsymbol{C}_{v}\left(\epsilon \boldsymbol{M}'(\boldsymbol{\theta}_{0})
IF\left(\boldsymbol{y}_{t},\boldsymbol{U}_{\alpha},H_{\boldsymbol{\theta}_{0}}\right), \left(
\boldsymbol{M}'(\boldsymbol{\theta}_{0})\boldsymbol{\Sigma}_{\alpha}(\boldsymbol{\theta}_{0})\boldsymbol{M}(\boldsymbol{\theta}_{0})\right)^{-1}
\right) P\left(\chi^{2}_{r+2v} > \chi^{2}_{r,\tau}\right).
\nonumber
\end{eqnarray}
\end{corollary}

Now, we can easily obtain the LIF and PIF of the proposed Wald-type test statistics via standard differentiation,
using Theorem \ref{Theo:10} and Corollary \ref{Remark:13}, which are presented in the following theorem. 
See the Appendix for a detailed proof.

\begin{theorem}\label{THM:PIF}
Assume that the conditions of Theorem \ref{Theo:10} is satisfied and that the IF of the MDPDE used is bounded. 
Then, for the proposed Wald-type test, the LIF of any order becomes identically zero 
and the PIF at the significance level $\tau$ has the form  
\begin{eqnarray}
PIF\big(\boldsymbol{y}_{t}, W_{n}, {{H}_{\boldsymbol{\theta}_{0}}}\big) 
=C_{r}^{*}\left(\boldsymbol{d}'\boldsymbol{N}\big(\boldsymbol{\theta}_{0}\big)\boldsymbol{d}\right)
\boldsymbol{d}'\boldsymbol{N}\big(\boldsymbol{\theta}_{0}\big)IF(\boldsymbol{y}_{t}, \boldsymbol{U}_{\alpha},{H}_{\boldsymbol{\theta}_{0}}),
\label{4.8}
\end{eqnarray} 
where  
$C_{r}^{*}\big(s\big)=e^{-\frac{s}{2}}\sum_{v=0}^{\infty}{s^{v-1}}{2^{-v}}(2v-s)P\big(\chi^{2}_{r+2v}>\chi^{2}_{r,\tau}\big)/v!$.
\end{theorem}

Note that the PIF of our proposed MDPDE-based Wald-type test is a linear function of the IF of the corresponding MDPDE as given in (\ref{4.1}). 
As a result, the PIF would be bounded for all $\alpha>0$, which indicates the desired robustness 
for the power of the proposed Wald-type test based on (\ref{2.10}).

\subsection{An example: one covariate and exponential baseline}
\label{Sec 4.3}

In order to better understand the implications of the general results derived above, let us now discuss, 
in detail, a particular example of the testing problem under the parametric Cox regression model. 
For simplicity, consider the exponential baseline hazard $\lambda\left(t, \boldsymbol{\gamma}\right)= \gamma$, with $\gamma >0$, 
and only one covariate, so that the full parameter vector is given by   $\boldsymbol{\theta}=\left(\beta,\gamma\right)$, 
with $\beta \in \mathbb{R}, \gamma \in \mathbb{R}^{+}$. Suppose that we are interested in testing
 \begin{eqnarray}
 H_{0} : \beta =\beta_{0} \hspace{2mm} \mbox{against} \hspace{2mm} 
 H_{1}: \beta \neq \beta_{0},
 \label{4.9}
 \end{eqnarray}
for a pre-fixed real number $\beta_{0}$. Note that, here $\gamma$ is an unknown nuisance parameter. 
If we consider $\beta_{0}=0$, the corresponding testing problem becomes that of testing for the significance of the covariate. 
Clearly, the simpler testing problem in (\ref{4.9}) belongs to the general class of hypotheses in (\ref{2.7}), with $r=1$,
$\boldsymbol{\Theta}_{0}=\big\{\boldsymbol{\theta}=(\gamma, \beta)^T : \beta = \beta_{0}, \gamma \in \mathbb{R}^{+} \big \}$,
$\boldsymbol{m}(\boldsymbol{\theta})=\beta-\beta_{0}$ and $M(\boldsymbol{\theta})= (0,1)'$.

If we denote the MDPDE with tuning parameter $\alpha$ by 
$\widehat{\boldsymbol{\theta}}_{n,\alpha}=\big(\widehat{\gamma}_{n,\alpha},\widehat{\beta}_{n,\alpha}\big)$,
then our proposed Wald-type test statistic (\ref{2.10}) for testing the hypothesis in (\ref{4.9}) simplifies  to 
 \begin{eqnarray}
 W_{n}= W_n(\widehat{\boldsymbol{\theta}}_{n,\alpha})= \frac{n\left(\widehat{\beta}_{n,\alpha} - \beta_{0}\right)^{2}}{\boldsymbol{\Sigma}_{n,\alpha}^{(22)}(\widehat{\boldsymbol{\theta}}_{n,\alpha})},
 \label{4.10}
 \end{eqnarray}
where $\boldsymbol{\Sigma}_{n,\alpha}^{(22)}(\boldsymbol{\theta})$ is the $(2,2)$-th element of 
the $2 \times 2$ covariance matrix  $\boldsymbol{\Sigma}_{n,\alpha}(\boldsymbol{\theta})$ of the MDPDE used. 
Then, under the null hypothesis in (\ref{4.9}), $W_{n}$ asymptotically follows $\chi_{1}^{2}$ distribution
and the test can be performed by comparing $W_{n}$ with the corresponding critical value. 
Further, the approximate expression of power function at the contiguous alternative hypotheses of the form 
$H_{1,n}: \beta= \beta_{0}+ n^{-1/2}d$, with $d\in \mathbb{R}$,
is given by  
$$
\beta_{\alpha}\big(\boldsymbol{\theta}_{n}\big)=1-G_{\chi_{1,\delta}^{2}}\big(\chi_{1,\tau}^{2}\big),
 ~~~~~\mbox{with}~~\delta=\frac{d^{2}}{\boldsymbol{\Sigma}_{n,\alpha}^{(22)}(\boldsymbol{\theta}_0)}, ~\boldsymbol{\theta}_0\in\boldsymbol{\Theta}_{0},
$$ 
where $G_{\chi_{1,\delta}^{2}}$ is the cumulative distribution function of a non-central $\chi_1^{2}(\delta)$ random variable. 
We have numerically computed the asymptotic contiguous powers for testing the hypothesis in (\ref{4.9}) at $5 \%$ level of significance, 
taking  $\beta_{0}=1$ and $\boldsymbol{\theta}_0 =(1,1)^T$,  for different values of $\alpha$;
the results are presented in Table \ref{TAB:Power}. 
It is clear that the asymptotic contiguous power of our proposed Wald-type tests (under pure data) decreases as $\alpha$ increases 
but this loss is not significant at smaller $\alpha> 0$.
 
\begin{table}[h]
 	\caption{Asymptotic contiguous power of the MDPDE based Wald-type tests for testing $\beta=0$ at $5\%$ level of significance 
 		with $\boldsymbol{\theta}_{0}=(1,1)^T$ and different values of $\alpha$ and $d$}
 	\centering	
 	\begin{tabular}{c | c c c c c c c}
 		\hline
 		& \multicolumn{7}{|c}{$\alpha$}\\
 		$d$ & 0 & 0.05 & 0.1 & 0.2 & 0.3 & 0.4 & 0.5\\
 		\hline
 		0.5	&0.7190	&0.7107	&0.7011	&0.6769	&0.6433	&0.5961	&0.5318\\
 		0.7	&0.9447	&0.9408	&0.9361	&0.9231	&0.9029	&0.8696	&0.8147\\
 		0.9	&0.9955	&0.9949	&0.9941	&0.9917	&0.9871	&0.9776	&0.9565\\
 		1.1	&0.9999	&0.9998	&0.9998	&0.9996	&0.9992	&0.9981	&0.9943\\
 		1.3	&1.0000	&1.0000	&1.0000	&1.0000	&1.0000	&0.9999	&0.9996\\
 		1.5	&1.0000	&1.0000	&1.0000	&1.0000	&1.0000	&1.0000	&1.0000\\
 		\hline
 		\end{tabular}
    \label{TAB:Power}
\end{table}

\begin{figure}[!h]
	\centering
	\begin{tabular}{c c}
		\subfloat[$\alpha = 0$]{
			\includegraphics[width=0.4\textwidth]{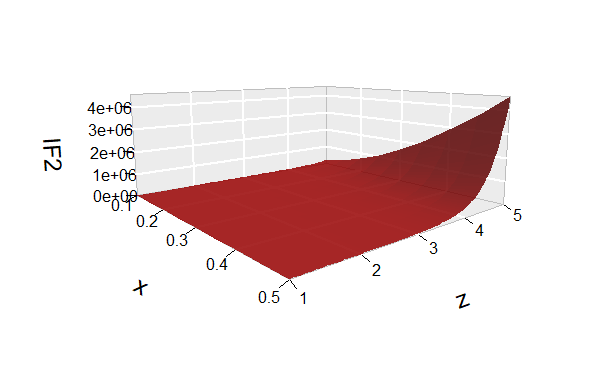}
			\label{FIG:IF2_1_1}}			
		& \subfloat[$\alpha = 0.05$]{
			\includegraphics[width=0.4\textwidth]{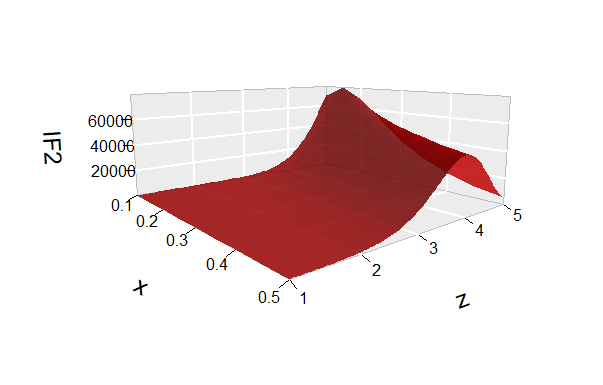}
			\label{FIG:IF2_1_2}}\\			
		\subfloat[$\alpha = 0.1$]{
			\includegraphics[width=0.4\textwidth]{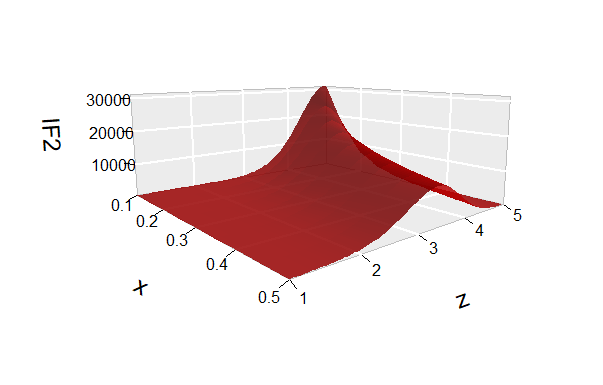}
			\label{FIG:IF2_1_3}}			
		& \subfloat[$\alpha = 0.3$]{
			\includegraphics[width=0.4\textwidth]{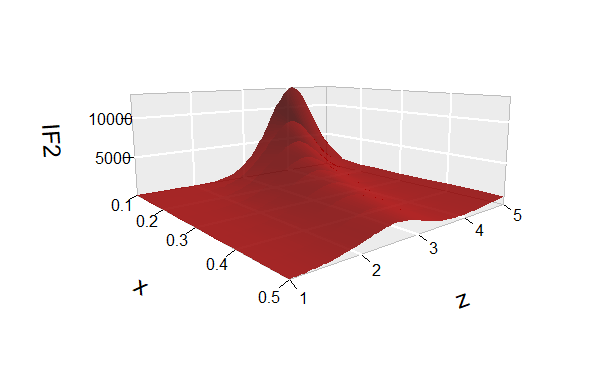}
			\label{FIG:IF2_1_4}}
	\end{tabular}
	\caption{Second order influence function for the MDPDE-based Wald-type test statistics for testing $\beta = 1$, 
		under the parametric Cox regression model with $\boldsymbol{\theta}_{0}=(1,1)^T$, when $\delta_{t} = 1$.}
	\label{FIG:WIF_1}
\end{figure}

\begin{figure}[!h]
	\centering
	\begin{tabular}{c c}
		\subfloat[$\alpha = 0$]{
			\includegraphics[width=0.4\textwidth]{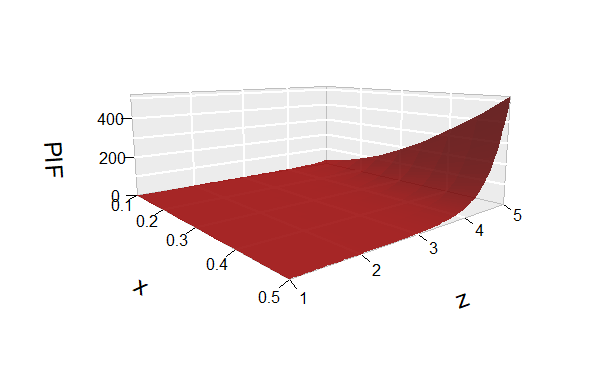}
			\label{FIG:PIF_1_1}}			
		& \subfloat[$\alpha = 0.05$]{
			\includegraphics[width=0.4\textwidth]{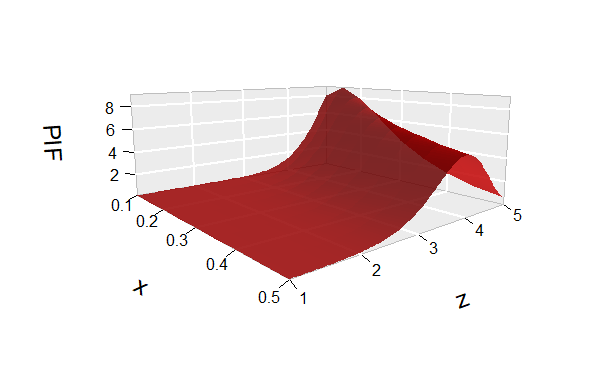}
			\label{FIG:PIF_1_2}}\\			
		\subfloat[$\alpha = 0.1$]{
			\includegraphics[width=0.4\textwidth]{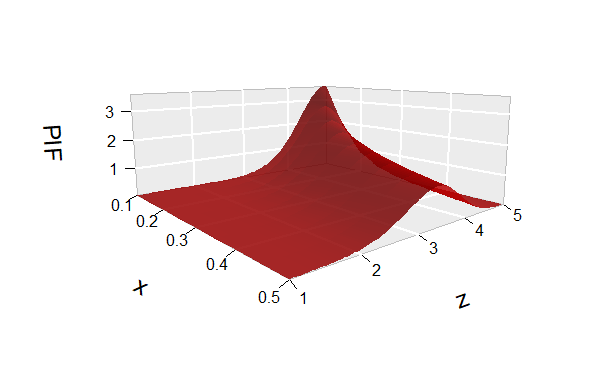}
			\label{FIG:PIF_1_3}}			
		& \subfloat[$\alpha = 0.3$]{
			\includegraphics[width=0.4\textwidth]{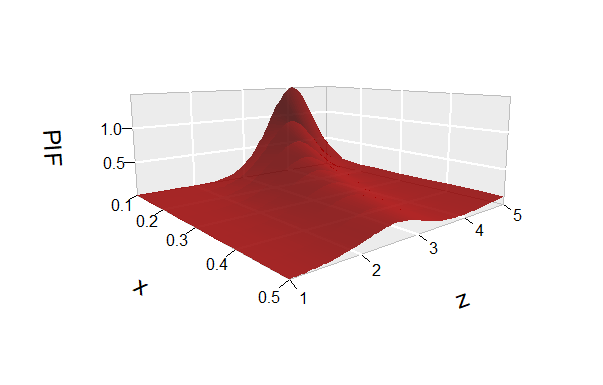}
			\label{FIG:PIF_1_4}}
	\end{tabular}
	\caption{Power influence function for the MDPDE-based Wald-type tests for testing $\beta = 1$, 
		under the parametric Cox regression model with $\boldsymbol{\theta}_{0}=(1,1)^T$ and $d=0.001$, when $\delta_{t} = 1$.}
	\label{FIG:PIF_1}
\end{figure}

Next we study the robustness of the Wald-type tests, based on (\ref{4.10}), for testing the hypothesis in (\ref{4.9}) 
with the help of the second order IF of the Wald-type test statistics as well as its PIF. 
From the general formulas presented in previous sections, we can easily calculate the simplified version of 
these measures for the present case for testing the hypothesis in (\ref{4.9}), which are given by 
 \begin{eqnarray}
 IF_2\big(\boldsymbol{y}_{t},W_{n},\boldsymbol{H_{\theta_{0}}}\big)
 &=&2 IF\big(\boldsymbol{y}_{t},\boldsymbol{U}_{\alpha}^{(\beta)},H_{\boldsymbol{{\theta}_{0}}}\big)^{2}/
 \boldsymbol{\Sigma}_{n,\alpha}^{(22)}(\boldsymbol{\theta}_{0}).
 \label{4.11}
 \\
 PIF \left(\boldsymbol{y}_{t}, W_{n}, \boldsymbol{H_{\theta_{0}}} \right) 
 &=&C_{1}^{*}\left({d}^{2}/\boldsymbol{\Sigma}_{n,\alpha}^{(22)}(\boldsymbol{\theta}_0)\right)
 IF(\boldsymbol{y}_{t}, \boldsymbol{U}_{\alpha}^{(\beta)}, \boldsymbol{H}_{\boldsymbol{\theta}_{0}})d/
 \boldsymbol{\Sigma}_{n,\alpha}^{(22)}(\boldsymbol{\theta}_{0}),
 \label{4.12}
 \end{eqnarray} 
where $\boldsymbol{y}_{t}$ is the contamination point,  
$\boldsymbol{U}_{\alpha}^{(\beta)}$ is the MDPDE functional corresponding to $\beta$ and  its IF
$IF(\boldsymbol{y}_{t}, \boldsymbol{U}_{\alpha}^{(\beta)}, \boldsymbol{H_{\theta_{0}}})$ is given by the second component of 
the $2$-dimensional IF vector of the MDPDE  in (\ref{4.1}).
For illustration, we have plotted these IF$_2$ and PIF over the contamination points $(x_t, z_t)$ within $\boldsymbol{y}_{t}$ 
for different values of tuning parameter $\alpha$ and $\delta_t=1$ in Figures \ref{FIG:WIF_1} and \ref{FIG:PIF_1}, respectively;
the results for $\delta_t=0$ have similar patterns and, hence, they are not presented for brevity. 
For plotting these IFs, we have taken $\beta_0=1$, $\boldsymbol{\theta}_{0}=(1,1)^T$, $d=0.001$, 
and $n=50$ observations are drawn from $\mathcal{N}(1,1)$ distribution as covariate, along with the incorporation of $10 \%$ uniform censoring. 
It is clearly evident that the classical MLE based Wald test statistic has unbounded second order IF and its PIF is also unbounded
indicating its extreme non-robust nature even under infinitesimal contamination. 
But, the MDPDE based Wald-type test statistic (\ref{4.10}) has bounded second order IF 
as well as bounded PIF for all $\alpha>0$, which justifies their desired robustness properties. 
Further, the extent of robustness increases with the value of $\alpha$,
indicated by the re-descending natures of all these IFs.

\section{A robust model selection criterion}
\label{Sec 5}


We now discuss a robust model selection procedure under the parametric proportional hazard model (\ref{EQ:2}), 
based on our MDPDE discussed in Section \ref{Sec2}.
In particular, we extend the idea of the divergence information criterion (DIC), 
which was originally discussed in \cite{Mattheou/Lee:2009,Mantalos et al:2010} for IID complete data
and recently extended for non-homogeneous (complete) data in \cite{Kurata/Hamada:2019}. 

Recall that, in our present case, given the values of $\boldsymbol{z}_{i}$ and $\delta_{i}$, 
the observed lifetime variable $X_{i}$ has true densities $h_{i}(x)$, for $i=1,\ldots,n$,
which we model respectively by the parametric density $f_{i,\boldsymbol{\theta}}(x)$ given in (\ref{2.1}). 
Now, suppose that there are $l$ many candidate models available for a given set of observed data
where the $s$-th model is denoted as $\{M_{1}^{\left(s\right)},M_{2}^{\left(s\right)},...,M_{n}^{\left(s\right)}\}$ for $s =1, \ldots, l$.
Also, assume that, for each $s$, the corresponding model of the $i$-th observation, namely $M_{i}^{\left(s\right)}$ 
is parametrized by a density function of the form $f_{i,\boldsymbol{\theta}}(x)$, as in (\ref{2.1}), 
with possibly different $\boldsymbol{\theta}$ for different $s$. 

Under such a set-up, the computation of the usual model selection criterion AIC uses the KL divergence 
to quantify the discrepancy between the true distribution $h_{i}$ and parametric family of models $f_{i,\boldsymbol{\theta}}$. 
Here, we will extend it by considering the DPD measure $d_{\alpha}$ in place of the KL divergence to define the DIC. 
Note that, as discussed in Section \ref{Sec2}, minimizing the DPD measure between the data and the postulated model 
amount to the minimization of the simpler objective function $H_{n,\alpha}\left(  \boldsymbol{\theta}\right)$ as defined in (\ref{EQ:H_na}).
Thus, our robust model selection criterion should then select the parametric model 
for which  $E_{X_{1}, \ldots, X_{n}}[H_{n,\alpha}(\widehat{\boldsymbol{\theta }}_{n,\alpha})]$ will be minimized, 
where $\widehat{\boldsymbol{\theta }}_{n,\alpha}$ denotes the corresponding MDPDE with tuning parameter $\alpha$. 
Since exact computation is not generally possible, we first find an asymptotic unbiased estimator of 
$E_{X_{1}, \ldots, X_{n}}[H_{n,\alpha}(\widehat{\boldsymbol{\theta }}_{n,\alpha})]$  and 
then minimize this estimator to select the optimum model robustly. 
This leads to the following definition of the DIC, the resulting information criterion for MDPDE-based robust model selection.

\begin{definition}
Let, $\{M_{1}^{\left(s\right)},M_{2}^{\left(s\right)},...,M_{n}^{\left(s\right)}\}$, $s = 1, \ldots, l$, 
be the set of $l$ candidate models for the observations $x_{i}$ for $i=1, \ldots, n$, 
where each of them are the parametric proportional hazards model having densities as in (\ref{EQ:2}). 
Then,  we select models $\left(M_{1}^{*}, M_{2}^{*},..., M_{n}^{*}\right)$ that satisfy 
\[\left(M_{1}^{*}, M_{2}^{*},..., M_{n}^{*}\right)=\underset{s}{min}~~DIC_{n,\alpha} \left(M_{1}^{\left(s\right)},M_{2}^{\left(s\right)},...,M_{n}^{\left(s\right)}\right)\] 
where the divergence information criterion (DIC) is defined as
\begin{equation}
\label{5.1}
DIC_{n,\alpha} \left(M_{1}^{\left(s\right)},M_{2}^{\left(s\right)},...,M_{n}^{\left(s\right)}\right)
=H_{n,\alpha}\left(\widehat{\boldsymbol{\theta}}_{n,\alpha}^{(s)}\right) + \frac{\alpha+1}{n}trace\left(K_{n,\alpha}\left(\widehat{\boldsymbol{\theta}}_{n,\alpha}^{(s)}\right)
J_{n,\alpha}^{-1}\left(\widehat{\boldsymbol{\theta}}_{n,\alpha}^{(s)}\right)\right),
\end{equation} 
withe $\widehat{\boldsymbol{\theta}}_{n,\alpha}^{(s)}$ being the MDPDE of the common parameter $\boldsymbol{\theta}$ under the $s$-th models
$\left(M_{1}^{\left(s\right)},M_{2}^{\left(s\right)},...,M_{n}^{\left(s\right)}\right)$, 
and $H_{n,\alpha}\left(\boldsymbol{\theta}\right)$, $K_{n,\alpha}\left(\boldsymbol{\theta}\right)$ and $J_{n,\alpha}\left(\boldsymbol{\theta}\right)$ 
are as defined in Section \ref{Sec2}.
\end{definition}

That the DIC, as defined in the above definition, is indeed an asymptotically unbiased estimator of 
$E_{X_{1}, \ldots, X_{n}}[H_{n,\alpha}(\widehat{\boldsymbol{\theta }}_{n,\alpha})]$ can be easily shown
by extending the arguments from \cite{Mattheou/Lee:2009,Kurata/Hamada:2019}; 
the following theorem summarizes the final rigorous result.

\begin{theorem}\label{THM:DIC} 
Suppose that the assumptions of Theorem \ref{Theo:1} hold for the randomly censored observations $(x_i, \delta_i, \boldsymbol{z}_i)$
for $i=1, \ldots, n$, and under the model	$\left(M_{1},M_{2}, \cdots, M_{n}\right)$,
where $M_i$ corresponds to the parametric proportional hazards regression model having the conditional density as in (\ref{EQ:2}), $i=1, \ldots, n$.
Denote the MDPDE of the parameter $\boldsymbol{\theta}$ under this specified model by  $\widehat{\boldsymbol{\theta }}_{n,\alpha}$ 
Then, the DIC measure, $DIC_{n,\alpha} \left(M_{1},M_{2}, \cdots, M_{n}\right)$ as defined in (\ref{5.1}), 
is asymptotically unbiased for $E_{X_{1}, \ldots, X_{n}}[H_{n,\alpha}(\widehat{\boldsymbol{\theta }}_{n,\alpha})]$.
\end{theorem}

\begin{remark}
When $\alpha \rightarrow 0$, then the MDPDE $\widehat{\boldsymbol{\theta}}_{n,\alpha}^{(s)}$ computed under the $s$-th model $\left(M_{1}^{\left(s\right)},M_{2}^{\left(s\right)},...,M_{n}^{\left(s\right)}\right)$ coincides with the corresponding MLE, 
say  $\widehat{\boldsymbol{\theta}}^{(s)}$. Thus, we have 
$$
\lim\limits_{\alpha\rightarrow 0} DIC_{n,\alpha} \left(M_{1}^{\left(s\right)},M_{2}^{\left(s\right)},...,M_{n}^{\left(s\right)}\right) 
= -\frac{1}{n}\log L_{n}\left(\widehat{\boldsymbol{\theta}}^{(s)}\right)
+\frac{1}{n}trace\left(K_{n,0}\left(\widehat{\boldsymbol{\theta}}^{(s)}\right)J_{n,0}^{-1}\left(\widehat{\boldsymbol{\theta}}^{(s)}\right)\right),
$$ 
which is nothing but a constant multiple of the Takeuchi information criterion (TIC), 
a generalized version of the AIC \cite{Takeuchi:1976}. 
\end{remark}


 \section{Simulations: Performance of the MDPDE based Wald-type tests}
 \label{Sec 6}

Here we present some interesting findings from an extensive simulation study in order to examine 
the finite-sample power and level of the proposed MDPDE based Wald-type tests in some specific cases under the parametric Cox regression model. 
For the purpose of computation of the MDPDEs, in all our numerical illustrations, we have implemented a two stage optimization technique. 
At the first stage, for a fixed baseline parameter value ($\boldsymbol{\gamma}$), 
we have used the Newton-Raphson method to estimate  the regression coefficients ($\boldsymbol{\beta}$) and then, in the second stage, 
we used the inbuilt R function \textsf {optim} to estimate $\boldsymbol{\gamma}$ for the estimated (fixed) value of $\boldsymbol{\beta}$. 
These two steps are repeated iteratively until the values of both the estimates converges.
Subsequent implementation of the testing and model selection proposals are done in R; 
the unoptimized codes are available from the authors upon request.

In all simulations, we have compared our proposed robust MDPDE-based Wald-type tests for different $\alpha>0$ 
with the classical MLE-based Wald test (which is the same as the MDPDE-based test at $\alpha=0$).
Additionally, we compare our proposal, when testing for the regression coefficients, 
with the Wald-type tests based on two semi-parametric estimates, namely the standard partial likelihood estimate (PLE) 
and the robust estimator of Bednarski \cite{Bednarski:1993} (denoted as BRE), of the regression coefficient. 
These semi-parametric estimates and their standard errors are computed using the R package \textsf{coxrobust} \cite{Bednarski/Borowicz:2006} 
which are then used to perform the appropriate tests in our simulations.

\begin{table}[!b]
	\caption{Empirical levels of the MDPDE based Wald-type tests, and its semi-parametric competitors, 
		for testing $H_{0}: \beta_{2}=1$ in a parametric Cox regression model with exponential baseline}
	\centering	
	\resizebox{0.8\textwidth}{!}{
		\begin{tabular}{c l |c c c c c c c|c c}
			\hline
			Cens.&  & \multicolumn{7}{c|}{Parametric MDPDE with $\alpha$} & \multicolumn{2}{|c}{Semiparametric}\\
			Prop. & $\epsilon$ & 0(MLE) & 0.05 & 0.1 & 0.2 & 0.3 & 0.4 & 0.5 & PLE & BRE\\
			\hline 	
			\multicolumn{11}{c}{$n=50$}\\
			\hline
			5$\%$ & 0 & 0.079 &	0.082 &	0.086 &	0.091 &	0.095 &	0.1	& 0.106 & 0.083 & 0.088\\
			& 0.05 & 0.745 & 0.617 & 0.503 & 0.394 & 0.236 &	0.149 &	0.093 &0.209&	0.155\\
			& 0.1 & 0.758 &	0.629 &	0.515 &	0.401 &	0.288 &	0.166 &	0.102 &0.231 & 0.169\\
			10$\%$ & 0 & 0.081 & 0.083 & 0.087 &	0.093 &	0.097 &	0.102 &	0.109 &	0.09&0.094\\
			& 0.05 & 0.774 & 0.631 & 0.515 &	0.407 &	0.269 &	0.174 &	0.104 &	0.243&0.162\\	
			& 0.1 & 0.78 & 0.642 & 0.528 & 0.412 &	0.309 &	0.187 &	0.108 &	0.269&0.175\\
			\hline
			\multicolumn{11}{c}{$n=100$}\\
			\hline
			5$\%$ & 0 & 0.067 &	0.068 &	0.07 & 	0.071 &	0.074 &	0.079 &	0.084 &	0.07&0.073\\
			& 0.05 & 0.768 & 0.634 & 0.513 &	0.403 &	0.225 &	0.127 &	0.088 &0.20	&0.146\\
			& 0.1 & 0.772 &	0.637 &	0.527 &	0.415 &	0.274 &	0.15 &	0.093 &0.223&0.152\\
			10$\%$ & 0 & 0.069 & 0.071 &	0.073 &	0.075 &	0.079 &	0.083 &	0.089 &0.074	&0.077\\
			& 0.05 & 0.782 & 0.652 & 0.524 & 	0.421 &	0.241 &	0.134 &	0.092 &0.209	&0.15\\
			& 0.1 & 0.785 &	0.646 &	0.548 &	0.426 &	0.293 &	0.169 &	0.098 &0.248	&0.16\\
			\hline
			\multicolumn{11}{c}{$n=200$}\\
			\hline
			5$\%$ & 0 & 0.056 &	0.057 &	0.059 &	0.062 &	0.066 &	0.071 &	0.076 &0.065	&0.069\\
			& 0.05 & 0.781 & 0.649 & 0.536 & 	0.417 & 0.219 &	0.112 &	0.081 &0.174	&0.139\\
			& 0.1 & 0.783 &	0.644 &	0.531 &	0.42 &	0.269 &	0.143 &	0.086 &0.186	&0.145\\
			10$\%$ & 0 & 0.058 & 0.06 & 0.062 & 	0.065 &	0.069 &	0.074 &	0.079 &0.068	&0.073\\
			& 0.05 & 0.792 & 0.662 & 0.55 &	0.428 &	0.243 &	0.121 &	0.086 &0.187	&0.148\\
			& 0.1 & 0.796 & 0.673 &	0.544 &	0.433 &	0.287 &	0.162 &	0.094 &0.211	&0.154\\   
			\hline
	\end{tabular}}
	\label{TAB:1}
\end{table}

\subsection{Testing for a regression coefficient when baseline hazard is a constant}
\label{Sec 6.1}

For our first illustrations, we consider the simple exponential baseline hazard, in the parametric Cox regression model (\ref{2.1}), 
given by $\lambda\left(x,\gamma  \right) = \gamma \in [0, \infty)$.
So, our parameter of interest is the $p+1$ dimensional vector $\boldsymbol{\theta=}\left(\gamma, \boldsymbol{\beta}'\right)'$. 
We have simulated samples of size $n=50, 100$ and $200$ from this model, 
with the covariates being drawn from the standard normal distribution, and with uniform censoring of proportions $5\%$ and $10\%$. 
We have considered three covariates ($p=3$) with the true parameters $\gamma=1$ and 
$\boldsymbol{\beta}=\left(\beta_{1},\beta_{2},\beta_{3}\right)'=\left(1,1,1\right)'$.
Then, we test for the hypothesis $H_{0}: \beta_{2}=1$, using the proposed $W_{n}\left(\widehat{\boldsymbol{\theta}}_{n,\alpha}\right)$, 
defined in (\ref{2.10}), as the test statistic. 
Based on $1000$ replications, we have computed the empirical levels of the tests as the proportion of
test statistics (among the $1000$ replications) exceeding the chi-square critical value (proportion of rejection).
Next we wish to compute the empirical power of the tests by repeating the above process
but under the contiguous alternative hypothesis $H_{1,n}: \beta_{2}=\beta_{2n}$, where $\beta_{2n}=1+\frac{d}{\sqrt{n}}$, for different  $d>0$.
In practice, however, we test the null hypothesis $\beta_2 = \beta_{2n}$ against the alternative $\beta_2 = 1$. 
Switching these hypothesis in this way has two advantages. 
Firstly, the power under pure data can be estimated with the same data  generated from $\beta_2 = 1$. 
Secondly, the same type of contamination now helps us to study the inflation in the level and the drop in power as we will shortly see. 
In case of power, the results for $d=4$ and $6$ are reported here. 
To illustrate the desired robustness, we recalculate the level and power of the Wald-type tests 
after the introduction of $100\epsilon\%$ contamination in each sample considered in the study with $\epsilon= 0.05,0.1$. 
The contaminating observations are generated from an exponential distribution with mean $31$. 
The same simulations are also repeated for the semi-parametric competitions, namely the Wald-type tests based on PLE and BRE.
The resulting values of the empirical levels and powers obtained from the different simulation scenarios are reported in 
Tables \ref{TAB:1} and \ref{TAB:2}.

\begin{table}[!h]
	\caption{Empirical powers of the MDPDE based Wald-type tests, and its semi-parametric competitors, 
		for testing the null hypothesis $\beta_2 = 1 + \frac{d}{\sqrt{n}}$ in a parametric Cox regression model with exponential baseline, 
		calculated at the alternative $\beta_2 = 1$, which is contiguous to the null. }
	\centering	
	\resizebox{0.77\textwidth}{!}{
	\begin{tabular}{c c l |c c c c c c c|c c}
		\hline
		& Cens.&  & \multicolumn{7}{c|}{Parametric MDPDE with $\alpha$} & \multicolumn{2}{c}{Semi-parametric}\\
		$d$ & Prop. & $\epsilon$ & 0(MLE) & 0.05 & 0.1 & 0.2 & 0.3 & 0.4 & 0.5 & PLE & BRE\\
		\hline 	
		\multicolumn{12}{c}{$n=50$}\\
		\hline
		4 & 5$\%$ & 0 & 0.914 &	0.906 &	0.877 &	0.838 &	0.826 &	0.779&	0.726&0.869	&0.828\\
		& & 0.05 & 0.282&	0.367&	0.478&	0.564&	0.643&	0.748&	0.814&0.742	&0.779\\
		& & 0.1 & 0.289	&0.375	&0.484	&0.573	&0.655	&0.751	&0.824&0.728	&0.763\\
		& 10$\%$ & 0 & 0.901	&0.899	&0.87	&0.815	&0.798	&0.759	&0.72&0.845	&0.809\\
		& & 0.05 & 0.264	&0.338	&0.459	&0.534	&0.627	&0.732	&0.807&0.739	&0.764 \\
		& & 0.1 & 0.277	&0.364	&0.467	&0.551	&0.643	&0.739	&0.817&0.727	&0.754\\
		6 & 5$\%$ & 0 & 0.956	&0.94	&0.935	&0.915	&0.903	&0.857	&0.814&0.918	&0.889\\
		& & 0.05 & 0.391	&0.472	&0.584	&0.694	&0.783	&0.842	&0.896&0.843	&0.871\\
		& & 0.1 & 0.402	&0.493	&0.591	&0.706	&0.79	&0.856	&0.9&0.856	&0.878\\
		& 10$\%$ & 0 & 0.945	&0.929	&0.921	&0.911	&0.887	&0.829	&0.792&0.896	&0.877\\
		& & 0.05 & 0.386	&0.461	&0.576	&0.682	&0.757	&0.827	&0.887&0.819	&0.852\\
		& & 0.1 & 0.395	&0.484	&0.582	&0.688	&0.771	&0.834	&0.892&0.831	&0.857\\
		\hline
		\multicolumn{12}{c}{$n=100$}\\
		\hline
		4 & 5$\%$ & 0 & 0.962	&0.957	&0.924	&0.897	&0.884	&0.835	&0.775&0.898	&0.874\\
		& & 0.05 & 0.353	&0.428	&0.541	&0.626	&0.714	&0.806	&0.869&0.779	&0.802\\
		& & 0.1 & 0.364	&0.437	&0.548	&0.633	&0.724	&0.811	&0.872&0.79	&0.807\\
		& 10$\%$ & 0 & 0.96	&0.946	&0.919	&0.889	&0.865	&0.827	&0.771&0.885	&0.864\\
		& & 0.05 & 0.346	&0.413	&0.529	&0.612	&0.701	&0.794	&0.858&0.764	&0.792\\
		& & 0.1 & 0.351	&0.42	&0.535	&0.622	&0.71	&0.803	&0.865&0.778	&0.796\\
		6 & 5$\%$ & 0 & 0.985	&0.98	&0.977	&0.971	&0.961	&0.915	&0.881&0.941	&0.924\\
		& & 0.05 & 0.454	&0.538	&0.649	&0.731	&0.823	&0.902	&0.948&0.893	&0.918\\
		& & 0.1 & 0.468	&0.547	&0.661	&0.742	&0.838	&0.91	&0.957&0.906	&0.927\\
		& 10$\%$ & 0 & 0.979	&0.976	&0.969	&0.961	&0.949	&0.907	&0.853&0.932	&0.918\\
		& & 0.05 & 0.447	&0.521	&0.636	&0.717	&0.809	&0.891	&0.942&0.884	&0.907\\
		& & 0.1 & 0.453	&0.529	&0.644	&0.732	&0.821	&0.898	&0.943&0.897	&0.913\\
		\hline
		\multicolumn{12}{c}{$n=200$}\\
		\hline
		4 & 5$\%$ & 0 & 0.981	&0.976	&0.937	&0.922	&0.915	&0.871	&0.826&0.934	&0.907\\
		& & 0.05 & 0.416 &0.497	&0.587	&0.694	&0.768	&0.837	&0.885&0.804	&0.831\\
		& & 0.1 & 0.419	&0.504	&0.592	&0.706	&0.782	&0.846	&0.889&0.785	&0.828 \\
		& 10$\%$ & 0 &0.979	&0.975	&0.925	&0.913	&0.901	&0.851	&0.814&0.925	&0.902 \\
		& & 0.05 & 0.409	&0.491	&0.569	&0.682	&0.752	&0.828	&0.867&0.795	&0.824\\
		& & 0.1 & 0.414	&0.499	&0.584	&0.687	&0.761	&0.824	&0.871&0.78	&0.817\\
		6 & 5$\%$ & 0 & 0.999	&0.998	&0.997	&0.992	&0.978	&0.959	&0.928&0.97	&0.949\\
		& & 0.05 & 0.519	&0.617	&0.728	&0.812	&0.893	&0.947	&0.989&0.932	&0.951\\
		& & 0.1 & 0.527	&0.623	&0.734	&0.822	&0.902	&0.951	&0.992&0.944	&0.958\\
		& 10$\%$ & 0 & 0.992	&0.99	&0.987	&0.982	&0.964	&0.948	&0.916&0.961	&0.943\\
		& & 0.05 & 0.492	&0.596	&0.705	&0.798	&0.877	&0.932	&0.975&0.912	&0.937\\
		& & 0.1 & 0.508	&0.615	&0.725	&0.813	&0.884	&0.937	&0.987&0.928	&0.949\\
		\hline
	\end{tabular}}
\label{TAB:2}
\end{table}

The observed levels in Table \ref{TAB:1} are all higher than the nominal level of $0.05$. 
The levels of the proposed tests exhibit an increasing relationship with $\alpha$. 
With increasing sample size, all the observed levels tend towards the nominal level. 
In case of power (Table \ref{TAB:2}), the empirical power decreases with increase in $\alpha$, 
but increases with increase in the sample size. 
Also the powers are higher for $d = 6$ than those for $d = 4$, as would be expected. 
Under pure data the power climbs up to be practically equal to $1$ at $d = 6$ and $n = 200$. 
Under contamination, the level of the classical Wald test is significantly inflated, 
but they climb down to more acceptable levels with increasing $\alpha$. 
On  the other hand, the contamination leads to a substantial drop in power for the classical Wald test, 
but the powers increase with increasing $\alpha$ and, for moderate values of $\alpha$, there is little or no loss in power.

The performance of the \textsf{semi-parametric} tests, based on the PLE and BRE,  fall somewhere in the middle of the range of our proposed tests. 
Between them the PLE generates slightly more efficient tests while the BRE leads to more robust outcomes. 
On the whole it appears that the level and the power of the these semi-parametric tests (using PLE and BRE) 
are dominated by the MDPDE based tests for low values of $\alpha$ under pure data, 
and relatively larger values of $\alpha$ for contaminated data (Here, in the context of power, to dominate is to have a higher value of the power, 
while in the context of level it indicates that the observed level is closer to the nominal value). 
Thus, with a suitable tuning parameter selection strategy which lets the user choose the optimal tuning parameter $\alpha$ 
depending on the amount of anomaly in the data, our proposal can beat these competitors in each situation.

\subsection{Testing for a regression coefficient when the baseline is Weibull}
\label{Sec 6.2}

Next we have considered the Weibull baseline hazard in the parametric Cox regression model (\ref{2.1}), 
given by $\lambda\left(x,\boldsymbol{\gamma}  \right) = \gamma_{2} \gamma_{1}^{\gamma_{2}}   x^{\gamma_{2}-1}$, 
where $\boldsymbol{\gamma }= \left(\gamma_{1},\gamma_{2}\right)' \in$ [0, $\infty$)$^{2}$. 
Then, our parameter of interest is the $p+2$ dimensional vector $\boldsymbol{\theta=}\left(  \boldsymbol{\gamma,\beta}'\right)'.$ 
We again simulate random samples of size $n=50, 100, 200$ and $300$ from this model 
with the covariates being drawn from the standard normal distribution, 
along with uniform censoring proportions of $5\%$ and $10\%$, as in the exponential baseline case. 
We have taken $p=3$, where the true parameters are $\gamma_{1}=1, \gamma_{2}=2$ and 
$\beta=\left(0.2,0.3,0.5\right)'=\left(\beta_{1},\beta_{2},\beta_{3}\right)'$
and considered the problem of testing the hypothesis $H_{0}: \beta_{2}=0.3$ 
against the contiguous alternative hypothesis $H_{1,n}: \beta_{2}=\beta_{2n}$, where $\beta_{2n}=0.3+\frac{d}{\sqrt{n}}$ with $d=2$. 
We have used $W_{n}\left(\widehat{\boldsymbol{\theta}}_{n,\alpha}\right)$  defined in (\ref{2.10}) as the test statistic. 
Again using $1000$ replications, we have computed the empirical levels and powers of the tests as described in Section \ref{Sec 6.1}. 
Finally, to illustrate the claimed robustness, we recalculate the level and power of the Wald-type tests 
after introducing $100\epsilon \%$ contamination in each sample of the previous simulation exercise with $\epsilon=0.05,0.1$. 
The contaminating observations are generated from exponential distribution with mean $8.5$, for the level and power calculations. 
We report all the resulting empirical levels and powers obtained from different simulation scenarios, 
along with the result obtained from the competitive semi-parametric tests, in Tables \ref{TAB:3} and \ref{TAB:4}.

\begin{table}[!h]
	\caption{Empirical levels of the MDPDE based Wald-type tests, and its semi-parametric competitors, 
		for testing $H_{0}: \beta_{2}=0.3$ in a parametric Cox regression model with Weibull baseline}
	\centering	
	\resizebox{0.8\textwidth}{!}{
		\begin{tabular}{c l |c c c c c c c|cc}
			\hline
			Cens.&  & \multicolumn{7}{c|}{Parametric MDPDE with $\alpha$} & \multicolumn{2}{c}{Semi-parametric}\\
			Prop. & $\epsilon$ & 0(MLE) & 0.05 & 0.1 & 0.2 & 0.3 & 0.4 & 0.5 &PLE & BRE\\
			\hline 	
			\multicolumn{11}{c}{$n=50$}\\
			\hline
			5$\%$ & 0 &0.088	&0.089	&0.092	&0.096	&0.101	&0.105	&0.112&0.094	&0.103\\
			& 0.05 &0.759	&0.618	&0.529	&0.394	&0.298	&0.191	&0.116&0.208	&0.167\\
			& 0.1 &0.778	&0.631	&0.543	&0.416	&0.313	&0.205	&0.125&0.224	&0.184\\
			10$\%$ & 0 &0.09	&0.092	&0.095	&0.099	&0.104	&0.109	&0.117&0.098	&0.108\\
			& 0.05 &0.771	&0.634	&0.548	&0.409	&0.31	&0.201	&0.122&0.227	&0.178\\
			& 0.1 &0.787	&0.651	&0.562	&0.434	&0.328	&0.219	&0.134&0.241	&0.195\\
			\hline 	
			\multicolumn{11}{c}{$n=100$}\\
			\hline
			5$\%$ & 0 &0.076	&0.078	&0.08	&0.083	&0.086	&0.091	&0.097&0.079	&0.085\\
			& 0.05 &0.776	&0.649	&0.537	&0.402	&0.287	&0.172	&0.105&0.189	&0.154\\
			& 0.1 &0.797	&0.687	&0.569	&0.438	&0.301	&0.192	&0.116&0.204	&0.162\\
			10$\%$ & 0 &0.078	&0.08	&0.083	&0.086	&0.09	&0.094	&0.101&0.085	&0.089\\
			& 0.05 &0.792	&0.679	&0.554	&0.423	&0.299	&0.186	&0.113&0.209	&0.166\\
			& 0.1 &	0.809	&0.704	&0.581	&0.453	&0.315	&0.206	&0.124&0.233	&0.179\\
			\hline 	
			\multicolumn{11}{c}{$n=200$}\\
			\hline
			5$\%$ & 0 &0.061	&0.062	&0.064	&0.067	&0.071	&0.076	&0.084&0.067	&0.072\\
			& 0.05 &0.794	&0.668	&0.544	&0.417	&0.269	&0.154	&0.089&0.168	&0.136\\
			& 0.1 &0.813	&0.697	&0.584	&0.456	&0.293	&0.176	&0.101&0.187	&0.148\\
			10$\%$ & 0 &0.063	&0.065	&0.067	&0.071	&0.075	&0.08	&0.088&0.071	&0.077\\
			& 0.05 &0.806	&0.682	&0.569	&0.429	&0.284	&0.168	&0.096&0.191	&0.145\\
			& 0.1 &0.829	&0.711	&0.596	&0.47	&0.308	&0.191	&0.109&0.212	&0.159\\
			\hline 	
			\multicolumn{11}{c}{$n=300$}\\
			\hline
			5$\%$ & 0 &0.054	&0.056	&0.058	&0.06	&0.063	&0.067	&0.075&0.062	&0.066\\
			& 0.05 &0.811	&0.694	&0.572	&0.434	&0.261	&0.14	&0.077&0.142	&0.117\\
			& 0.1 &0.834	&0.719	&0.602	&0.469	&0.282	&0.162	&0.085&0.163	&0.131\\
			10$\%$ & 0 &0.057	&0.059	&0.062	&0.065	&0.069	&0.074	&0.081&0.068	&0.073\\
			& 0.05 &0.828	&0.706	&0.586	&0.451	&0.278	&0.153	&0.086&0.154	&0.129\\
			& 0.1 &0.85	&0.732	&0.618	&0.483	&0.297	&0.175	&0.094&0.168	&0.144\\
			\hline
	\end{tabular}}
	\label{TAB:3}
\end{table}

 \begin{table}[!h]
 	\caption{Empirical powers of the MDPDE based Wald-type tests, and its semi-parametric competitors, 
 		for testing the null hypothesis $\beta_2 = \beta_{2n}=0.3+\frac{2}{\sqrt{n}}$ in a parametric Cox regression model with Weibull baseline, calculated at the alternative $\beta_2 = 0.3$, which is contiguous to the null. }
 	\centering	
\resizebox{0.8\textwidth}{!}{
 	\begin{tabular}{c l |c c c c c c c|c c}
 		\hline
 		Cens.&  & \multicolumn{7}{c|}{Parametric MDPDE with $\alpha$} & \multicolumn{2}{c}{Semi-parametric}\\
 		Prop. & $\epsilon$ & 0(MLE) & 0.05 & 0.1 & 0.2 & 0.3 & 0.4 & 0.5 & PLE& BRE\\
 		\hline 	
 		\multicolumn{11}{c}{$n=50$}\\
 		\hline
 		5$\%$ & 0 &0.917	&0.914	&0.906	&0.891	&0.849	&0.785	&0.752&0.807	&0.773\\
 		& 0.05 &0.396	&0.458	&0.521	&0.597	&0.654	&0.701	&0.743&0.729	&0.739\\
 		& 0.1 &0.409	&0.469	&0.538	&0.608	&0.677	&0.713	&0.756&0.743	&0.752\\
 		10$\%$ & 0 &0.907	&0.897	&0.886	&0.875	&0.836	&0.762	&0.738&0.786	&0.756\\
 		& 0.05 &0.374	&0.421	&0.496	&0.563	&0.621	&0.676	&0.73&0.678	&0.713\\
 		& 0.1 &0.389	&0.437	&0.514	&0.582	&0.653	&0.698	&0.741&0.695	&0.733\\
 		\hline 	
 		\multicolumn{11}{c}{$n=100$}\\
 		\hline
 		5$\%$ & 0 &0.969	&0.963	&0.958	&0.947	&0.905	&0.873	&0.798&0.840	&0.815\\
 		& 0.05 &0.431	&0.489	&0.542	&0.624	&0.683	&0.728	&0.789&0.745	&0.782\\
 		& 0.1 &0.447	&0.506	&0.557	&0.643	&0.698	&0.744	&0.803&0.769	&0.794\\
 		10$\%$ & 0 & 0.953	&0.944	&0.932	&0.92	&0.881	&0.854	&0.779&0.824	&0.793\\
 		& 0.05 &0.409	&0.462	&0.523	&0.603	&0.658	&0.706	&0.773&0.714 	&0.76\\
 		& 0.1 &0.423	&0.485	&0.532	&0.619	&0.675	&0.722	&0.782&0.736	&0.775\\
 		\hline 	
 		\multicolumn{11}{c}{$n=200$}\\
 		\hline
 		5$\%$ & 0 &0.989	&0.987	&0.983	&0.972	&0.945	&0.918	&0.867&0.905	&0.879\\
 		& 0.05 &0.48	&0.539	&0.584	&0.653	&0.717	&0.793	&0.857&0.821	&0.843\\
 		& 0.1 &0.498	&0.551	&0.602	&0.668	&0.73	&0.808	&0.869&0.839	&0.858\\
 		10$\%$ & 0 &0.974	&0.961	&0.952	&0.942	&0.921	&0.904	&0.851&0.886	&0.862\\
 		& 0.05 &0.459	&0.508	&0.562	&0.629	&0.695	&0.766	&0.838&0.809	&0.827\\
 		& 0.1 &0.473	&0.527	&0.581	&0.649	&0.709	&0.784	&0.852&0.824	&0.839\\
 		\hline 	
 		\multicolumn{11}{c}{$n=300$}\\
 		\hline
 		5$\%$ & 0 &0.995	&0.989	&0.989	&0.982	&0.963	&0.944	&0.903&0.936	&0.912\\
 		& 0.05 &0.546	&0.602	&0.654	&0.713	&0.786	&0.841	&0.892&0.868	&0.885\\
 		& 0.1 &0.563	&0.615	&0.671	&0.735	&0.797	&0.854	&0.904&0.882	&0.896\\
 		10$\%$ & 0 	&0.982	&0.974	&0.965	&0.958	&0.943	&0.931	&0.891&0.917	&0.897\\
 		& 0.05 &0.524	&0.581	&0.637	&0.696	&0.759	&0.817	&0.88&0.857	&0.871\\
 		& 0.1 &	0.539	&0.594	&0.653	&0.718	&0.772	&0.836	&0.893&0.87	&0.882\\
 		\hline
 	\end{tabular}}
 	\label{TAB:4}
 \end{table}
 
In general, the findings are similar to the exponential baseline case. 
The levels of the proposed tests (as well as those of PLE and BRE) are all higher than the nominal level under pure data, 
but tend towards the nominal value with increasing sample size. Also the inflation in the level increases with $\alpha$. 
On the other hand, the stability of the level is better maintained by the tests with larger values of $\alpha$. 
For pure data, power drops with increasing $\alpha$, but for larger $\alpha$ power is more stable under contamination. 
Tests with lower values of $\alpha$ provide better performance compared to PLE and BRE under pure data, 
and tests with larger values of $\alpha$ dominate the PLE and BRE under contamination. 
Thus, a suitable tuning parameter selection strategy can provide a suitable candidate depending on the situation.

\begin{table}[!h]
	\caption{Empirical levels of the MDPDE based Wald-type tests for testing the exponentiality of the  baseline
		hazard in a parametric Cox regression model}
	\centering	
\resizebox{0.6\textwidth}{!}{
	\begin{tabular}{c l |c c c c c c c}
		\hline
		Cens.&  & \multicolumn{7}{c}{Parametric MDPDE with $\alpha$}\\
		Prop. & $\epsilon$ & 0(MLE) & 0.05 & 0.1 & 0.2 & 0.3 & 0.4 & 0.5 \\
		\hline 	
		\multicolumn{9}{c}{$n=50$}\\
		\hline
		5$\%$ & 0 &0.081	&0.083	&0.086	&0.09	&0.093	&0.098	&0.107\\
		& 0.05 &0.728	&0.609	&0.514	&0.396	&0.287	&0.195	&0.119\\
		& 0.1 &0.757	&0.635	&0.548	&0.421	&0.309	&0.218	&0.131\\
		10$\%$ & 0 &0.085	&0.087	&0.091	&0.094	&0.098	&0.104	&0.114\\
		& 0.05 &0.749	&0.638	&0.536	&0.421	&0.306	&0.207	&0.13\\
		& 0.1 &0.757	&0.635	&0.548	&0.421	&0.309	&0.218	&0.131\\
		\hline 	
		\multicolumn{9}{c}{$n=100$}\\
		\hline
		5$\%$ & 0 &0.073	&0.075	&0.077	&0.08	&0.084	&0.088	&0.096\\
		& 0.05 &0.743	&0.633	&0.526	&0.408	&0.275	&0.178	&0.103\\
		& 0.1 &0.771	&0.665	&0.567	&0.435	&0.296	&0.198	&0.117\\
		10$\%$ & 0 &0.078	&0.081	&0.084	&0.088	&0.091	&0.095	&0.104\\
		& 0.05 &0.766	&0.662	&0.553	&0.429	&0.294	&0.191	&0.115\\
		& 0.1 &0.798	&0.692	&0.589	&0.457	&0.314	&0.215	&0.128\\
		\hline 	
		\multicolumn{9}{c}{$n=200$}\\
		\hline
		5$\%$ & 0 &0.062	&0.064	&0.066	&0.07	&0.073	&0.077	&0.086\\
		& 0.05 &0.759	&0.654	&0.544	&0.423	&0.264	&0.167	&0.093\\
		& 0.1 &0.786	&0.687	&0.591	&0.462	&0.289	&0.191	&0.106\\
		10$\%$ & 0 &0.067	&0.069	&0.072	&0.075	&0.079	&0.084	&0.094\\
		& 0.05 &0.785	&0.678	&0.565	&0.442	&0.281	&0.184	&0.103\\
		& 0.1 &0.809	&0.721	&0.614	&0.48	&0.316	&0.209	&0.118\\
		\hline 	
		\multicolumn{9}{c}{$n=300$}\\
		\hline
		5$\%$ & 0 &0.053	&0.055	&0.057	&0.059	&0.062	&0.066	&0.074\\
		& 0.05 &0.778	&0.675	&0.563	&0.436	&0.249	&0.155	&0.084\\
		& 0.1 &0.812	&0.711	&0.592	&0.476	&0.276	&0.178	&0.097\\
		10$\%$ & 0 &0.059	&0.061	&0.063	&0.066	&0.069	&0.073	&0.083\\
		& 0.05 & 0.803	&0.697	&0.584	&0.455	&0.268	&0.173	&0.096\\
		& 0.1 &0.836	&0.739	&0.621	&0.491	&0.298	&0.195	&0.108\\
		\hline
	\end{tabular}}
	\label{TAB:5}
\end{table}

\subsection{Testing for baseline function: exponentiality against monotone hazards} 
\label{Sec 6.3}

We now present another type of important testing example, 
where we test for ``\textit{Exponentiality against monotone hazardness}" for the baseline hazard formulation.
We know that when $\gamma_{2}=1$, Weibull distribution becomes exponential and it has constant hazard; 
for $\gamma_{2} \neq 1$, it has monotone hazard. So, equivalently our objective can be formulated as the problem of testing the hypothesis 
$H_{0}: \gamma_{2}=1$ with a Weibull family of alternative distributions (although it restricts the class of alternatives to the Weibull family only).
Note that, the semi-parametric tests (based on PLE or BRE) can not be  performed in this case, 
but we can still apply our proposed MDPDE-based Wald-type tests.

\begin{table}[!h]
	\caption{Empirical powers of the MDPDE based Wald-type tests for testing the exponentiality of the  baseline
	hazard in a parametric Cox regression model, calculated at $\gamma_{2n}=1+\frac{3.5}{\sqrt{n}}$}
	\centering	
\resizebox{0.6\textwidth}{!}{
	\begin{tabular}{c l |c c c c c c c}
		\hline
		Cens.&  & \multicolumn{7}{c}{Parametric MDPDE with $\alpha$}\\
		Prop. & $\epsilon$ & 0(MLE) & 0.05 & 0.1 & 0.2 & 0.3 & 0.4 & 0.5\\
		\hline 	
		\multicolumn{9}{c}{$n=50$}\\
		\hline
		5$\%$ & 0 &0.952	&0.949	&0.947	&0.943	&0.939	&0.933	&0.921\\
		& 0.05 &0.419	&0.492	&0.557	&0.634	&0.738	&0.827	&0.91\\
		& 0.1 &0.407	&0.479	&0.538	&0.612	&0.724	&0.813	&0.896\\
		10$\%$ & 0 & 0.937	&0.935	&0.932	&0.927	&0.922	&0.917	&0.906\\
		& 0.05 &0.394	&0.469	&0.534	&0.619	&0.712	&0.809	&0.897\\
		& 0.1 &0.381	&0.453	&0.514	&0.589	&0.696	&0.789	&0.881\\
		\hline 	
		\multicolumn{9}{c}{$n=100$}\\
		\hline
		5$\%$ & 0 &0.971	&0.968	&0.966	&0.962	&0.958	&0.953	&0.943\\
		& 0.05 &0.474	&0.541	&0.603	&0.682	&0.781	&0.859	&0.93\\
		& 0.1 &0.457	&0.523	&0.581	&0.663	&0.768	&0.842	&0.919\\
		10$\%$ & 0 &0.956	&0.952	&0.949	&0.945	&0.94	&0.934	&0.924\\
		& 0.05 &0.433	&0.509	&0.564	&0.641	&0.746	&0.831	&0.915\\
		& 0.1 &0.419	&0.487	&0.549	&0.624	&0.722	&0.819	&0.904\\
		\hline 	
		\multicolumn{9}{c}{$n=200$}\\
		\hline
		5$\%$ & 0 &0.987	&0.984	&0.977	&0.974	&0.971	&0.966	&0.957\\
		& 0.05 &0.529	&0.593	&0.668	&0.737	&0.828	&0.897	&0.946\\
		& 0.1 &0.512	&0.579	&0.647	&0.724	&0.813	&0.881	&0.935\\
		10$\%$ & 0 &0.973	&0.97	&0.966	&0.963	&0.959	&0.953	&0.941\\
		& 0.05 &0.492	&0.57	&0.642	&0.711	&0.807	&0.873	&0.928\\
		& 0.1 &0.473	&0.551	&0.619	&0.698	&0.787	&0.858	&0.917\\
		\hline 	
		\multicolumn{9}{c}{$n=300$}\\
		\hline
		5$\%$ & 0 &0.999	&0.998	&0.997	&0.997	&0.994	&0.99	&0.976\\
		& 0.05 &0.573	&0.628	&0.702	&0.773	&0.858	&0.919	&0.967\\
		& 0.1 &0.556	&0.613	&0.684	&0.751	&0.839	&0.903	&0.954\\
		10$\%$ & 0 &0.987	&0.983	&0.98	&0.977	&0.973	&0.968	&0.955\\
		& 0.05 &0.547	&0.603	&0.678	&0.749	&0.823	&0.889	&0.949\\
		& 0.1 &	0.533	&0.587	&0.659	&0.728	&0.805	&0.878	&0.937\\
		\hline
	\end{tabular}}
	\label{TAB:6}
\end{table}

We again perform a simulation study, as in Section \ref{Sec 6.2}, with $p=3$, $\gamma_{1}=\gamma_{2}=1$ and 
$\beta=\left(0.2,0.6,0.4\right)'=\left(\beta_{1},\beta_{2},\beta_{3}\right)'$. 
We generate samples of size $n=50,100,200$ and $300$ from this model with $5\%$ and $10\%$ censored observations as in the previous case. 
Based on the $1000$ replications, we calculate the empirical level in the same manner, as discussed in Section \ref{Sec 6.1},
but now for testing $H_{0}: \gamma_{2}=1$. 
For calculating power, we have replicated $1000$ samples from the model where true value of the parameter in question is 
$\gamma_{2n}=1+\frac{3.5}{\sqrt{n}}$,  and we calculate the proportion of times $H_{0}$ is rejected, which gives us the empirical power. 
Then, we recalculate the level and power of the Wald-type tests after  introducing $100 \epsilon \%$ contamination 
in each sample with $\epsilon= 0.05,0.1$. The contaminating observations are generated from Weibull distributions with parameters 
($\gamma_{1} =1,\gamma_{2} =0.8$) and ($\gamma_{1} =1, \gamma_{2} =0.4$), respectively, for the level and power calculations. 
We have summarized the results in Tables \ref{TAB:5} and \ref{TAB:6}.

Here we again observe patterns similar to our previous simulation studies. 
The level stabilizes to the nominal level with increasing $n$ for all $\alpha$, 
but have higher observed values and slower convergence for larger values of $\alpha$ for pure data. 
Similarly, under pure data, power decreases with increasing $\alpha$ and increases with increasing sample size. 
Under contamination, lower values of $\alpha$ produce drastically inflated levels, 
but the degree of inflation drops with increasing $\alpha$. 
Similarly, the contamination leads to a huge drop in power for lower values of $\alpha$, 
most of which is slowly recovered with increasing $\alpha$.

On the whole we feel that the MDPDE based robust Wald-type tests have the potential 
to become really useful tools for the applied scientist for the situations described here.

\section{On the choice of the robustness tuning parameter $\alpha$}
\label{Sec 8}

Our proposed MDPDE-based Wald-type tests and the model selection criterion, DIC, crucially depends on the choice of $\alpha$.
In simulation studies, we have seen that, an appropriately chosen tuning parameter $\alpha$ leads to better inference 
than both the MLE based inference of the semi-parametric approach in most situations. 
In fact, the tuning parameter $\alpha$ controls the trade-off between asymptotic efficiency and robustness of the MDPDE 
under the parametric proportional hazards model (\ref{EQ:2}) as well. 
At $\alpha= 0$, the MDPDE coincides with MLE, which is the most efficient (asymptotically) under pure data but has no robustness property. 
As $\alpha$ increases, the robustness of the corresponding MDPDE increases under contaminated data 
but, on the other hand, its efficiency gradually decreases under the pure data. 
Generally, for smaller positive $\alpha$, this loss in efficiency is not so much significant. 
The robustness properties of the proposed Wald-type tests depend directly on that of the MDPDE used to construct the tests-statistics,
as we have shown theoretically in Section \ref{Sec 4}. 
Additionally, in our simulations also, the MDPDE-based Wald-type tests with lower values of $\alpha$ 
are seen to provide better performance compared to the tests based PLE and BRE under pure data, 
whereas they are seen to dominate the PLE and BRE based tests under contaminated data for larger values of $\alpha$. 
The same can be shown, both theoretically and empirically, for our model selection criterion (DIC) as well.
Since the amount of contamination is often unknown, it is extremely important 
to choose an appropriate tuning parameter $\alpha$ based on the given data to apply our proposed inference methodologies to any real dataset.

Based on the above discussions, we have a direct one-to-one relationship between the inference procedures 
and the underlying MDPDE in terms of the effect of $\alpha$ on them. 
So, the optimum value of $\alpha$ for a given dataset can equivalently be chosen based on either of the statistical procedures; 
in particular, it may be chosen based on controlling the trade-offs between the efficiency and robustness of the MDPDEs. 
Since this phenomenon is not only restricted to the current proportional hazard models, 
there have been a few attempts in the literature to choose an appropriate data-driven value of $\alpha$ for the MDPDEs 
under general parametric set-ups with complete data. 
The most popular one is the work of Warwick and Jones \cite{Warwick/Jones:2005},
who proposed to select this tuning parameter $\alpha$ in the MDPDEs, under the IID set up, 
by minimizing an estimate of their asymptotic mean square error (MSE). 
Another method had been proposed in \cite{Hong/Kim:2001} in a similar line but minimizing the estimate of asymptotic variance rather than the MSE. 
It has latter been seen that the use of bias in considering the MSE, as done in \cite{Warwick/Jones:2005}, often lead to better performance,
and this method has subsequently used for MDPDEs under more complex parametric models. 
In particular, Ghosh and Basu \cite{Ghosh/Basu:2015,Ghosh/Basu:2013} extended this method, based on minimizing the MSE, 
for non-homogeneous setups along with a detailed investigation of their usefulness in analyzing real datasets. 
The same process was also applied for the MDPDEs under the parametric proportional hazards model (\ref{EQ:2}) by \cite{ghosh}.
So, here, to find an optimum $\alpha$ for our proposed MDPDE-based Wald-type tests and DIC, we propose 
to use the same procedure of minimizing  the estimated (asymptotic) MSE as described in \cite{ghosh}. 
For the sake of completeness, let us briefly describe the full procedure.

Suppose that the observed dataset is obtained from a true distribution, 
where the conditional density of $x_i$ given $\delta_i$ and $\boldsymbol{z}_i$, 
is contaminated as $h_{i}=(1-\epsilon)f_{i,\boldsymbol{\theta}^{*}}+ \epsilon \Delta_{i}$, for each $i$
and for some contaminating densities $\Delta_{i}$. Then, $\boldsymbol{\theta}^{*}$ is the target parameter value. 
Let the MDPDE of $\boldsymbol{\theta}$ with tuning parameter $\alpha$ is denoted as $\widehat{\boldsymbol{\theta}}_{n,\alpha}$,
which is a consistent estimator of the corresponding functional value, say $\boldsymbol{\theta}_0$, at the true distribution. 
Note that, $\boldsymbol{\theta}^{*}$  and $\boldsymbol{\theta}_0$ may not be the same. 
Then, the asymptotic MSE (AMSE) of the MDPDE $\widehat{\boldsymbol{\theta}}_{n,\alpha}$ with respect to the target $\boldsymbol{\theta}^{*}$ 
is given by 
\begin{equation}
AMSE\left(\widehat{\boldsymbol{\theta}}_{n,\alpha}\right)
= \left(\boldsymbol{\theta}_0-\boldsymbol{\theta}^{*}\right)' \left(\boldsymbol{\theta}_0-\boldsymbol{\theta}^{*}\right)
+ \frac{1}{n} \mbox{trace}\left[\boldsymbol{J}_{\alpha}^{-1}\left(\boldsymbol{\theta}_0\right)
\boldsymbol{K}_{\alpha}\left(\boldsymbol{\theta}_0\right)  \boldsymbol{J}_{\alpha}^{-1}\left(  \boldsymbol{\theta}_0\right)\right].
\label{8.1}
\end{equation}
Now, we have provided a consistent estimator of the asymptotic variance matrix $\boldsymbol{J}_{\alpha}^{-1}\left(\boldsymbol{\theta}_0\right)
\boldsymbol{K}_{\alpha}\left(\boldsymbol{\theta}_0\right)  \boldsymbol{J}_{\alpha}^{-1}\left(  \boldsymbol{\theta}_0\right)$
as $\boldsymbol{\Sigma}_{n,\alpha}$ defined in (\ref{EQ:Asym_var_est}).
We also know that $\widehat{\boldsymbol{\theta}}_{n,\alpha}$ is consistent for $\boldsymbol{\theta}_0$ by Theorem \ref{Theo:1}. 
However, since we do not have any straightforward estimate of $\boldsymbol{\theta}^{*}$, 
we need to use an appropriate pilot estimator, say $\widehat{\boldsymbol{\theta}}_{P}$, for this purpose. 
Thus, an estimate of the AMSE at tuning parameter $\alpha$ is given by 
\begin{equation}
\widehat{AMSE}\left(\widehat{\boldsymbol{\theta}}_{n,\alpha}\right)
= \left(\widehat{\boldsymbol{\theta}}_{n,\alpha}-\widehat{\boldsymbol{\theta}}_{P}\right)' 
\left(\widehat{\boldsymbol{\theta}}_{n,\alpha}-\widehat{\boldsymbol{\theta}}_{P}\right)
+ \frac{1}{n} \mbox{trace} \left[\boldsymbol{\Sigma}_{n,\alpha}\right].
\label{8.2}
\end{equation}
Note that, for a given dataset, the estimated MSE, $\widehat{AMSE}\left(\widehat{\boldsymbol{\theta}}_{n,\alpha}\right)$, 
is a function of the tuning parameter $\alpha$ only; so, we can minimize it over $\alpha \in [0,1]$ to get the optimal tuning parameter 
for the given dataset. This can be verified by standard grid-search over $[0,1]$ to obtain the global minimum. 

However, it is important to note that the final choice of optimal $\alpha$ often depend on the pilot estimator $\widehat{\boldsymbol{\theta}}_{P}$. 
This issue has been studied via detailed simulations in previous works. 
Under the IID set-up, its choice has been recommended in \cite{Warwick/Jones:2005} as the MDPDE at $\alpha= 1$,
whereas the MDPDE at $\alpha=0.5$ has been recommended as the good pilot estimator under non-homogeneous set-ups in \cite{Ghosh/Basu:2015}.
Recently, it has been suggested in \cite{Basak/etc:2020} to iterate the process with successive values of the pilot estimator 
being chosen as the MDPDE at the optimum  $\alpha$ from the previous step until convergence.
Since the present case of parametric Cox regression model (\ref{EQ:2}) is considered as a non-homogeneous set-up in Section \ref{Sec2}, 
we also suggest to simply use the MDPDE at $\alpha=0.5$ as the pilot estimator while choosing the optimal value of $\alpha$ for a given dataset.
The iterative approach of \cite{Basak/etc:2020} can also be considered in the present case, but it is often not required in practice.


\section{Real data applications}
\label{Sec 7}

Let us now apply the proposed parametric proportional hazards regression model to analyze three interesting survival data examples. 
In each example, we have fitted the model using both Weibull and exponential baselines. 
For each baseline model, we have calculated our robust model selection criterion, DIC, for different $\alpha$ 
and used these values to compare two baseline models. 
In all three examples, we have observed that for $\alpha$ less than $0.2$, 
the exponential model gives a better result than the Weibull model, 
but for higher $\alpha$, the Weibull model outperforms exponential model in terms of the DIC. 
Further, we have  used the DIC for covariate selection as well; for each baseline, 
we have considered all possible subsets of the full model (except the intercept model) 
and compared the DIC values to find the best subset of covariates.
Due to the dependence on $\alpha$, based on the discussions of Section \ref{Sec 8},
the  optimum model for each dataset is selected through the procedure described in the following.

Given a particular dataset, for each possible sub-model (both in terms of available covariates and the two baseline hazards), 
we first compute the `optimum' value of $\alpha$ that provides the best trade-off between 
the efficiency and robustness of the MDPDEs  of the model parameters, 
following the procedure described in Section \ref{Sec 8} with the pilot choice of $\alpha=0.5$.
The DIC for each sub-model is then calculated at the respective optimum $\alpha$-values 
and compared across all possible sub-models; we choose the model having minimum DIC as the best robust model 
for the given dataset which is then used for subsequent inference. 
In this final step, we have applied the proposed MDPDE-based Wald-type tests, at the corresponding optimum $\alpha$,
to test for the significance of each of the covariates present in the selected model. 
For brevity, only the results for the best selected model are presented in the main text for each of the three datasets;
the detailed results for additional competitive models, that are used in our computations, are provided at the end in Appendix \ref{App:numerical}.

The classical MLE based Wald tests were done to perform the same inference, and, due to the presence of the outliers in the data, 
the results were seen to be significantly different from those of our proposed tests.  For the purpose of comparison, 
we also analyzed these data with the corresponding semi-parametric models,
where the test of significance of the model parameters are performed based on the PLE and the BRE.
These semi-parametric results are also presented  along with our proposed MDPDE based results 
(with optimum $\alpha$) under the fully parametric model for each of the three datasets.

\subsection{Small cell lung cancer data}
\label{Sec 7.1}

Our first example is a lung cancer study dataset available in the R package \textsf{emplik}. 
The data contain $121$ observations on patients with small cell lung cancer (SCLC) from a clinical study \cite{Maksymiuk/etc:1994}
which was designed to evaluate two regimens: Arm A - cisplatin (C) followed by etoposide (E); 
and Arm B - etoposide (E) followed by cisplatin (C). 
Actually, for patients with small cell lung cancer (SCLC), 
the standard therapy is to use a combination of etoposide (E) and cisplatin (C). 
However, the optimal sequencing and administration schedule have not been established. 
In this study, $121$ patients with limited-stage SCLC were randomly assigned to these two arms 
($62$ patients to A and $59$ patients to B). 
At the time of the analysis, there was no loss to follow-up. Each death time was either observed or administratively censored. 
Therefore, the censoring variable does not depend on the two covariates - treatment indicator (\textit{arms}) 
and patient's entry age (\textit{entry-age}). 
Previously, these data were analyzed in \cite{Ying/Jung:1995}, 
where these data were fitted with a semi-parametric median regression model 
by regressing the logarithm of the survival time over the covariates \textit{arms} and \textit{entry-age}.

Here, we have fitted the parametric Cox regression model, 
with the two covariates \textit{arms} and \textit{entry-age} as covariates,  using our proposed method. 
The final results corresponding to the best model selected via the DIC at the optimum $\alpha$-values 
are reported in Table \ref{TAB:smallcell}; 	
refer to Table \ref{TAB:smallcell_selectoon} in Appendix \ref{App:numerical} for the model selection details.
Here, we have reported the estimate, the standard error (SE), and the p-value for testing 
the significance of each covariate included in the model, i.e., testing $H_{0}:\beta_{j}=0$ as described in Section \ref{Sec 6}. 
Note that, our proposal selects a model with exponential baseline hazard and only one covariate \textit{arm}; 
the other covariate (\textit{entry-age}) is left outside as insignificant. 
However, the variable \textit{entry-age} is seen to be significant when the fully parametric model with the same baseline hazard
is analyzed by the MLE-based Wald test or the corresponding semi-parametric model is analyzed by the PLE based approach.
That this is an erroneous results due to the presence of outliers can partially be observed 
by the semi-parametric BRE-based results (Table \ref{TAB:smallcell}), 
where the covariate \textit{entry-age} is not marginally significant even at 10\% level of significance.
The regression coefficient for \textit{arm} obtained by MDPDE in our best selected model is also very close to that obtained by BRE,
which further justifies the correctness of our results along with the advantages of the fully parametric modeling.
Thus our proposed method is found very useful in choosing proper model in terms of robustness against any outliers present in the data.


\begin{table}[h]
\caption{The MDPDE-based results for the best selected model at an optimum $\widehat{\alpha}=0.7$ for Small cell data,
		along with the results obtained by the MLE-based approach and the semi-parametric PLE or BRE based approaches. 
		The baseline hazard for the selected model is exponential with parameter $\gamma$.
}
\centering
\begin{tabular}{l|  c c c }
\hline
 &\textit{arm} & \textit{entry-age} & $\gamma$ \\\hline
\multicolumn{4}{l}{\underline{Best model selected by our proposal (DIC$(\widehat{\alpha})=87.483$)}}	\\
Estimate & 0.685 & - & 0.034\\
SE		 &(0.359)&-&(0.115)\\
p-value	 &(0.014)&-&\\
\hline
\multicolumn{4}{l}{\underline{Fully parametric MLE-based results}}\\ 
Estimate & 0.336& 0.036 &0.019 \\
SE	     &(0.597)&(0.005)&(0.002)\\
p-value  &(0.073)&(9.56e-12)&\\
\hline
\multicolumn{4}{l}{\underline{Semi-parametric PLE-based results}}\\ 
Estimate   	&0.513&0.028&-\\
SE			&(0.204)&(0.013)&-\\
p-value		&(0.012)&(0.029)&-\\
\hline
\multicolumn{4}{l}{\underline{Semi-parametric PLE-based results}}\\ 
Estimate  	&0.757&0.026&-\\
SE			&(0.231)&(0.016)&-\\
p-value 	&(0.001)&(0.101)&-\\
	\hline
		\end{tabular}
\label{TAB:smallcell}
\end{table}

Even if we have not gone through the model selection procedures, 
we can obtain the results similar to the best selected model by using the MDPDEs at some appropriate $\alpha$. 
We have summarized the MDPDE at different $\alpha$, obtained under the full model, 
along with their standard errors in Table \ref{TAB:7} of Appendix \ref{App:numerical}.
It is clear that the MDPDE with $\alpha\approx 0.4$ gives us the results close to the best model.

\subsection{Veterans administration lung cancer data}
\label{Sec 7.2}
 
As our second example, which illustrates the effectiveness of the fully parametric approach over the semi-parametric one, 
we consider the data on $137$ advanced lung cancer patients as collected by the Veterans Administration Lung Cancer Study Group. 
Patients were randomized according to one of two chemotherapeutic agents (1-standard; 2-test). 
Tumors are classified into one of four broad groups (squamous, small cell, adeno, and large). 
The covariates recorded are performance status, time from diagnosis to starting on study (months), age, 
and previous therapy (0 for no; 10 for yes). 
These data are now available in R package \textsf{survival} and were previously used by R. J. Prentice \cite{Prentice:1973} 
to determine which covariates have an important relation with survival 
and to compare efficacy of treatments with respect to longevity by fitting a Cox model of survival time over the covariates.
 
After fitting the standard semi-parametric model using the \textsf{coxrobust} package, 
we have found that among several covariates, karno (Karnofsky performance score) and three cell types
- adeno, large and small-cell - are significant. 
Using these covariates, we have next fitted the fully parametric model using the proposed MDPE-based approach,
and tabulated the MDPDEs at different $\alpha$ in Table \ref{TAB:9} in Appendix \ref{App:numerical}.
We can see that, these MDPDEs are quite close to the MLE indicating that there is no significant outlier-effect present 
in these data with respect to the fully parametric model.


\begin{table}[h]
	\caption{The MDPDE-based results for the best selected model at an optimum $\widehat{\alpha}=0.62$ for Veteran data,
		along with the results obtained by the MLE-based approach and the semi-parametric PLE or BRE based approaches. 
		The baseline hazard for the selected model is exponential with parameter $\gamma$.
	}
\centering
\begin{tabular}{l|  c c c c c }
\hline
&\textit{karno}&\textit{adeno}&\textit{large}&\textit{smallcell} & $\gamma$ \\\hline
\multicolumn{6}{l}{\underline{Best model selected by our proposal (DIC$(\widehat{\alpha})=86.498$)}}	\\
Estimate 	&-0.096&-&-&0.469&0.037\\
SE 			&(0.028)&-&-&(0.295)&(0.019)\\
p-value 	&(0.019)&-&-&(0.418)&\\
\hline
\multicolumn{6}{l}{\underline{Fully parametric MLE-based results}}\\ 
Estimate 	&-0.030&-0.093&-0.311&-0.101&0.033\\
SE 			&(2.6e-3)&(2.77e-1)&(7.16e-1)&(2.16e-1)&(0.033)\\
{p-value}	&(8.44e-31)&(0.737)&(0.664)&(0.640)&\\
\hline
\multicolumn{6}{l}{\underline{Semi-parametric PLE-based results}}\\ 
Estimate	&-0.031&-1.15&-0.831&-0.438&-\\
SE			&(0.0052)&(0.293)&(0.2935)&(256)&-\\
p-value		&(2.05e-9)&(8.68e-5)&(0.249)&(4.83e-3)&-\\
\hline
\multicolumn{6}{l}{\underline{Semi-parametric BRE-based results}}\\ 
Estimate 	&-0.042&-1.237&-1.027&-0.164&-\\
SE			&(0.0065)&(0.528)&(0.329)&(0.285)&-\\
p-value		&(9.39e-11)&(0.0192)&(0.655)&(0.0183)&-\\
\hline
	\end{tabular}
	\label{TAB:veteran}
\end{table}

We next proceed to select the best model for these data, as per our proposed methodology
and the final results are reported in Table \ref{TAB:veteran}, 
with details in Table \ref{TAB:veteran_selection} in Appendix \ref{App:numerical}.
The results obtained by the MLE based approach under the fully parametric set-up 
and the PLE or BRE based approaches under the corresponding semi-parametric model are also reported in Table \ref{TAB:veteran}.
We can see that, under the semi-parametric model, variable \textit{large} is insignificant while using PLE or BRE;
the similarity in the PLE an BRE also indicates no outlier-effect. 
However, if we look at the results obtained by the MLE under the fully parametric model 
all covariates except \textit{karno} becomes insignificant, 
indicating the significant change in our inference while using the proposed fully parametric modelling. 
The model selected by our proposed procedure also provide the same inference
with only the covariate \textit{karno} being significant for the patient's lifetime;
it illustrates that the proposed MDPDE based procedure also give the results 
that are consistent with the most efficient MLE based inference in most cases when there is no significant outlier-effect.
Additionally, our proposed model selection strategy automatically leaves out the two unimportant covariates
providing us a better model with only one insignificant covariate, along with the significant one.

\subsection{Criminal recidivism data}
\label{Sec 7.3}

Our final example is from a completely different applied domain where our procedure is shown to provide 
new insights significantly different from those obtained by the existing method. 
These data were obtained from \cite{Rossi/Berk:1980}, and were also used in \cite{Allison:1995,Fox/Carvalho:2012}. 
The data relate to $432$ convicts who were released from the state institutions of Georgia and Texas in the 1970s 
and who were followed up for one year after release. 
Some of the prisoners were offered eligibility for unemployment insurance payments for periods of up to 6 months 
or until they managed to locate employment. 
These ex-prisoners were carefully selected for conducting an experiment run by the Department of Labor 
in collaboration with the two states. 
Other prisoners who were not offered unemployment benefits also participated in the experiment to serve as controls. 
The ex-prisoners who were offered unemployment insurance benefits were compared with others released around the same time who were not
 made the same offer.
The purpose of the experiment was to test a new way of helping persons
who had completed their sentences or were released on parole to bring  themselves into civilian life. 
In broadest terms, it is the social problem of crime that is the center of concern of that study. 
Specifically, the researchers focused on recidivism, the unfortunate tendency of persons convicted of crime at one point in
time to be arrested and convicted again, sometimes to repeat this sequence  over and over. 
Note that here censoring happens if a prisoner does not commit a crime within one year after release.

\begin{table}[h]
	\caption{The MDPDE-based results for the best selected model at an optimum $\widehat{\alpha}=0.57$ for Criminal recidivism data,
		along with the results obtained by the MLE-based approach and the semi-parametric PLE or BRE based approaches. 
		The baseline hazard for the selected model is Weibull$(\gamma_1, \gamma_{2})$.
	}
	\centering
	\begin{tabular}{l|  c c c c c }
		\hline
		&\textit{fin}&\textit{wexp}&\textit{prio}&$\gamma_{1}$&$\gamma_{2}$ \\\hline
		\multicolumn{6}{l}{\underline{Best model selected by our proposal (DIC$(\widehat{\alpha})=132.897$)}}	\\
		Estimate 	&0.138&0.086&-0.029&0.001&0.314\\
		SE 			&(0.696)&(0.865)&(0.084)&(0.001)&(0.946)\\
		p-value 	&(0.178)&(0.271)&(0.094)&&\\
		\hline
		\multicolumn{6}{l}{\underline{Fully parametric MLE-based results}}\\ 
		Estimate 	&-0.425&-0.472&0.109&0.004&0.472\\
		SE 			&(0.228)&(0.241)&(0.009)&(0.0001)&(0.543)\\
		{p-value}	&(0.026)&(0.24)&(9.61e-29)&&\\
		\hline
		\multicolumn{6}{l}{\underline{Semi-parametric PLE-based results}}\\ 
		Estimate	&-0.346&-0.223&0.092&-&-\\
		SE			&(0.190)&(0.209)&(0.028)&-&-\\
		p-value		&(0.069)&(0.285)&(0.001)&-&-\\
		\hline
		\multicolumn{6}{l}{\underline{Semi-parametric BRE-based results}}\\ 
		Estimate 	&-0.352&-0.512&0.077&-&-\\
		SE			&(0.247)&(0.242)&(0.034)&-&-\\
		p-value		&(0.154)&(0.034)&(0.022)&-&-\\
		\hline
	\end{tabular}
	\label{TAB:criminal}
\end{table}

Based on an initial analysis using semi-parametric Cox model via \textsf{coxrobust}, 
we observed that the significant covariates are ``\textit{wexp}" (full-time work experience before incarceration; 1 for yes), 
``\textit{fin}" (financial aid provided or not; 1 for yes) and ``\textit{prio}" (number of convictions prior to current incarceration). 
So, we next fit a fully parametric Cox model using these covariates and summarize the final results in Table \ref{TAB:criminal} 
(see Tables \ref{TAB:11} and \ref{TAB:criminal_selection} in Appendix \ref{App:numerical} for detailed analyses).
Here, we note that, the semi-parametric results based on PLE or BRE, respectively, 
indicates the variable \textit{wexp} or \textit{fin} to be insignificant; the other two remain significant in each.
The difference between the PLE and BRE based inference indicates the presence of outliers in the data,
and hence the MLE based results obtained under the fully parametric model is also unreliable. 
So, under this fully parametric approach, we can use the model obtained via the proposed MDPDE-based model selection procedure 
to be the optimum one. Interestingly, the p-values obtained based on the MDPDE based Wald-type tests at this optimum model 
indicates that both the covariates  \textit{fin} and \textit{wexp} are insignificant 
and the third variable \textit{prio} is only marginally significant at 90\% level of significance. 
These completely new results provide a new aspect to the underlying research problem 
with the proposed robust  and efficient fully parametric inference procedures.


\section{Concluding remarks}
\label{Sec 9}

In this paper we have considered a parametric proportional hazards model for randomly censored response 
as an alternative to the traditional semi-parametric Cox regression model. 
Due to the non-robust nature of the traditional likelihood based inference under the presence of outliers, 
we have, in order to provide a viable alternative, developed a robust generalized Wald-type tests based on the MDPDE of \cite{ghosh}. 
Our work provides a thorough theoretical evaluation of the proposed robust Wald-type tests 
for testing the general composite hypothesis in the proposed parametric Cox model establishing the claimed robustness 
and its advantages over the classical Wald-test. We have provided simulation studies along with three real data examples 
to demonstrate how these theoretical advantages help  in practice in real situations and 
how this method sometimes performs better than testing procedure based on traditional semi-parametric model. 
Besides that, we have provided the influence function (IF) of robust Wald-type test statistic and its asymptotic power. 
From the figures we also observe that the IF is unbounded for $\alpha=0$ but becomes bounded for positive $\alpha$. 
A robust model selection criterion (DIC) is proposed using the MDPDE and 
is used in the real data applications to chose the optimum parametric model
successfully resisting the  effects of outliers present in the data. 

It may possibly be argued that the method for selecting the best model in terms of the combination of covariates, 
baseline functions and tuning parameter for the estimation of the model parameter as illustrated in Table 11 is a bit ad-hoc.
A more satisfactory approach would involve a simultaneous optimization over the different factors, 
rather than the sequential optimization over the tuning parameter followed by 
the optimization over the combinations of covariates and baseline functions.
This issue possibly needs more attention and we plan to develop a more comprehensive strategy 
for finding the most appropriate model in our future research.

On the whole, even based on the demonstrations provided in the present paper, 
we can observe the proposed MDPDE-based robust tests and model selection criterion to 
become extremely useful tools for robust inference in practical applications with survival data. 
So, it would be important future works to extend this MDPDE and the associated testing procedure for Cox regression with frailty components,
with time varying covariates, and even with multivariate censored responses.
We hope to take up some of these extensions in the future.

\bigskip\bigskip
\noindent\textbf{Acknowledgment:}\\
This research is supported by Ministerio de Ciencia, Innovación y Universidades (Spain) under the Grant PGC2018-095194-B-I00.
Additionally, the research of AN and AG is partially supported by an internal grant from Indian Statistical Institute, Kolkata,
and that of AG is also supported partially by the INSPIRE Faculty Research Grant from  Department of Science and Technology, Government of India.

\appendix
\appendixpage
\addappheadtotoc

\section{Proof of the results}

\subsection{Proof of Theorem \ref{THM:asymp_null}}

Take any $\boldsymbol{\theta}_0 \in \boldsymbol{\Theta_{0}}$; then we have 
$\boldsymbol{m}(\boldsymbol{\theta}_0)=0$.
Also, by Theorem \ref{Theo:1}, we have 	
and $\sqrt{n}\left(  \widehat{\boldsymbol{\theta}}_{n,\alpha}-\boldsymbol{\theta}_0\right)  
\underset{n\longrightarrow\infty}{\overset{\mathcal{L}}{\longrightarrow}}
\mathcal{N}\left(\boldsymbol{0}_{q+p},\boldsymbol{\Sigma}_{\alpha} (\boldsymbol{\theta}_0)\right)$. 
Then using the delta method, we get
\[
\sqrt{n}\boldsymbol{m} \left(\widehat{\boldsymbol{\theta}}_{n,\alpha}\right) 
\underset{n\longrightarrow\infty}{\overset{\mathcal{D}}{\longrightarrow}}
\mathcal{N}\left(  \boldsymbol{0}_{q+p},\boldsymbol{M}'(\boldsymbol{\theta}_0) \boldsymbol{\Sigma}_{\alpha} (\boldsymbol{\theta}_0) 
\boldsymbol{M}(\boldsymbol{\theta}_0)\right).
\]
But, by assumptions $\boldsymbol{M}(\boldsymbol{\theta})$ and $\boldsymbol{\Sigma}_{\alpha}(\boldsymbol{\theta}_0)$ 
are continuous around $\boldsymbol{\theta}=\boldsymbol{\theta}_0$.
Also, by Theorem \ref{Theo:1}, $\widehat{\boldsymbol{\theta}}_{n,\alpha}$ is consistent for $\boldsymbol{\theta}_0$.
Hence, $\boldsymbol{M}'(\widehat{\boldsymbol{\theta}}_{n,\alpha})\boldsymbol{\Sigma}_{\alpha}(\widehat{\boldsymbol{\theta}}_{n,\alpha})
\boldsymbol{M}(\widehat{\boldsymbol{\theta}}_{n,\alpha})$ is consistent for 
$\boldsymbol{M}'(\boldsymbol{\theta}_0) \boldsymbol{\Sigma}_{\alpha} (\boldsymbol{\theta}_0) \boldsymbol{M}(\boldsymbol{\theta}_0)$. 
Therefore, by applying Slutsky's theorem, we get that the asymptotic distribution of $W_{n}(\widehat{\boldsymbol{\theta}}_{n,\alpha})$ 
as $\chi^{2}_{r}$.
      
\subsection{Proof of Theorem \ref{THM:app_power}}

Here, $\widehat{\boldsymbol{\theta}}_{n,\alpha}\underset{n\longrightarrow\infty}{\overset{\mathcal{P}}{\longrightarrow}}{\boldsymbol{\theta}}^{\ast}.$ 
So, asymptotic distribution of $l^{*}(\widehat{\boldsymbol{\theta}}_{n,\alpha},\widehat{\boldsymbol{\theta}}_{n,\alpha})$ 
and $l^{*}(\widehat{\boldsymbol{\theta}}_{n,\alpha},\boldsymbol{\theta}^{*})$ will be the same. 
Now, a first order Taylor series expansion of $l^{*}(\widehat{\boldsymbol{\theta}}_{n,\alpha},\boldsymbol{\theta}^{*})$ 
at $\widehat{\boldsymbol{\theta}}_{n,\alpha}$ around $\boldsymbol{\theta}^{*}$ gives 
\[
l^{*}(\widehat{\boldsymbol{\theta}}_{n,\alpha},\boldsymbol{\theta}^{*})-l^{*}(\boldsymbol{\theta}^{*},\boldsymbol{\theta}^{*})
=\frac{\partial l^{*}\left(\boldsymbol{\theta},\boldsymbol{\theta}^{*}\right)}{
	\partial \boldsymbol{\theta}'}|_{\boldsymbol{\theta}=\boldsymbol{\theta}^{*}}
\left(\widehat{\boldsymbol{\theta}}_{n,\alpha}-\boldsymbol{\theta}^{*}\right)
+o_{p}\left(||\widehat{\boldsymbol{\theta}}_{n,\alpha}-\boldsymbol{\theta}^{*}||\right).
\]
Thus we get,
\[
\sqrt{n}\left(  \ell^{\ast}(\widehat{\boldsymbol{\theta}}_{n,\alpha},\widehat{\boldsymbol{\theta}}_{n,\alpha})
		-\ell^{\ast}({\boldsymbol{\theta}}^{\ast},{\boldsymbol{\theta}}^{\ast})\right)  
		\underset{n\longrightarrow\infty}{\overset{\mathcal{D}}{\longrightarrow}}
		\mathcal{N}\left(\boldsymbol{0},\sigma_{W_{n}}^{2}\left(  {\boldsymbol{\theta}}^{\ast}\right)\right),
\]
where
\[
\sigma_{W_{n}}^{2}\left(  {\boldsymbol{\theta}}^{\ast}\right)  =\left(
 \frac{\partial\ell^{\ast}({\boldsymbol{\theta}},{\boldsymbol{\theta}}^{\ast}%
 	)}{\partial{\boldsymbol{\theta}}}\right)  _{{\boldsymbol{\theta}%
 	}={\boldsymbol{\theta}}^{\ast}}'\boldsymbol{\Sigma}_{\alpha}({\boldsymbol{\theta}%
 }^{\ast})\left(  \frac{\partial\ell^{\ast}({\boldsymbol{\theta}}%
 	,{\boldsymbol{\theta}}^{\ast})}{\partial{\boldsymbol{\theta}}}\right)
 _{{\boldsymbol{\theta}}={\boldsymbol{\theta}}^{\ast}}'.
\]
 
\subsection{Proof of Theorem \ref{THM:Cont_power}}

From a Taylor series expansion of $\boldsymbol{m}(\widehat{\boldsymbol{\theta}}_{n,\alpha})$ around $\boldsymbol{\theta}_{n}$, we get
\[
\boldsymbol{m}(\widehat{\boldsymbol{\theta}}_{n,\alpha})=\boldsymbol{m}(\boldsymbol{\theta}_n)
+\boldsymbol{M}'(\boldsymbol{\theta}_n)(\widehat{\boldsymbol{\theta}}_{n,\alpha}-\boldsymbol{\theta}_{n})
+o(||\widehat{\boldsymbol{\theta}}_{n,\alpha}-\boldsymbol{\theta}_{n}||).
\]
But, from (\ref{22.21}), we get 
\[
\boldsymbol{m}(\widehat{\boldsymbol{\theta}}_{n,\alpha})=\boldsymbol{m}(\boldsymbol{\theta}_n)
+\boldsymbol{M}'(\boldsymbol{\theta}_n)(\widehat{\boldsymbol{\theta}}_{n,\alpha}-\boldsymbol{\theta}_{n})
+n^{-1/2}\boldsymbol{M}'(\boldsymbol{\theta}_{0})\boldsymbol{d}
+o\left(  \left\Vert\boldsymbol{\theta}_{n}-\boldsymbol{\theta}_{0}\right\Vert \right)
+o(||\widehat{\boldsymbol{\theta}}_{n,\alpha}-\boldsymbol{\theta}_{n}||).
\]
Now we can write
\[
\boldsymbol{\widehat{\theta}}_{n,\alpha}-\boldsymbol{\theta}_{0}
=\boldsymbol{\widehat{\theta}}_{n,\alpha}-\boldsymbol{\theta}_{n}+\boldsymbol{\theta}_{n}-\boldsymbol{\theta}_{0}
=(\boldsymbol{\widehat{\theta}}_{n,\alpha}-\boldsymbol{\theta}_{n})+n^{-1/2}\boldsymbol{d}.
\]

Next, using the theory of contiguity (Le Cam's third lemma) and Theorem \ref{Theo:1} we get that, under $H_{1,n}$ given in (\ref{22.12}), 
\[
\sqrt{n}(\boldsymbol{\widehat{\theta}}_{n,\alpha}-\boldsymbol{\theta}_{0})
\underset{n\longrightarrow\infty}{\overset{\mathcal{D}}{\longrightarrow}}
\mathcal{N}(\boldsymbol{d},\boldsymbol{\Sigma}_{\alpha}(\boldsymbol{\theta}_{0})).
\]
Thus, under $H_{1,n}$, we get
\[
\sqrt{n}\left(\widehat{\boldsymbol{\theta}}_{n,\alpha}-\boldsymbol{\theta}_{n}\right)
\underset{n\longrightarrow\infty}{\overset{\mathcal{D}}{\longrightarrow}}
\mathcal{N}\left(\boldsymbol{0},\boldsymbol{\Sigma}_{\alpha}\left(\boldsymbol{\theta}_{0}\right)\right),
\]
and also  
\[
\sqrt{n}(o\left(  \left\Vert\boldsymbol{\theta}_{n}-\boldsymbol{\theta}_{0}\right\Vert \right)
+o(||\widehat{\boldsymbol{\theta}}_{n,\alpha}-\boldsymbol{\theta}_{n}||))=o_{p}(\boldsymbol{1}).
\]
Hence, combining, we have
\[
\sqrt{n} \boldsymbol{m}(\widehat{\boldsymbol{\theta}}_{n,\alpha})
\underset{n\longrightarrow	\infty}{\overset{\mathcal{D}}{\longrightarrow}}
\mathcal{N}_{r}(\boldsymbol{M}'(\boldsymbol{\theta}_{0})\boldsymbol{d},
\boldsymbol{M}'(\boldsymbol{\theta}_{0}) \boldsymbol{\Sigma}_{\alpha}(\boldsymbol{\theta}_{0})\boldsymbol{M}(\boldsymbol{\theta}_{0})). 
\]

Finally note that, if $\boldsymbol{X}\sim \mathcal{N}_{k}(\boldsymbol{\mu},\boldsymbol{\Sigma})$ with $rank(\boldsymbol{\Sigma})= k$, 
then $\boldsymbol{X}' \boldsymbol{\Sigma}^{-1} \boldsymbol{X}$ follows a $\chi^{2}$ distribution with degrees of freedom $k$ 
and non-centrality parameter $\boldsymbol{\mu}' \boldsymbol{\Sigma}^{-1}\boldsymbol{\mu}$. 
Thus, $W_{n}(\boldsymbol{\widehat{\theta}}_{n,\alpha})$, defined in (\ref{22.23}), must follow a $\chi^{2}$ distribution 
with degrees of freedom $r$ and non-centrality parameter 
$$ \boldsymbol{d}'\boldsymbol{M}(\boldsymbol{\theta}_{0})
\left(  \boldsymbol{M}'(\boldsymbol{\theta}_{0})\boldsymbol{\Sigma}_{\alpha}(\boldsymbol{\theta}_{0})
\boldsymbol{M}(\boldsymbol{\theta}_{0})\right)^{-1}\boldsymbol{M}'(\boldsymbol{\theta}_{0})\boldsymbol{d}.
$$

The second part of the theorem follows from the equivalence of the two contiguous hypotheses via the relationship $\boldsymbol{\delta}=\boldsymbol{M}'(\boldsymbol{\theta}_{0})\boldsymbol{d}$.

\subsection{Proof of Theorem \ref{Theo:10}}

Let us denote $\boldsymbol{\theta_{n}}^{*}= \boldsymbol{U}_{\alpha}(\boldsymbol{H_{\epsilon, \boldsymbol{y_{t}}}^{P}})$. Then we get

\begin{align}
W_{n}(\widehat{\boldsymbol{\theta}}_{n,\alpha})   
& =n\boldsymbol{m}'%
(\widehat{\boldsymbol{\theta}}_{n,\alpha})\left(  \boldsymbol{M}%
'(\widehat{\boldsymbol{\theta}}_{n,\alpha})\boldsymbol{\Sigma
}_{n,\alpha}(\widehat{\boldsymbol{\theta}}_{n,\alpha})\boldsymbol{M}%
(\widehat{\boldsymbol{\theta}}_{n,\alpha})\right)  ^{-1}\boldsymbol{m}%
(\widehat{\boldsymbol{\theta}}_{n,\alpha})\nonumber \\
 & = n\boldsymbol{m}'%
(\boldsymbol{\theta_{n}}^{*})\left(  \boldsymbol{M}%
'(\widehat{\boldsymbol{\theta}}_{n,\alpha})\boldsymbol{\Sigma
}_{n,\alpha}(\widehat{\boldsymbol{\theta}}_{n,\alpha})\boldsymbol{M}%
(\widehat{\boldsymbol{\theta}}_{n,\alpha})\right)  ^{-1}\boldsymbol{m}%
(\boldsymbol{\theta_{n}}^{*}) \nonumber \\
& + n (\boldsymbol{m} (\widehat{\boldsymbol{\theta}}_{n,\alpha})-\boldsymbol{m}
(\boldsymbol{\theta_{n}}^{*}))'\left(  \boldsymbol{M}%
'(\widehat{\boldsymbol{\theta}}_{n,\alpha})\boldsymbol{\Sigma
}_{n,\alpha}(\widehat{\boldsymbol{\theta}}_{n,\alpha})\boldsymbol{M}%
(\widehat{\boldsymbol{\theta}}_{n,\alpha})\right)  ^{-1}(\boldsymbol{m} (\widehat{\boldsymbol{\theta}}_{n,\alpha})-\boldsymbol{m}
(\boldsymbol{\theta_{n}}^{*})) \nonumber \\
& + 2n(\boldsymbol{m} (\widehat{\boldsymbol{\theta}}_{n,\alpha})-\boldsymbol{m}
(\boldsymbol{\theta_{n}}^{*}))'\left(  \boldsymbol{M}%
'(\widehat{\boldsymbol{\theta}}_{n,\alpha})\boldsymbol{\Sigma
}_{n,\alpha}(\widehat{\boldsymbol{\theta}}_{n,\alpha})\boldsymbol{M}%
(\widehat{\boldsymbol{\theta}}_{n,\alpha})\right)  ^{-1}\boldsymbol{m}
(\boldsymbol{\theta_{n}}^{*}) \nonumber \\
& = T_{1,n}+T_{2,n}+T_{3,n}, ~~~~~~~~~~~\text{ say}
\label{Theo 10.1}
\end{align}

 Now, let us consider $\boldsymbol{\theta_{n}}^{*}$ as a function of $\epsilon_{n}=\frac{\epsilon}{\sqrt{n}}$ i.e. $f(\epsilon_{n})$. Then a Taylor series expansion of $f(\epsilon_{n})$ at $\epsilon_{n}= 0$ gives 
 \begin{align*}
 f(\epsilon_{n}) & = \sum_{i=0}^{\infty}\frac{1}{i!}\frac{\epsilon^{i}}{n^{\frac{i}{2}}}\frac{\partial^{i}f(\epsilon_{n})}{\partial \epsilon_{n}^{i}}|_{\epsilon_{n}= 0}\\
 & = \boldsymbol{\theta_{n}}+ \frac{\epsilon}{\sqrt{n}} IF(\boldsymbol{y_{t}},\boldsymbol{U}_{\alpha},\boldsymbol{H_{\theta_{0}}})+ o_{p}(\boldsymbol{1}/\sqrt{n}).
 \end{align*}
From this, we get 
 \begin{align*}
 \sqrt{n}(\boldsymbol{\theta_{n}}^{*}-\boldsymbol{\theta_{n}}) & = \sqrt{n}(\boldsymbol{\theta_{n}}^{*}-\boldsymbol{\theta_{0}}-n^{-1/2}\boldsymbol{d})\\
 & = \epsilon IF(\boldsymbol{y_{t}},\boldsymbol{U}_{\alpha},\boldsymbol{H_{\theta_{0}}})+ o_{p}(\boldsymbol{1}),
 \end{align*}
 and thus 
 \begin{align}
 \sqrt{n}(\boldsymbol{\theta_{n}}^{*}-\boldsymbol{\theta_{0}}) & =\boldsymbol{d}+ \epsilon IF(\boldsymbol{y_{t}},\boldsymbol{U}_{\alpha},\boldsymbol{H_{\theta_{0}}})+ o_{p}(\boldsymbol{1}) \nonumber\\
 & = \boldsymbol{d}_{\epsilon,\boldsymbol{y_{t}},\alpha}\left(\boldsymbol{\theta_{0}}\right)+ o_{p}(\boldsymbol{1}).
 \label{Theo 10.2}
 \end{align}
Additionally, using a Taylor series expansion of $\boldsymbol{m}(\boldsymbol{\theta_{n}}^{*})$ around $\boldsymbol{\theta_{0}}$, we can get
 \begin{align}
 \sqrt{n}\boldsymbol{m}'(\boldsymbol{\theta_{n}}^{*}) & = \boldsymbol{M}'(\boldsymbol{{\theta}_{0}})\boldsymbol{d}_{\epsilon,\boldsymbol{y_{t}},\alpha}\left(\boldsymbol{\theta_{0}}\right)+ o_{p}(\boldsymbol{1}).
 \label{Theo 10.3}
 \end{align}
Now, under $\boldsymbol{H}^{P}_{\epsilon,\boldsymbol{y_{t}}}$, the asymptotic distribution of MDPDE yields 
\begin{equation}
\sqrt{n}\left(\widehat{\boldsymbol{\theta}}_{n,\alpha}-\boldsymbol{\theta_{n}}^{*}\right)
\underset{n\longrightarrow 	\infty}{\overset{\mathcal{D}}{\longrightarrow}}
\mathcal{N}\left(\boldsymbol{0},\boldsymbol{\Sigma}_{\alpha}\left(\boldsymbol{\theta}_{0}\right)\right).
\label{Theo 10.4}
\end{equation}
Thus we get 
$T_{2,n} \underset{n\longrightarrow	\infty}{\overset{\mathcal{D}}{\longrightarrow}} \chi^{2}_{r}.$
Now, combining (\ref{Theo 10.1}), (\ref{Theo 10.2}) and (\ref{Theo 10.3}), we get 
\[
W_{n}(\widehat{\boldsymbol{\theta}}_{n,\alpha})
=\boldsymbol{V_{n}}'\left(  \boldsymbol{M}'(\boldsymbol{\theta_{0}})\boldsymbol{\Sigma}_{\alpha}(\boldsymbol{\theta_{0}})
\boldsymbol{M}(\boldsymbol{\theta_{0}})\right)  ^{-1}\boldsymbol{V_{n}}+ o_{p}(\boldsymbol{1}),
\]
where
\[
\boldsymbol{V_{n}}=\sqrt{n} (\boldsymbol{m} (\widehat{\boldsymbol{\theta}}_{n,\alpha})-\boldsymbol{m}(\boldsymbol{\theta_{n}}^{*})) 
+ \boldsymbol{M}'(\boldsymbol{{\theta_{0}}})\boldsymbol{d}_{\epsilon,\boldsymbol{y_{t}},\alpha}\left(\boldsymbol{\theta_{0}}\right).
\]
But, by (\ref{Theo 10.4}), we have
\[
\boldsymbol{V_{n}}\underset{n\longrightarrow\infty}{\overset{\mathcal{D}}{\longrightarrow}} 
\mathcal{N} \left(\boldsymbol{M}'(\boldsymbol{{\theta_{0}}})\boldsymbol{d}_{\epsilon,\boldsymbol{y_{t}},\alpha}\left(\boldsymbol{\theta_{0}}\right), 
\left(  \boldsymbol{M}'(\boldsymbol{\theta_{0}})\boldsymbol{\Sigma}_{\alpha}(\boldsymbol{\theta_{0}})\boldsymbol{M}(\boldsymbol{\theta_{0}})
\right)\right),
\]
and hence we finally get that
\[
W_{n}(\widehat{\boldsymbol{\theta}}_{n,\alpha}) \underset{n\longrightarrow\infty}{\overset{\mathcal{D}}{\longrightarrow}} \chi_{r}^{2}(d^{*}),
\]
where $d^{*}$ is as defined in the statement of the theorem.

\subsection{Proof of Theorem \ref{THM:PIF}}

We start with the expression of $\beta_{W_{n}}(\boldsymbol{\theta}_{n},\epsilon, \boldsymbol{y}_{t})$ from Theorem \ref{Theo:10}. 
Clearly, by definition of PIF and using the chain rule of derivatives, we get
\begin{align*}
&PIF\left(\boldsymbol{y}_{t},W_{n},H_{\boldsymbol{\theta}_{0}}\right) 
= \left. \frac{\partial}{\partial \epsilon} \beta_{W_{n}}(\boldsymbol{\theta}_{n},\epsilon, \boldsymbol{y}_{t})\right|_{\epsilon=0}
\\
&\cong \left. \sum_{v=0}^{\infty} \frac{\partial}{\partial \epsilon}\boldsymbol{C}_{v}\left(
\boldsymbol{M}'(\boldsymbol{\theta}_{0})\boldsymbol{d}_{\epsilon,\boldsymbol{y}_{t},\alpha}\left(\boldsymbol{\theta}_{0}\right),
\left(\boldsymbol{M}'(\boldsymbol{\theta}_{0})\boldsymbol{\Sigma}_{\alpha}(\boldsymbol{\theta}_{0})\boldsymbol{M}(\boldsymbol{\theta}_{0})\right)^{-1}
\right)\right|_{\epsilon= 0} P\left(\chi^{2}_{r+2v} > \chi^{2}_{r,\tau}\right)
\\
& \cong \left.\sum_{v=0}^{\infty} \frac{\partial}{\partial\boldsymbol{t}'}\boldsymbol{C}_{v}\left(
\boldsymbol{M}'(\boldsymbol{\theta}_{0})\boldsymbol{t},
\left(\boldsymbol{M}'(\boldsymbol{\theta}_{0})\boldsymbol{\Sigma}_{\alpha}(\boldsymbol{\theta}_{0})\boldsymbol{M}(\boldsymbol{\theta}_{0})\right)^{-1}
\right)\right|_{\boldsymbol{t}= \boldsymbol{d}_{0,\boldsymbol{y}_{t},\alpha}\left(\boldsymbol{\theta}_{0}\right)}
\times \left. \frac{\partial}{\partial \epsilon} \boldsymbol{d}_{\epsilon,\boldsymbol{y}_{t},\alpha}\left(\boldsymbol{\theta}_{0}\right)
\right|_{\epsilon= 0} P\left(\chi^{2}_{r+2v} > \chi^{2}_{r,\tau}\right).
\end{align*}

But, $\boldsymbol{d}_{0,\boldsymbol{y}_{t},\alpha}\left(\boldsymbol{\theta}_{0}\right) = \boldsymbol{d}$ 
and the standard derivatives give
\[
\frac{\partial}{\partial \epsilon} \boldsymbol{d}_{\epsilon,\boldsymbol{y}_{t},\alpha}\left(\boldsymbol{\theta}_{0}\right)
= IF(\boldsymbol{y}_{t},\boldsymbol{U}_{\alpha}, H_{\boldsymbol{\theta}_{0}}),
\]
and
\[
\frac{\partial}{\partial \boldsymbol{t}} \boldsymbol{C}_{v}(\boldsymbol{t},\boldsymbol{A})
= \frac{(\boldsymbol{t}'\boldsymbol{A}\boldsymbol{t})^{v-1}}{v!2^{v}}(2v-\boldsymbol{t}'\boldsymbol{A}\boldsymbol{t}) 
\boldsymbol{A} \boldsymbol{t} e^{-\frac{1}{2}\boldsymbol{t}'\boldsymbol{A}\boldsymbol{t}}.
\]
Combining the above results and simplifying, we get the required expression of the PIF as given in the theorem.

Next, to calculate the LIF, we may start from the expression of $\mathcal{T}_{W_{n}}(\epsilon, \boldsymbol{y}_{t})$,
as given in Remark \ref{Remark:13}, and proceed as in the calculation of the PIF. 
Alternatively, we may also obtain the LIF just by substituting $\boldsymbol{d}= \boldsymbol{0}$ in the expression of PIF.
Through either way, we  get that 
\[
LIF(\boldsymbol{y}_{t}, W_{n}, H_{\boldsymbol{\theta}_{0}})
= \frac{\partial}{\partial \epsilon} \mathcal{T}_{W_{n}}(\epsilon,\boldsymbol{y}_{t})|_{\epsilon=0} = 0.
\]
Further, the derivative of $\mathcal{T}_{W_{n}}(\epsilon, \boldsymbol{y_{t}})$ of any order with respect to $\epsilon$ 
will also be zero at $\epsilon = 0$, which implies that the LIF of any order will be identically zero.

\newpage
\section{Additional numerical results for real data applications}
\label{App:numerical}

\begin{table}[h]
\caption{The MDPDE-based results for all possible sub-model in Small cell lung cancer data,
along with the results obtained by the competitors under the full model.}
\centering
\resizebox{1.0\textwidth}{!}{
\begin{tabular}{l| c c c c c | c c c c c c}
\hline
& optimal $\alpha$ & DIC &\textit{arm} & \textit{entry-age} & $\gamma$ & optimal $\alpha$ & DIC &\textit{arm} & \textit{entry-age} 
& $\gamma_{1}$ &$\gamma_{2}$\\\hline
&\multicolumn{5}{|c}{\textit{Exponential baseline }}&\multicolumn{6}{|c}{\textit{Weibull baseline }}\\ 
\hline
\multicolumn{4}{l}{\underline{Fully parametric MDPDE-based results}}\\ 
Estimate& 0.7 & \textbf{87.483}&0.685&-&0.034& 0.67&90.815&0.626&-&0.115&0.013\\
SE&&&(0.359)&-&(0.115)&&&(0.109)&-&(0.027)&(0.008)\\
{p-value}&&&(0.014)&-&&&&(9.29e-9)&-&&\\
Estimate&0.55&96.217&-&0.095&0.042&0.52&99.074&-&0.076&0.098&0.019\\
SE&&&-&(0.138)&(0.098)&&&-&(0.154)&(0.042)&(0.007)\\
{p-value}&&&-&(0.477)&&&&-&(0.622)&&\\
Estimate&0.59&92.632&0.931&0.046&0.026&0.57&95.385&0.882&0.021&0.101&0.006\\
SE			&&&(1.961)&(0.107)&(0.094)&&&(1.648)&(0.094)&(0.016)&(0.042)\\
{p-value}&&&(0.635)&(0.667)&&&&(0.593)& (0.823)&&\\
\hline
\hline
\multicolumn{4}{l}{\underline{Fully parametric MLE-based results ($\alpha=0$)}}\\ 
Estimate&&734.235& 0.336&0.036&0.019&&824.514&0.542&0.032&0.165&0.007\\
SE	&&&(0.597)&(0.005)&(0.002) &&&(0.202)&(0.021)&(0.002)&(0.0008)\\
{p-value}&&&(0.073)&(9.56e-12)&&&&(0.007)&(0.127)&&\\
			\hline
\multicolumn{4}{l}{\underline{Semi-parametric PLE-based results}}\\ 
Estimate &&&0.513&0.028&-&\\
SE	&&&(0.204)&(0.013)&-&\\
{p-value}&&&(0.012)&(0.029)&-&\\
			\hline
\multicolumn{4}{l}{\underline{Semi-parametric BRE-based results}}\\ 
Estimate &&&0.757&0.026&-&\\
SE	&&&(0.231)&(0.016)&-&\\
{p-value}&&&(0.001)&(0.101)&-&\\
			\hline
		\end{tabular}
	}
\label{TAB:smallcell_selectoon}
\end{table}


\begin{table}[!h]
	\caption{Estimates and SE (in parenthesis) of different parameters for Small cell ling cancer data}
	\centering	
\resizebox{0.85\textwidth}{!}{
	\begin{tabular}{c |c c c c c c c|c c}
		\hline
		Variable  &	\multicolumn{7}{c|}{Parametric MDPDE with $\alpha$} & \multicolumn{2}{c}{Semi-parametric}\\
		&  0(MLE) & 0.05 & 0.1 & 0.2 & 0.3 & 0.4 & 0.5 & PLE & BRE\\
		\hline
		\multicolumn{10}{c}{\textit{Exponential baseline}}\\
		\textit{\textbf{arm}} & 0.336&0.399&0.385&0.43&0.553&0.651&0.844&0.513&0.757\\
		& (0.597)&(0.726)&(1.056)&(1.274)&(1.560)&(1.614)&(1.764)&(0.204)&(0.231)\\
		\textit{\textbf{entry-age}} & -0.036&0.012&0.021&0.027&0.030&0.037&0.038&0.028&0.026\\
		&  (0.005)&(0.019)&(0.024)&(0.034)&(0.044)&(0.054)&(0.062)&(0.013)&(0.016)\\
		\textit{\textbf{$\gamma$}} & 0.019&0.02&0.02&0.02&0.024&0.024&0.024&-&-\\
		&  (0.002)&(0.021)&(0.035)&(0.045)&(0.058)&(0.072)&(0.087)&-&-\\
		\hline
		\multicolumn{10}{c}{\textit{Weibull baseline}}\\
		\textit{\textbf{arm}} & 0.542&0.552&0.622&0.602&0.641&0.709&0.746\\
		& (0.202) & (0.416) & (0.668) & (0.759)&(0.894)&(0.992)&(1.143)\\
		
		\textit{\textbf{entry-age}} & 0.032&0.035&0.037&0.032&0.030&0.028&0.025\\
		& (0.021)&(0.027)&(0.032)&(0.040)&(0.057)&(0.074)&(0.085)\\
		\textit{\textbf{$\gamma_{1}$}} &  0.165 & 0.149&0.134&0.122&0.119&0.114&0.106&-&-\\
		& (0.002)&(0.005)&(0.006)&(0.007)&(0.007)&(0.009)&(0.010)&-&-\\
		\textit{\textbf{$\gamma_{2}$}}  & 0.007&0.007&0.008&0.008&0.008&0.007&0.007&-&-\\
		&  (0.0008)&(0.001)&(0.0014)&(0.002)&(0.0021)&(0.003)&(0.0033)&-&-\\
		\hline
	\end{tabular}}
	\label{TAB:7}
\end{table}

%

 \begin{table}[!h]
	\caption{Estimates and SE (in parenthesis) of different parameters for Veteran data}
	\centering	
	\resizebox{1.05\textwidth}{!}{
		\begin{tabular}{c |c c c c c c c|c c}
			\hline
			Variable & 	\multicolumn{7}{c|}{Parametric MDPDE with $\alpha$} & \multicolumn{2}{c}{Semiparametric}\\
			&  0(MLE) & 0.05 & 0.1 & 0.2 & 0.3 & 0.4 & 0.5 & PLE & BRE\\
			\hline
			\multicolumn{10}{c}{\textit{Exponential baseline}}\\
			\textit{\textbf{karno}} &-0.030&-0.028&-0.024&-0.030&-0.044&-0.044&-0.045&-0.031&-0.042\\
			&(2.6e-03)&(2.62e-03)&(2.86e-03)&(3.14e-03)&(3.53e-03)&(3.98e-03)&(0.0046)&(0.0052)&(0.0065)\\
			\textit{\textbf{celltype adeno}} &-0.093&-0.225&-0.748&-0.987&-1.236&-1.293&-1.328&-1.15&-1.237\\
			&(2.77e-01)&(2.92e-01)&(3.19e-01)&(3.83e-01)&(5.46e-01)&(5.91e-03)&(0.6232)&(0.293)&(0.528)\\
			\textit{\textbf{celltype large}} &-0.311&-0.510&-0.621&-0.827&-0.952&-1.077&-1.310&-0.831&-1.027\\
			&(7.16e-01)&(3.17e-01)&(3.52e-01)&(5.50e-01)&(8.55e-01)&(8.84e-01)&(0.9422)&(0.2935)&(0.329)\\
			\textit{\textbf{celltype smallcell}} &-0.101&-0.167&-0.120&-0.193&-0.246&-0.238&-0.183&-0.438&-0.164\\
			& (2.16e-01)&(2.34e-01)&(2.68e-01)&(4.61e-01)&(5.27e-01)&(5.51e-01)&(0.608)&(0.256)&(0.285)\\
			\textit{$\gamma$} &0.033&0.023&0.026&0.033&0.022&0.026&0.032&-&-\\ &(2.31e-06)&(8.30e-06)&(8.78e-06)&(1.45e-05)&(1.50e-05)&(1.67e-05)&(1.77e-05)&-&-\\
			\hline
			\multicolumn{10}{c}{\textit{Weibull baseline}}\\
			\textit{\textbf{karno}} &-0.025&-0.027&-0.029&-0.026&-0.032&-0.036&-0.039\\ &(4.21e-03)&(5.52e-03)&(5.59e-03)&(6.48e-03)&(7.12e-03)&(8.82e-03)&(9.28e-03)\\
			\textit{\textbf{celltype adeno}} &-1.101&-1.12&-1.159&-1.144&-1.159&-1.188&-1.214\\ &(6.88e-01)&(7.23e-01)&(8.19e-01)&(8.90e-01)&(9.50e-01)&(1.00)&(1.12)\\
			\textit{\textbf{celltype large}} &-0.742&-0.723&-0.772&-0.842&-0.892&-0.954&-1.086\\ &(6.24e-01)&(6.59e-01)&(7.05e-01)&(7.42e-01)&(8.11e-01)&(9.48e-01)&(1.04)\\
			\textit{\textbf{celltype smallcell}} &-0.459&-0.423&-0.325&-0.244&-0.215&-0.190&-0.176\\ &4.55e-01&5.29e-01&6.18e-01&6.99e-01&7.58e-01&8.11e-01&9.07e-01\\
			\textit{$\gamma_{1}$} &0.112&0.123&0.134&0.116&0.185&0.134&0.123&-&-\\ &(4.75e-03)&(5.62e-03)&(6.18e-03)&(7.24e-03)&(8.29e-03)&(8.90e-03)&(9.59e-03)&-&-\\
			\textit{$\gamma_{2}$} &0.058&0.054&0.013&0.010&0.006&0.008&0.009&-&-\\ &(2.75e-05)&(3.49e-05)&(4.23e-05)&(5.59e-05)&(6.42e-05)&(7.71e-05)&(8.88e-05)&-&-\\
			\hline
	\end{tabular}}
	\label{TAB:9}
\end{table}

\begin{table}[h]
	\caption{The MDPDE-based results for all possible sub-model in Veteran administration lung cancer data,
		along with the results obtained by the competitors under the full model.}
	\centering
	\resizebox{1.1\textwidth}{!}{	
		\begin{tabular}{l|ccccccc|cccccccc}
			\hline
			& optimal $\alpha$&DIC&\textit{karno}&\textit{adeno}&\textit{large}&\textit{smallcell}&$\gamma$& optimal $\alpha$&DIC&\textit{karno}&\textit{adeno}&\textit{large}&\textit{smallcell}&$\gamma_{1}$&$\gamma_{2}$\\
			\hline
			&\multicolumn{7}{|c}{\textit{Exponential baseline }}&\multicolumn{8}{|c}{\textit{Weibull baseline }}\\ 
			\hline
			\multicolumn{4}{l}{\underline{Fully parametric MDPDE-based results}}\\
			
			Estimate&0.56&92.368&-0.167&-&-&-&0.051&0.52&87.348&-0.089&-&-&-&0.021&0.469\\
			SE&&&(0.019)&-&-&-&(0.008)&&&(0.047)&-&-&-&(0.006)&(0.287)\\
			{p-value}&&&(0.007)&-&-&-&&&&(0.167)&-&-&-&&\\
			
			Estimate&0.48&114.975&-&0.260&-&-&0.047&0.47&111.837&-&0.179&-&-&0.107&0.384\\
			SE&&&-&(0.248)&-&-&(0.0096)&&&-&(0.143)&-&-&(0.038)&(0.485)\\
			{p-value}&&&-&(0.216)&-&-&&&&-&(0.196)&-&-&&\\
			
			Estimate&0.5&105.945&-&-&-0.767&-&0.069&0.48&104.982&-&-&-0.015&-&0.028&0.237\\
			SE&&&-&-&(0.686)&-&(0.016)&&&-&-&(0.003)&-&(0.019)&(0.315)\\
			{p-value}&&&-&-&(0.238)&-&&&&-&-&(0.018)&-&&\\
			
			Estimate&0.55&111.837&-&-&-&0.656&0.074&0.49&103.784&-&-&-&0.439&0.076&0.317\\
			SE&&&-&-&-&(0.449)&(0.021)&&&-&-&-&(0.382)&(1.2e-3)&(0.093)\\
			{p-value}&&&-&-&-&(0.319)&&&&-&-&-&(0.537)&&\\
			
			Estimate&0.49&100.19&-0.206&0.714&-&-&0.106&0.45&94.997&-0.158&0.596&-&-&0.085&0.523\\
			SE&&&(0.056)&(0.564)&-&-&(0.009)&&&(0.008)&(0.431)&-&-&(1.5e-4)&(0.005)\\
			{p-value}&&&(0.004)&(0.427)&-&-&&&&(4.2e-4)&(0.397)&-&-&&\\
			
			Estimate&0.54&90.462&-0.108&-&-0.549&-&0.048&0.56&90.089&-0.184&-&-0.462&-&0.131&0.452\\
			SE&&&(0.008)&-&(0.468)&-&(0.007)&&&(0.027)&-&(0.233)&-&(0.019)&(0.284)\\
			{p-value}&&&(1.15e-4)&-&(0.349)&-&&&&(0.002)&-&(0.259)&-&&\\
			
			Estimate&0.62&\textbf{86.498}&-0.096&-&-&0.469&0.037&0.64&90.882&-0.074&-&-&0.325&0.045&0.829\\
			SE&&&(0.028)&-&-&(0.295)&(0.019)&&&(0.017)&-&-&(0.313)&(1.8e-4)&(1.76e-3)\\
			{p-value}&&&(0.019)&-&-&(0.418)&&&&(0.036)&-&-&(0.529)&&\\
			
			Estimate&0.48&112.382&-&-0.196&-0.506&-&0.089&0.51&102.498&-&-0.086&-0.395&-&0.14&0.681\\
			SE&&&-&(0.088)&(0.382)&-&(0.018)&&&-&(0.022)&(0.278)&-&(0.089)&(0.191)\\
			{p-value}&&&-&(0.119)&(0.361)&-&&&&-&(0.194)&(0.483)&-&&\\
			
			Estimate&0.65&100.018&-&0.78&-&0.857&0.051&0.62&92.858&-&0.623&-&0.778&0.097&0.421\\
			SE&&&-&(0.367)&-&(0.601)&(0.021)&&&-&(0.247)&-&(0.491)&(2.1e-4)&(1.54e-3)\\
			{p-value}&&&-&(0.103)&-&(0.439)&&&&-&(0.219)&-&(0.426)&&\\
			
			Estimate&0.46&107.892&-&-&-0.486&0.431&0.039&0.46&102.38&-&-&-0.344&0.288&0.231&0.529\\
			SE&&&-&-&(0.478)&(0.371)&(0.008)&&&-&-&(0.173)&(0.221)&(3.48e-4)&(2.3e-3)\\
			{p-value}&&&-&-&(0.621)&(0.594)&&&&-&-&(0.412)&(0.515)&&\\
			
			Estimate&0.52&94.589&-0.158&0.626&-0.593&-&0.037&0.55&91.995&-0.134&0.473&-0.689&-&0.062&0.498\\
			SE&&&(0.068)&(0.464)&(0.285)&-&(0.004)&&&(0.118)&(0.307)&(0.229)&-&(0.077)&(0.043)\\
			{p-value}&&&(0.098)&(0.484)&(0.037)&-&&&&(0.256)&(0.124)&(0.003)&-&&\\
			
			Estimate&0.47&97.212&-0.162&1.142&-&0.945&0.096&0.5&91.991&-0.068&1.481&-&0.822&0.1&0.372\\
			SE&&&(0.494)&(0.562)&-&(0.604)&(0.011)&&&(0.366)&(0.723)&-&(0.529)&(9.2e-4)&(0.008)\\
			{p-value}&&&(0.743)&(0.042)&-&(0.118)&&&&(0.853)&(0.041)&-&(0.120)\\
			
			Estimate&0.52&90.008&-0.113&-&-0.625&0.387&0.106&0.53&87.682&-0.074&-&-0.748&0.295&0.164&0.639\\
			SE&&&(0.126)&-&(0.486)&(0.342)&(0.008)&&&(0.039)&-&(0.394)&(0.221)&(1.76e-3)&(0.027)\\
			{p-value}&&&(0.37)&-&(0.198)&(0.258)&&&&(0.058)&-&(0.057)&(0.182)&&\\
			
			Estimate&0.51&103.427&-&0.959&-0.102&0.849&0.058&0.49&103.181&-&0.848&-0.217&0.917&0.139&0.613\\
			SE&&&-&(0.614)&(0.089)&(0.701)&(0.096)&&&-&(0.574)&(0.114)&(0.647)&(2.44e-3)&(0.072)\\
			{p-value}&&&-&(0.118)&(0.252)&(0.226)&&&&-&(0.14)&(0.057)&(0.156)&&\\
			
			Estimate&0.58&90.182&-0.062&-1.385&-1.119&-0.164&0.049&0.56&91.465&-0.041&-1.273&-1.196&-0.171&0.132&0.011\\
			SE&&&(0.007)&(1.564)&(1.698)&(0.716)&(1.41e-4)&&&(9.94e-3)&(1.231)&(1.185)&(0.998)&(9.93e-3)&(9.76e-5)\\
			{p-value}&&&(1.93e-5)&(0.376)&(0.51)&(0.448)&&&&(3.71e-5)&(0.301)&(0.313)&(0.864)\\
			
			\hline
			\hline
			\multicolumn{4}{l}{\underline{Fully parametric MLE-based results($\alpha=0$)}}\\
			Estimate&&743.944&-0.030&-0.093&-0.311&-0.101&0.033&&878.482&-0.025&-1.101&-0.742&-0.459&0.112&0.058\\
			SE&&&(2.6e-3)&(2.77e-1)&(7.16e-1)&(2.16e-1)&(0.033)&&&(4.21e-3)&(6.88e-1)&(6.24e-1)&(4.55e-1)&(4.75e-3)&(2.75e-5)\\
			{p-value}&&&(8.44e-31)&(0.737)&(0.664)&(0.640)&&&&(2.88e-9)&(0.110)&(0.234)&(0.313)\\
			\hline
			\multicolumn{4}{l}{\underline{Semi-parametric PLE-based results}}\\
			Estimate&&&-0.031&-1.15&-0.831&-0.438&-\\
			SE&&&(0.0052)&(0.293)&(0.2935)&(256)&-\\
			{p-value}&&&(2.05e-9)&(8.68e-5)&(2.49e-1)&(4.83e-3)&-\\
			\hline
			\multicolumn{4}{l}{\underline{Semi-parametric BRE-based results}}\\
			Estimate&&&-0.042&-1.237&-1.027&-0.164&-\\
			SE&&&(0.0065)&(0.528)&(0.329)&(0.285)&-\\
			{p-value}&&&(9.39e-11)&(1.92e-2)&(6.55e-1)&(1.83e-2)&-\\
			\hline
		\end{tabular}
	}
	\label{TAB:veteran_selection}
\end{table}

\begin{table}[!h]
	\caption{Estimate and SE of different parameters of Criminal recidivism data}
	\centering	
	\begin{tabular}{c |c c c c c c c|c c}
		\hline
		Variable & 	\multicolumn{7}{c|}{Parametric MDPDE with $\alpha$} & \multicolumn{2}{c}{Semiparametric}\\
		&  0(MLE) & 0.05 & 0.1 & 0.2 & 0.3 & 0.4 & 0.5 & PLE & BRE\\
		\hline
		\multicolumn{10}{c}{\textit{Exponential baseline}}\\
		\textit{\textbf{fin yes}} &-0.337&-0.122&-0.117&0.083&0.167&0.182&0.161&-0.346&-0.352\\ &(0.153)&(0.155)&(0.189)&(0.215)&(0.247)&(0.321)&(0.427)&(0.190)&(0.247)\\
		\textit{\textbf{wexp yes}} &-0.217&-0.093&-0.025&-0.011&0.005&0.037&0.060&-0.223&-0.512\\ &(0.189)&(0.266)&(0.317)&(0.379)&(0.491)&(0.621)&(0.794)&(0.209)&(0.242)\\
		\textit{\textbf{prio}} &0.083&0.022&-0.007&-0.018&-0.031&-0.049&-0.054& 0.092&0.077\\ &(0.026)&(0.032)&(0.041)&(0.067)&(0.082)&(0.131)&(0.299)&(0.028)&(0.034)\\
		\textit{$\gamma$} &0.022&0.024&0.025&0.027&0.029&0.030&0.030&-&-\\ &(0.001)&(0.001)&(0.002)&(0.002)&(0.006)&(0.007)&(0.010)&-&-\\
		\hline
		\multicolumn{10}{c}{\textit{Weibull baseline}}\\
		\textit{\textbf{fin yes}} &-0.425&-0.342&-0.218&-0.109&-0.053&0.068&0.123\\ &(0.228)&(0.245)&(0.278)&(0.301)&(0.356)&(0.435)&(0.542)\\
		\textit{\textbf{wexp yes}} &-0.472&-0.382&-0.314&-0.235&-0.172&-0.108&0.003\\ &(0.241)&(0.314)&(0.365)&(0.423)&(0.481)&(0.596)&(0.742)\\
		\textit{\textbf{prio}} &0.109&0.081&0.025&0.011&0.003&-0.001&-0.009\\ &(0.009)&(0.014)&(0.019)&(0.025)&(0.031)&(0.045)&(0.058)\\
		\textit{$\gamma_{1}$} &0.004&0.005&0.005&0.004&0.002&0.001&0.001&-&-\\ &(0.0001)&(0.0002)&(0.0003)&(0.0003)&(0.0004)&(0.0005)&(0.0007)&-&-\\
		\textit{$\gamma_{2}$} &0.478&0.512&0.493&0.0.452&0.422&0.371&0.332&-&-\\ &(0.543)&(0.589)&(0.634)&(0.698)&(0.772)&(0.856)&(0.982)&-&-\\
		\hline
	\end{tabular}
	\label{TAB:11}
\end{table}

\begin{table}[h]
	\caption{The MDPDE-based results for all possible sub-model in Criminal recidivism data,
		along with the results obtained by the competitors under the full model.}
	\centering
	\resizebox{\textwidth}{!}{	
		\begin{tabular}{l|cccccc|ccccccc}
			\hline
			& optimal $\alpha$&DIC&\textit{fin}&\textit{wexp}&\textit{prio}&$\gamma$& optimal $\alpha$&DIC&\textit{fin}&\textit{wexp}&\textit{prio}&$\gamma_{1}$&$\gamma_{2}$\\
			\hline
			&\multicolumn{6}{|c}{\textit{Exponential baseline }}&\multicolumn{7}{|c}{\textit{Weibull baseline }}\\ 
			\hline
			\multicolumn{4}{l}{\underline{Fully parametric MDPDE-based results}}\\
			
			Estimate& 0.47&189.85&-0.285&-&-&0.048&0.48&181.216&-0.316&-&-&0.009&0.128\\
			SE&&&(0.197)&-&-&(0.056)&&&(0.372)&-&-&(0.001)&(0.215)\\
			{p-value}&&&(0.148)&-&-&&&&(0.396)&-&-&&\\
			
			Estimate& 0.53&154.496&-&-0.82&-&0.057&0.49&167.842&-&-0.057&-&0.012&0.212\\
			SE&&&-&(0.464)&-&(0.081)&&&-&(0.024)&-&(0.008)&(0.271)\\
			{p-value}&&&-&(0.077)&-&&&&-&(0.018)&-&&\\
			
			Estimate& 0.57&180.809&-&-&0.096&0.082&0.54&162.71&-&-&-0.026&0.024&0.085\\
			SE&&&-&-&(0.056)&(0.098)&&&-&-&(0.089)&(0.001)&(0.094)\\
			{p-value}&&&-&-&(0.086)&&&&-&-&(0.734)&&\\
			
			Estimate&0.55&145.834&-0.214&-0.056&-&0.049&0.61&148.423&-0.256&-0.047&-&0.008&0.137\\
			SE&&&(0.394)&(0.034)&-&(0.062)&&&(0.392)&(0.069)&-&(0.002)&(0.169)\\
			{p-value}&&&(0.066)&(0.099)&-&&&&(0.514)&(0.496)&-&&\\
			
			Estimate& 0.48&174.984&-0.286&-&0.094&0.062&0.47&166.828&-0.228&-&-0.016&0.017&0.187\\
			SE&&&(0.162)&-&(0.082)&(0.071)&&&(0.198)&-&(0.005)&(0.008)&(0.189)\\
			{p-value}&&&(0.077)&-&(0.252)&&&&(0.25)&-&(0.001)&\\
			
			Estimate& 0.52& 163.429&-&-0.659&0.917&0.037&0.51&169.398&-&-0.416&0.48&0.027&0.296\\
			SE&&&-&(0.614)&(0.351)&(0.059)&&&-&(0.011)&(0.51)&(0.004)&(0.142)\\
			{p-value}&&&-&(0.283)&(0.009)&&&&-&(5.71e-3)&(0.347)&\\
			
			Estimate&0.58&134.917&0.154&0.697&-0.102&0.038&0.57&\textbf{132.897}&0.138&0.086&-0.029&0.001&0.314\\
			SE&&&(0.516)&(0.832)&(0.343)&(0.021)&&&(0.696)&(0.865)&(0.084)&(0.001)&(0.946)\\
			{p-value}&&&(0.165)&(0.296)&(0.154)&&&&(0.178)&(0.271)&(0.094)&&\\
			
			\hline
			\hline
			\multicolumn{4}{l}{\underline{Fully parametric MLE-based results($\alpha=0$)}}\\
			
			Estimate&&694.124&-0.337&-0.217&0.083&0.022&&978.543&-0.425&-0.472&0.109&0.004&0.472\\
			SE&&&(0.153)&(0.189)&(0.026)&(0.001)&&&(0.228)&(0.241)&(0.009)&(0.0001)&(0.543)\\
			{p-value}&&&(0.023)&(0.251)&(0.002)&&&&(0.026)&(0.24)&(9.61e-29)\\
			
			\hline
			\multicolumn{4}{l}{\underline{Semi-parametric PLE-based results}}\\
			
			Estimate&&&-0.346&-0.223&0.092&-\\
			SE&&&(0.190)&(0.209)&(0.028)&-\\
			{p-value}&&&(0.069)&(0.285)&(0.001)&-\\
			
			\hline
			\multicolumn{4}{l}{\underline{Semi-parametric BRE-based results}}\\
			
			Estimate&&&-0.352&-0.512&0.077&-\\
			SE&&&(0.247)&(0.242)&(0.034)&-\\
			{p-value}&&&(0.154)&(0.034)&(0.022)&-\\
			
			\hline
		\end{tabular}	
	}
	\label{TAB:criminal_selection}
\end{table}

\cleardoublepage

\end{document}